\documentclass{article}%
\usepackage{amssymb}
\usepackage{amsmath}
\usepackage{amsfonts}
\usepackage{mathabx}
\usepackage{stmaryrd}
\usepackage{graphicx}%
\begin{document}

\title{Computations by fly-automata beyond monadic second-order logic}
\author{Bruno Courcelle and Ir\`{e}ne\ Durand\\LaBRI\thanks{ This work has been supported by the French National Research
Agency (ANR) in the IdEx Bordeaux program "Investments for the future", CPU,
ANR-10-IDEX-03-02.}, CNRS and Bordeaux \ University\\351\ Cours de la Lib\'{e}ration, 33405\ Talence, France\\courcell@labri.fr ; idurand@labri.fr}
\maketitle

\begin{abstract}
The validity of a monadic-second order (MS) expressible property can be
checked in linear time on graphs of bounded tree-width or clique-width given
with appropriate decompositions. This result is proved by constructing from
the MS sentence expressing the property and an integer that bounds the
tree-width or clique-width of the input graph, a finite automaton intended to
run bottom-up on the algebraic term representing a decomposition of the input
graph.\ As we cannot construct practically the transition tables of these
automata because they are huge, we use \emph{fly-automata} whose states and
transitions are computed "on the fly", only when needed for a particular
input.\ Furthermore, we allow infinite sets of states and we equip automata
with \emph{output functions}.\ Thus, they can check properties that are
\emph{not} MS expressible and compute values, for an example, the number of
$p$-colorings of a graph. We obtain XP and FPT\ graph algorithms,
parameterized by tree-width or clique-width. We show how to construct
easily\ such algorithms by combining predefined automata for basic functions
and properties. These combinations reflect the structure of the MS\ formula
that specifies the property to check or the function to compute.


\end{abstract}

\section*{Introduction}

\bigskip

Fixed-parameter tractable (FPT) algorithms can be built by many
techniques.\ In their recent book \cite{DF2}, Downey and Fellows distinguish
"elementary techniques" (bounded search trees, kernelization, color coding,
iterative compression etc.) and techniques based on well quasi-orders from
those "based on graph structure".\ The notion of graph structure includes the
graph decompositions from which tree-width, path-width, local tree-width,
clique-width etc. are defined. A central result is the following
\emph{algorithmic meta-theorem} \cite{CouEng,DF,DF2,FG}:

\begin{quote}
\emph{The validity of a monadic-second order (MS) expressible property can be
checked in linear time on graphs of bounded tree-width and on graphs of
bounded clique-width given with appropriate decompositions.}
\end{quote}

As in \cite{GroKre,Kre}, we call it a meta-theorem because it applies in a
uniform way to all graph properties expressed by MS\ sentences.\ An easy proof
of this result consists in defining by an algorithm a finite automaton
intended to run bottom-up on the labelled tree or the algebraic term that
represents the structure of the input graph (\cite{CouEng}, Chapter 6).\ This
automaton is built from the MS sentence expressing the property and an
integer, say $k$, that bounds the tree-width or clique-width of the input
graph.\ The number of states is a tower of exponentials of height essentially
equal to the number of quantifier alternations of the considered sentence,
with the bound $k$ at the top.\ In most cases we cannot construct practically
the sets of states and the transition tables of these automata.\ This obstacle
is intrinsic \cite{FriGro}, it is not due to the choice of finite automata to
implement the meta-theorem.

However, we can remedy this problem in many significant cases by using
\emph{fly-automata} (introduced in \cite{BCID}).\ They are automata whose
states and transitions are computed "on the fly", only when needed for a
particular input\footnote{Fly-automata are useful when the number of states is
large compared to the number of function symbols and the size of input terms.
\emph{Symbolic automata} \cite{VeaBjo} are defined for the case when the
number of states is manageable but the set of function symbols is very large,
and possibly infinite. In these automata, states are listed but function
symbols and transitions are described by logical formulas. Fly-automata also
admit infinite sets of function symbols.}.\ A deterministic fly-automaton
$\mathcal{A}$ having 2$^{1000}$ states only computes 100\ states on a tree or
term of size 100.\ Actually, the evaluation algorithm determines the smallest
subautomaton $\mathcal{A}\upharpoonright t$ of $\mathcal{A}$ able to process
the given term $t$.\ 

\bigskip

In this article we develop a theory of fly-automata and extend the notion
introduced in \cite{BCID}.\ In particular, we allow infinite sets of states
(e.g., a state may contain counters recording the unbounded numbers of
occurrences of particular symbols) and we equip automata with \emph{output
functions} that map the accepting states to some effective domain
$\mathcal{D}$ (e.g., the set of integers, or of pairs of integers, or the set
of words over a fixed alphabet).\ Thus, fly-automata can check properties that
are \emph{not} monadic second-order expressible, for example that a graph is
\emph{regular} (has all its vertices of same degree) and compute values, for
example, the number of $p$-colorings. We will construct fly-automata that
yield FPT and XP algorithms (definitions are reviewed in Section 1.5) for
tree-width or clique-width as parameter.\ We will combine basic fly-automata
by means of products, direct and inverse images in a way that reflects the
structure of the defining formula. For example, product of automata implements
conjunction and taking a direct image implements existential quantification.
We have implemented these constructions in the system AUTOGRAPH\footnote{See
http://dept-info.labri.u-bordeaux.fr/\symbol{126}idurand/autograph} and tested
them successfully on coloring and connectedness problems.

\bigskip

Our model-checking algorithms are intended for \emph{fixed graph properties}
and we are interested in analyzing their \emph{data complexity} formulated in
the framework of \emph{fixed-parameter tractability}.\ However, these
algorithms are constructed in uniform ways from logical expressions that cover
a large variety of problems. The constructions are easily extendable to
labelled graphs and relational structures.

\bigskip

\emph{Our computation model}.

We now motivate our choices.\ Fly-automata have several advantages: they
overcome in many significant cases the "size problem" met with usual finite
automata, they are not restricted to fixed bounds on clique-width, they allow
to check some properties that are not MS\ expressible and to compute values
attached to graphs and, last but not least, they offer a flexible framework: a
slight change in the formula that specifies the problem is quickly reflected
in the construction of a new automaton, performed by the system AUTOGRAPH.

We study fly-automata over the signature of graph operations from which
\emph{clique-width} is defined.\ We have chosen to deal with these operations
rather than\ those for \emph{tree-width} because the automata are much simpler
\cite{Cou12}. This choice also yields a gain in generality because the
clique-width of a simple graph $G$ is bounded in terms of its tree-width but
not vice-versa.\ All our FPT and XP algorithms parameterized by clique-width
are also FPT and XP\ respectively for tree-width.\ Furthermore, replacing a
graph $G$\ by its incidence graph allows to handle in our setting edge
quantifications, see the conclusion and \cite{CouLAGOS,CouLAGOSa}.

At the end of this introduction, we review methods for implementing the
verification of MS\ properties on graphs of bounded tree-width that are not
based on automata.

\bigskip

\emph{Overview of the main definitions}.

An automaton takes as input a term $t\in T(F)$ over a \emph{signature} $F$,
i.e., a set of operations, each given with a fixed arity.\ The graphs of
clique-width at most $k$ are those defined by a term over a finite signature
$F_{k}$, and $F_{\infty}$ is the union of the signatures $F_{k}$.\ We will
construct fly-automata over the infinite signature $F_{\infty}$, with which
all finite graphs can be defined.

We will construct fly-automata\ for basic properties and functions, for
example the regularity of a graph or the degree of a vertex, and we will also
use the automata constructed in \cite{BCID} for some basic MS\ properties.\ We
will combine these automata in order to define more complex properties and
functions, for example the possibility of partitioning the vertex set of a
graph into two sets inducing regular subgraphs.\ 

Here are some typical examples of decision problems and functions that we can
handle in this way:

(1) Is it possible to cover the edges of a graph with those of $s$ cliques?

(2) Does there exist an equitable $s$-coloring?\ \ \emph{Equitable} means that
the sizes of any two color classes differ by at most $1$ \cite{Fell2}. We can
express this property by $\ \exists X_{1},\ldots,X_{s}.(Partition(X_{1}%
,\ldots,X_{s})\wedge St[X_{1}]\wedge...\wedge St[X_{s}]\wedge"|X_{1}|=...=$

$|X_{i-1}|\geq|X_{i}|=...=|X_{s}|\geq|X_{1}|-1$ for some $i$") where $St[X]$
means that the induced subgraph $G[X]$ of the considered graph $G$ is
\emph{stable}, \textit{i.e.,} has no edge.\ 

(3) Assuming the graph $s$-colorable, what is the minimum size of the first
color class of an $s$-coloring?

(4) What is the minimum number of edges between $X$ and $Y$ for a partition
$(X,Y)$ of the vertex set such that $G[X]$ and $G[Y]$ are connected?

\bigskip

More generally, let \ $P(X_{1},...,X_{s})$ be a property of sets of vertices
$X_{1},...,X_{s}$ or of positions of a term; we will use\ $\overline{X}$ to
denote $(X_{1},...,X_{s})$ and $t\models P(\overline{X})$ to mean that
$\overline{X}$\ satisfies $P$ in the term $t$ or in the graph $G(t)$ defined
by $t$; this writing does not assume that $P$ is written in any particular
logical language.\ We are interested, not only to check the validity of
$\exists\overline{X}.P(\overline{X})$ in $t$ or in $G(t)$ for some given term
$t$, but also to compute (among others) the following objects associated with
$t$:

\begin{quote}
$\#\overline{X}.P(\overline{X}),$ defined as the number of assignments
$\overline{X}$ such that $t\models P(\overline{X}),$

$\mathrm{Sp}\overline{X}.P(\overline{X})$, the \emph{spectrum} of
$P(\overline{X})$, defined as the set of tuples $(|X_{1}|,\ldots,|X_{s}|)$
such that $t\models P(\overline{X})$,

$\mathrm{MSp}\overline{X}.P(\overline{X})$, the \emph{multispectrum} of
$P(\overline{X})$, defined as the multiset

\qquad\qquad of tuples $(|X_{1}|,\ldots,|X_{s}|)$ such that $t\models
P(\overline{X})$,

$\mathrm{MinCard}X_{1}.P(\overline{X})$ defined as $\min\{\left\vert
X_{1}\right\vert \mid t\models\exists X_{2},...,X_{s}.P(\overline{X})\},$

$\mathrm{Sat}\overline{X}.P(\overline{X})$, defined as the set of
tuples$\ \overline{X}$ such that $t\models P(\overline{X}).$
\end{quote}

\bigskip

We will say that the functions $\#\overline{X}.P(\overline{X}),$
$\mathrm{Sp}\overline{X}.P(\overline{X})$, $\mathrm{MSp}\overline
{X}.P(\overline{X})$, $\mathrm{MinCard}X_{1}.P(\overline{X})$%
\ and\ $\mathrm{Sat}\overline{X}.P(\overline{X})$\ taking terms as arguments
are \emph{MS expressible} if $P(\overline{X})$ is an MS expressible property.
Their values are numbers or sets of tuples of numbers in the first four cases
and our automata will give XP\ or FPT\ algorithms.\ Computing $\mathrm{Sat}%
\overline{X}.P(\overline{X})(t)$ is more difficult because the result may be
of exponential size in the size of $t$.

\bigskip

\emph{Our main results}

Here are the four main ideas and achievements. First, we recall that the state
of a deterministic bottom-up automaton collects, at each position $u$ of an
input term, some information about the subterm issued from $u$.\ This
information should be of size as small as possible so that the computation of
a run be efficient.\ An appealing situation is when the set of states is
finite, but finiteness alone does not guarantee efficient algorithms. This is
well-known for MS definable sets of terms over a finite signature: the number
of states of an automaton that implements an MS formula has a number of states
that cannot be bounded by an elementary function (an iterated exponentiation
of bounded height) in the size of the formula, see \cite{FriGro,Rei}. The
notion of \emph{fly-automaton} permits to construct usable algorithms based on
finite automata whose transitions cannot be compiled in manageable tables.

Second, as we do not insist on compiling transitions in tables, we have no
reason to insist on finiteness of the set of states.\ So we use fly-automata
whose states are integers, or pairs of integers or any information
representable by a finite word over a fixed finite alphabet. These automata
yield \emph{polynomial-time dynamic programing algorithms} if the computation
of each transition takes polynomial time in the size of the input term.

Third, fly-automata can run on terms over countably infinite signatures,
encoded in effective ways.\ In particular, we will define automata that run on
terms describing input graphs.\ These terms yield upper-bounds to the
clique-width of these graphs.\ As no finite set of such operations can
generate all graphs, the use of an infinite signature is
necessary\footnote{The corresponding constructed algorithms that are FPT (or
XP) for clique-width are immediately FPT (or XP) for tree-width.}. By
analyzing how these automata are constructed from logical descriptions, we can
understand (in part) why some algorithms constructed from automata are FPT
whereas others are\ only XP.\ 

Fourth, we go beyond MS logic in two ways.\ We adapt the classical
construction of finite automata from formulas exposed in Chapter 6 of
\cite{CouEng} and in \cite{BCID} to properties of terms and graphs that are
not MS expressible (for example the regularity of a graph) and we build
automata that compute functions (for example, the largest size of an induced
subgraph that is regular). Such properties and functions are defined by
formulas using new atomic formulas, such as $Reg[X]$ expressing that the
induced subgraph $G[X]$ of the considered graph $G$ is regular, and new
constructions, such as $\#\overline{X}.P(\overline{X})$ or $\mathrm{Sp}%
\overline{X}.P(\overline{X}),$ that can be seen as generalized
quantifications, as they bind the variables of $\overline{X}$ while delivering
more information than $\exists\overline{X}.P(\overline{X})$. From the usual
case of MS\ logic, we keep the inductive construction of an automaton based on
the structure of the defining formula.

We generalize results from \cite{ALS,CMR,CouMos} that build algorithms for
properties of terms or of graphs of bounded tree-width or clique-width that
are of the form $\exists X_{1},\ldots,X_{s}.(\varphi(X_{1},\ldots,X_{s})\wedge
R(|X_{1}|,\ldots,|X_{s}|))$ \ where $\varphi(X_{1},\ldots,X_{s})$ is an
MS\ formula and $R$ is an $s$-ary arithmetic relation that can be checked in
polynomial time.\ However, we cannot allow such atomic formulas $R(|X_{1}%
|,\ldots,|X_{s}|)$ to occur everywhere in formulas. We discuss this issue in
Section 4.6.\ 

Our automata that compute functions generalize the \emph{automata with cost
functions} of \cite{Sei} and the \emph{weighted automata} of \cite{Dro},
Chap.\ 9.\ However, these automata do not allow infinite signatures that we
use for handling graphs, and the results they prove for finite automata are
not related with our constructions.

We do not investigate here the parsing problem: graphs are given by terms over
the signature of graph operations $F_{\infty}$ from which clique-with is defined.\ 

\bigskip

To summarize, we provide logic based methods for constructing FPT and
XP\ \emph{dynamic programming graph algorithms} by means of fly-automata on
terms. The system AUTOGRAPH,\ currently under development, implements the
presented constructions.\ 

\bigskip

\emph{Alternative tools}.

There are other methods intended to overcome the "size problem" that is
unavoidable with finite automata \cite{FriGro,Rei}.\ Kneis \emph{et
al.}\ \cite{KLR,Lan} use games in the following way.\ From a graph $G$ given
with a tree-decomposition $T$ and an MS\ sentence $\varphi$ to check, they
build a \emph{model-checking game} $G(T,\varphi)$ that is actually a tree.\ An
alternating automaton running on this tree can decide if the graph $G$
satisfies $\varphi$.\ The game $G(T,\varphi)$ is of bounded size because
equivalent subgames are merged by taking into account the fact that MS
formulas of bounded quantifier height have a limited power of distinguishing
structures.\ It depends on $G$, and not only on $\varphi$ and on a bound on
its tree-width.\ This is similar to our use of fly-automata where a
subautomaton $\mathcal{A}\upharpoonright t$ of a "huge" automaton
$\mathcal{A}$ is computed for the input term $t$. (A precise comparison of the
states of $\mathcal{A}\upharpoonright t$ and these games would be interesting
but is beyond the scope of the present article.) This game approach extends to
optimization problems such as computing $\mathrm{MinCard}X_{1}.P(\overline
{X})$ or more generally, those considered in \cite{CouMos}. It has been
implemented and works for several problems on graphs with about 200\ vertices.
However, the correctness proof of the method and the programming task are by
far much more complex than those for our fly-automata.

Another proposal consists in using Datalog \cite{FK,GPW}.\ However, it seems
to be nothing but a translation of automata on terms into monadic Datalog
programs together with some manual optimization.\ It is unclear whether and
how the "size problem" is avoided. This approach is discussed in detail in
\cite{Lan}.

\bigskip

\emph{Summary of article}: Section 1 reviews notation and definitions relative
to terms, graphs and computability notions. Section 2\ reviews the definitions
concerning fly-automata. Section 3\ gives the main algorithms to build
fly-automata by transforming or combining previously constructed automata.
Section 4 details some applications to graphs.\ It gives also some direct
constructions of automata smaller than those obtained by the general
construction based on logic. Section 5 gives an overview of the software
AUTOGRAPH and reports some experiments.\ Section 6 is a conclusion. An
appendix recalls definitions concerning MS logic and establishes a technical
lemma about terms that define graphs.

Fly-automata and AUTOGRAPH have been presented in conferences; see
\cite{BCID13,BCID13a,FREC2014}.

\section{Definitions}

\bigskip

We review all necessary definitions, mostly from \cite{BCID}, and we give some
new ones. Monadic second-order logic on graphs is reviewed in the appendix.

\bigskip

\subsection{General notation}

$\bigskip$

We denote by $\mathbb{N}$ the set of natural numbers, by $\mathbb{N}_{+}$ the
set of positive ones, by $[n,m]$ the interval $\{i\mid n\leq i\leq m\}$ and by
$[n]$ the interval $[1,n]$. \ We denote by $w[i]$ the $i$-th element of a
sequence or the $i$-th letter of a word $w$. As usual, logarithms are in base
2 and $\log(x)$ stands for $\max\{1,\log_{2}(x)\}$. \emph{Countable} means
countably infinite.

The cardinality of a set $A$ is denoted by $|A|$. An encoding of a finite set
is larger than its cardinality. For example, a set of $m$ integers in $[n]$
can be encoded in size $O(m.\log(n)+1)$ by a word over a fixed finite
alphabet. We denote by $\Vert A\Vert$ the size of such an encoding.

We denote by $A-B$ the difference of two sets $A,B$ and by $B^{c}$ (the
\emph{complement} of $B$ in $A$) if $B\subseteq A$ and $A$\ is clear from the
context. We denote by $[A\rightarrow B]$ the set of mappings (i.e., of total
functions): $A\rightarrow B$. If $C\subseteq A$ and $f\in\lbrack A\rightarrow
B],$ we denote by $f\upharpoonright C$ the restriction of $f$ to $C$ and we
will consider that $[C\rightarrow C]$ is a subset of $[A\rightarrow A]$ by
identifying $h:C\rightarrow C$ with its extension $h^{\prime}$ to $A$ such
that $h^{\prime}(a):=a$ \ if \ $a\in A-C.$ However, no such identification is
needed if we represent $h$ by the set of pairs $(a,h(a))$ such that $a\in C$
and $h(a)\neq a$, because in this case the set of pairs corresponding to
$h^{\prime}$ is exactly the same. This observation yields a way to implement
$h$ if $C$ is finite and $A$ is infinite.

If $A$ is any set $\mathcal{P}(A)$, $\mathcal{P}_{f}(A)$, $\mathcal{P}_{n}%
(A)$, $\mathcal{P}_{\leq n}(A)$ denote respectively its set of subsets, of
finite subsets, of subsets of cardinality $n$ and of subsets of cardinality at
most $n$, and $\uplus$ denotes the union of disjoint sets ($B\uplus C$ is
undefined if $B$ and $C$ are not disjoint).

A \emph{multiset over} a finite or countable set $A$ is a mapping $\alpha$:
$A\rightarrow\mathbb{N}\cup\{\omega\}$ where $\alpha(a)$ is the number of
occurrences of $a\in A$ in the multiset $\alpha$.\ We denote by $\emptyset$
the empty multiset ($\alpha(a)=0$ for all $a$) and by $\sqcup$ the union of
two multisets (we have $(\alpha\sqcup\beta)(a)=\alpha(a)+\beta(a)$).\ The
cardinality of $\alpha$ is $\left\vert \alpha\right\vert :=\Sigma_{a\in
A}\alpha(a)$ and this gives a notion of finite multiset. If furthermore $A$ is
a commutative monoid with an addition $+$ and a zero $\boldsymbol{0}$, we
define $\alpha$ as \emph{finite} if $\Sigma_{a\in A-\{\boldsymbol{0}\}}%
\alpha(a)$ is finite. Then we define $\Sigma\alpha:=\Sigma_{a\in
A-\{\boldsymbol{0}\}}\alpha(a).a$.\ We have $\Sigma\emptyset=\boldsymbol{0}$
and $\Sigma(\alpha\sqcup\beta)=\Sigma\alpha+\Sigma\beta.$ Furthermore,
$\Sigma\alpha=\boldsymbol{0}$ if $\alpha$ consists of countably many
occurrences of $\boldsymbol{0}$. We denote respectively by $\mathcal{M}(A)$
and $\mathcal{M}_{f}(A)$ the sets of multisets and of finite multisets over
$A$.\ 

Let $f\in\lbrack A\rightarrow B]$ and $X\subseteq A$\ where $A$ is finite or
countable.\ We denote by $\llbracket f(x)\mid x\in X\rrbracket$ the multiset
$\beta$ over $B$ such that $\beta(b):=\left\vert f^{-1}(b)\cap X\right\vert $
and by $\{f(x)\mid x\in X\},$ or $f(X)$ as usual, the corresponding set.\ Let
for an example $A:=\{a,b,c,d\},$ $B:=\{1,2,3\}$, $f(a):=f(b):=f(c):=1$,
$f(d):=2$ and $X:=\{a,b,d\}$.\ Then\ $\{f(x)\mid x\in X\}=\{1,2\},\llbracket
f(x)\mid x\in X\rrbracket$\ is the multiset $\{1,1,2\}$ and $\Sigma
\llbracket
f(x)\mid x\in X\rrbracket=4$.

\bigskip

The set of finite words over an alphabet $Z$ is denoted by $Z^{\ast}$ and the
empty word is $\varepsilon$.

\bigskip

\emph{Terms and their syntactic trees}

A \emph{signature} $F$ is a finite or countable set of \emph{function
symbols}, each being given with a natural number called its \emph{arity}:
$\rho(f)$ denotes the arity of the symbol $f$ and $\rho(F)$ the maximal arity
of a symbol of $F,$ provided its symbols have bounded arity. We denote by
$T(F)$ the set of finite \emph{terms over} $F$ and by $\mathit{Pos}(t)$ the
set of positions of a term $t$. Each position is an occurrence of some symbol
and $\mathit{Pos}_{f}(t)$ is the set of occurrences of $f\in F$.\ Positions
are defined as Dewey words. For example, the positions of the term
$f(g(a,b),g(b,c))$ are denoted by the Dewey words $\varepsilon
,1,11,12,2,21,22$. For a term $t=h(t_{1},t_{2},t_{3})$, we have
$Pos(t)=\{\varepsilon\}\cup1.Pos(t_{1})\cup2.Pos(t_{2})\cup3.Pos(t_{3})$ where
$.$ denotes concatenation.\ (If function symbols have arity at most 9, we can
omit the concatenation marks, as in the above example). We denote by $Sig(t)$
the finite subsignature of $F$\ consisting of the symbols that have
occurrences in $t$.

The \emph{syntactic tree} of a term $t$ is a rooted, labelled and ordered tree
with set of nodes in bijection with $Pos(t)$; each node $u$ is labelled by a
symbol $f$ and has a sequence of $\rho(f)$ sons; its root denoted by
$root_{t}$ corresponds to the first position relative to the linear writing of
$t$, and the leaves to the occurrences of the nullary symbols.

We denote by $t/u$ the \emph{subterm of} $t$ \emph{issued from position} $u$
and by $Pos(t)/u$ \ the set of positions of \ $t$ below $u$ or equal to it. In
terms of Dewey words, we have $Pos(t)/u=u.Pos(t/u)$.\ Note that $Pos(t)/u\neq
Pos(t/u)$ unless $u=\varepsilon$ corresponding to the root. If $X$ is a set of
positions of $t$, then $X/u$\ denotes \ $X\cap(Pos(t)/u)$, hence is the set of
elements of $X$\ below or equal to $u$. The \emph{height} $ht(t)$ of a term
$t$ is 1 if $t$ is a nullary symbol and $1+\max\{ht(t_{1}),...,ht(t_{r})\}$ if
$t=f(t_{1},...,t_{r}).$

Let $H$ be a signature and $h:H\rightarrow F$ be an \emph{arity preserving
mapping}, i.e., a mapping such that $\rho(h(f))=\rho(f)$ for every $f\in H$.
For every $t\in T(H)$, we let $h(t)\in T(F)$ be the term obtained from $t$ by
replacing $f$ by $h(f)$ at each of its occurrences. Such a mapping is called a
\emph{relabelling}.

We denote by $|t|$ the number of positions of a term $t$. In order to discuss
algorithms taking terms as input, we must define the size of $t$. If $F$ is
finite, we can take $|t|$\ as its size.\ If $F$ is infinite, its symbols must
be encoded by words of variable length. We define the size $\Vert t\Vert$\ of
$t$ \ as the sum of lengths of the words that encode its symbols\footnote{If
$w_{x}$ $\neq\varepsilon$ encodes a symbol $x$, then $\Vert
f(g(a,b),g(b,c))\Vert=|w_{f}|+2.|w_{g}|+|w_{a}|+2.|w_{b}|+|w_{c}|.$ The length
of a LISP list implementing a term $t$ is between $\Vert t\Vert$ and $3.\Vert
t\Vert.$ (We use LISP\ to implement fly-automata, see Section 5.)}. In both
cases, we denote the size of $t$ by $\Vert t\Vert.$\ We have $|Sig(t)|\leq
|t|\leq\Vert t\Vert$. We say that an algorithm takes time $\mathrm{poly}(\Vert
t\Vert)$ if its computation time is bounded by $p(\Vert t\Vert)$\ for some
polynomial $p$ that we do not specify.

A \emph{language} is either a set of words or a set of terms.

\bigskip

\emph{F-algebras}

Let $F$ be a signature and $\mathcal{D}=\left\langle D,(f_{\mathcal{D}})_{f\in
F}\right\rangle $ be an $F$-algebra. The set $D$ is its \emph{domain}.\ We
denote by $val_{\mathcal{D}}$ the mapping: $T(F)\rightarrow D$ that yields the
\emph{value} of a term. We let then $F_{\squplus}$\ be the signature
$F\uplus\{\squplus,\mathbf{0}\}$ such that $\squplus$ is binary and
$\mathbf{0}$ is nullary.

A \emph{distributive F-algebra} is an \ $F_{\squplus}$-algebra $\mathcal{E}%
=\left\langle E,\squplus_{\mathcal{E}},\mathbf{0}_{\mathcal{E}}%
,(f_{\mathcal{E}})_{f\in F}\right\rangle $ such that $\squplus_{\mathcal{E}}%
$\ is associative and commutative with neutral element $\mathbf{0}%
_{\mathcal{E}}$, and the functions $f_{\mathcal{E}}$ satisfy the following
distributivity properties:

\begin{quote}
$f_{\mathcal{E}}(...,d\squplus _{\mathcal{E}}d^{\prime},...)=f_{\mathcal{E}%
}(...,d,...)\squplus _{\mathcal{E}}f_{\mathcal{E}}(...,d^{\prime},...),$

$f_{\mathcal{E}}(...,\mathbf{0}_{\mathcal{E}},...)=\mathbf{0}_{\mathcal{E}%
}\mathbf{.}$
\end{quote}

We extend $\squplus_{\mathcal{E}}$ to finite subsets of $E$\ by:

\begin{quote}
$\squplus_{\mathcal{E}}(A\uplus B):=(\squplus_{\mathcal{E}}%
A)\squplus_{\mathcal{E}}(\squplus_{\mathcal{E}}B)$ and

$\squplus_{\mathcal{E}}\emptyset:=\mathbf{0}_{\mathcal{E}},$
\end{quote}

and similarly for finite multisets.\ If $A$ is infinite and $g:A\rightarrow E$
is a mapping such that $g(a)\neq\mathbf{0}_{\mathcal{E}}$ for finitely many
$a\in A$, then $\squplus_{\mathcal{E}}\llbracket g(a)\mid a\in X$ and
$g(a)\neq\mathbf{0}_{\mathcal{E}}\rrbracket $ is well-defined and will be
denoted more shortly by $\squplus_{\mathcal{E}}\llbracket g(a)\mid a\in
X\rrbracket $.

The \emph{powerset algebra (of finite subsets)} \ of an $F$-algebra
$\mathcal{D}$ is\footnote{Powerset algebras are called powerset magmas in
\cite{Cou86}.}:

\begin{quote}
$\mathcal{P}_{f}(\mathcal{D}):=\left\langle \mathcal{P}_{f}(D),\cup
,\emptyset,(f_{\mathcal{P}_{f}(\mathcal{D})})_{f\in F}\right\rangle $

$f_{\mathcal{P}_{f}(\mathcal{D})}(A_{1},...,A_{r}):=\{f_{\mathcal{D}}%
(a_{1},...,a_{r})\mid a_{1}\in A_{1},...,a_{r}\in A_{r}\}.$
\end{quote}

We define also its \emph{multiset algebra (of finite multisets)}:

\begin{quote}
$\mathcal{M}_{f}(\mathcal{D}):=\left\langle \mathcal{M}_{f}(D),\sqcup
,\emptyset,(f_{\mathcal{M}_{f}(\mathcal{D})})_{f\in F}\right\rangle $ \ where

$f_{\mathcal{M}_{f}(\mathcal{D})}(\alpha_{1},...,\alpha_{r})$ is the multiset
$\beta$ such that $\beta(b)$ (the number of occurrences in $\beta$\ of $b\in
D)$ is the sum over all $r$-tuples $(a_{1},...,a_{r})$ such that $a_{1}%
\in\alpha_{1},...,a_{r}\in\alpha_{r}$ and $b=f_{\mathcal{D}}(a_{1},...,a_{r})$
of the numbers $\alpha_{1}(a_{1})\times...\times\alpha_{r}(a_{r}).$
\end{quote}

It is easy to check that $\mathcal{P}_{f}(\mathcal{D})$ \ and $\mathcal{M}%
_{f}(\mathcal{D})$ are distributive $F$-algebras.\ If $t\in T(F)$, then its
values in $\mathcal{P}_{f}(\mathcal{D})$ and in $\mathcal{M}_{f}(\mathcal{D})$
are $\{val_{\mathcal{D}}(t)\}$.

\subsection{Graphs and clique-width}

Notation and definitions are as in \cite{BCID,CouEng}. Some technical points
are developped in the appendix.

\bigskip

\emph{Graphs}

All graphs are finite, loop-free and simple (without parallel edges). A graph
$G$ is identified with the relational structure $\langle V_{G},edg_{G}\rangle$
where $edg_{G}$ is a binary relation representing the directed or undirected
adjacency. If $X\subseteq V_{G},$ we denote by $G[X]$ the induced subgraph of
$G$ with vertex set $X$, i.e., $G[X]:=\langle X,edg_{G}\cap(X\times X)\rangle
$. If $E\subseteq edg_{G},$ then $G[E]:=\langle V_{G},E\rangle$.

If $P$ is a property of graphs and $X\subseteq V_{G}$, then $P[X]$ expresses
that $G[X]$ satisfies $P$.\ A\ graph is \emph{stable} if it has no edge and we
denote this property, called \emph{stability}, by $St$.\ Hence, $St[X]$\ used
in\ the introduction says that $G[X]$ has no edge.\ 

In order to build graphs by means of graph operations, we use labels attached
to vertices. We let $L$ be a fixed countable set of \emph{port labels}. A
\emph{p-graph} (or \emph{graph with ports}) is a triple $G=\langle
V_{G},edg_{G},\pi_{G}\rangle$ where $\pi_{G}$ is a mapping: $V_{G}\rightarrow
L$. So, $\pi_{G}(x)$ is the label of $x$ and, if $\pi_{G}(x)=a,$ we say that
$x$ is an $a$-\emph{port}. If $X$ is a set of vertices, then $\pi_{G}(X)$ is
the set of its port labels. The set $\pi(G):=\pi_{G}(V_{G})$ is the
\emph{type} of $G$. \ A p-graph $G$ is identified with the relational
structure $\langle V_{G},edg_{G},(lab_{a\,G})_{a\in L}\rangle$ where $lab_{a}$
is a unary relation and $lab_{a\,G}$ is the set of $a$-ports of $G$. Since we
only consider simple graphs, two graphs or p-graphs $G$ and $H$ are isomorphic
if and only if the corresponding relational structures are isomorphic. In this
article, we will take port labels in $L:=\mathbb{N}_{+}$.

We denote by $G\approx G^{\prime}$ the fact that two p-graphs $G$ and
$G^{\prime}$ are isomorphic and by $G\simeq G^{\prime}$ that they are
isomorphic up to port labels.\ 

\bigskip

\emph{Operations on p-graphs}

We let $F_{k}$ consist of the following function symbols; they define
\emph{operations} on the p-graphs of type included in the set of port labels
$C:=[k]$ that we also define:

\begin{quote}
- the binary symbol $\oplus$ denotes the union of two \emph{disjoint} p-graphs
(i.e., $G\oplus H:=$ $\langle V_{G}\cup V_{H},edg_{G}\cup edg_{H}%
,(lab_{a\,G}\cup lab_{a\,H})_{a\in C}\rangle$ with $V_{G}\cap V_{H}%
=\emptyset),$

- the unary symbol $relab_{h}$ denotes the \emph{relabelling} that replaces in
the argument p-graph every port label $a$ by $h(a)$, where $h$ is a mapping
from $C$ to $C$ defined as a subset\footnote{For example, if $k=3$, then
$relab_{\{(1,2),(3,1)\}}=relab_{h}$ where $h(1):=2,h(2):=2$ and $h(3):=1$.\ We
denote also $relab_{\{(a,b)\}}$ by $relab_{a\rightarrow b}$. Each operation
$relab_{h}$ can be expressed as a composition of operations
$relab_{a\rightarrow b}$. See Proposition 2.118 of \cite{CouEng} for details.}
of $C\times C$, as explained in Section 1.1;

- the unary symbol $\overrightarrow{add}_{a,b}$, for $a\neq b$, denotes the
\emph{edge-addition} that adds an edge from every $a$-port $x$ to every
$b$-port $y$, unless there is already an edge $x\rightarrow y$ because graphs
are simple; this operation is idempotent,

- the nullary symbol $\mathbf{a}$, for $a\in C,$ denotes an isolated $a$-port,
and the nullary symbol $\boldsymbol{\varnothing}$ denotes the empty graph.
\end{quote}

We denote $\{\mathbf{a}\mid a\in C\}$ by $\mathbf{C}$. For constructing
undirected graphs, we use the operation $add_{a,b}$ where $a<b$ (the set $C$
is linearly ordered as it is of the form $[k]$) as an abbreviation of
$\overrightarrow{add}_{a,b}\circ\overrightarrow{add}_{b,a}$. For constructing
undirected graphs, we will use the signature $F_{k}^{\mathrm{u}}$ defined as
$F_{k}$ where the operations $\overrightarrow{add}_{a,b}$ are replaced by
$add_{a,b}$. Every operation of $F_{k}$ (resp. $F_{k}^{\mathrm{u}})$ is an
operation of $F_{k^{\prime}}$ (resp. $F_{k^{\prime}}^{\mathrm{u}})$ if
$k<k^{\prime}$ by our convention on mappings $h$ in $relab_{h}$. We let
$F_{\infty}$ (resp. $F_{\infty}^{\mathrm{u}}$) \ be the union of the
signatures $F_{k}$ (resp. $F_{k}^{\mathrm{u}})$. Hence, $F_{k}$ (resp.
$F_{k}^{\mathrm{u}})$ is the restriction of $F_{\infty}$ (resp. $F_{\infty
}^{\mathrm{u}}$) \ to the operations and constants involving labels in $[k]$.

Let $t\in T(F_{\infty})$.\ We say that a port label $a$ \emph{occurs in} $t$
if either $\mathbf{a}$, $\overrightarrow{add}_{a,b}$, $\overrightarrow
{add}_{b,a}$ or $relab_{h}$ such that\ $h(a)\neq a$ or $h(b)=a\neq b$ has an
occurrence in $t$.\ We denote by $\mu(t)$ the set of port labels that occur in
$t$ and by $\max\mu(t)$ its maximal element. We also denote by $\pi(t)$ the
set of port labels $\pi(G(t))$ and by $\max\pi(t)$ its maximal element.
Clearly, $\pi(t)\subseteq\mu(t)$.

\bigskip

\emph{Clique-width}

Every term $t$ in $T(F_{k})\cup T(F_{k}^{\mathrm{u}})$ denotes a p-graph
$G(t)$ that we now define formally. We let $Pos_{0}(t)$ be the set of
occurrences in $t$ of the symbols from $\mathbf{C}$. For each $u\in Pos(t)$,
we define a p-graph $G(t)/u$, whose vertex set is $Pos_{0}(t)/u$, the set of
leaves of $t$ below $u$ that are not occurrences of $\boldsymbol{\varnothing}%
$. The definition of $G(t)/u$ is by bottom-up induction on $u$.

\begin{quote}
If $u$ is an occurrence of $\boldsymbol{\varnothing}$, then $G(t)/u$ is the
empty graph,

if $u$ is an occurrence of $\mathbf{a}$ , then $G(t)/u$ has the unique vertex
$u$ that is an $a$-port,

if $u$ is an occurrence of $\oplus$ with sons $u_{1}$ and $u_{2}$, then
$G(t)/u:=$ $G(t)/u_{1}\oplus G(t)/u_{2}$; note that $G(t)/u_{1}$ and
$G(t)/u_{2}$ are disjoint,

if $u$ is an occurrence of $relab_{h}$ with son $u_{1}$, then
$G(t)/u:=relab_{h}(G(t)/u_{1})$,

if $u$ is an occurrence of $\overrightarrow{add}_{a,b}$ with son $u_{1}$, then
$G(t)/u:=$ $\overrightarrow{add}_{a,b}(G(t)/u_{1}),$

if $u$ is an occurrence of $add_{a,b}$ with son $u_{1}$, then $G(t)/u:=$
$add_{a,b}(G(t)/u_{1})$.
\end{quote}

\bigskip

Finally, $G(t):=G(t)/root_{t}$. Its vertex set is thus $Pos_{0}(t)$. Note the
following facts:

\begin{quote}
(1) up to port labels, $G(t)/u$ is a subgraph of $G(t)$: a port label of a
vertex of $G(t)/u$ can be modified by a relabelling occurring on the path in
$t$ from $u$ to its root;

(2) if \ $u$ and $w$ are incomparable positions under the ancestor relation,
then the graphs $G(t)/u$ and $G(t)/w$ are disjoint.
\end{quote}

If $t\in T(F_{k})\cup\,T(F_{k}^{\mathrm{u}})$, $X\subseteq Pos_{0}(t)$ and
$t^{\prime}$ is the term obtained by replacing, for each $u\in X$, the symbol
occurring there by $\boldsymbol{\varnothing}$, then $G(t^{\prime})$ is the
induced subgraph $G(t)[Pos_{0}(t)-X]$ of $G(t).$

\medskip

The \emph{clique-width} of a graph $G$, denoted by $cwd(G),$ is the least
integer $k$ such that $G\simeq G(t)$ for some term $t$ in $T(F_{k})$ (in
$T(F_{k}^{\mathrm{u}})$ if $G$ is undirected). A term $t$ in $T(F_{k})\cup
T(F_{k}^{\mathrm{u}})$ is \emph{optimal} if $k=cwd(G)$. Every graph $G$\ has
clique-width at most $|V_{G}|.\ $Two terms $t$ and $t^{\prime}$ are
\emph{equivalent}, denoted by $t\simeq t^{\prime}$, if $G(t)\simeq
G(t^{\prime}).$

\bigskip

All definitions and results stated below for $F_{k}$ and $F_{\infty}$\ apply
to $F_{k}^{\mathrm{u}}$ and $F_{\infty}^{\mathrm{u}}$.\ Let $t\in T(F_{k}).$
Each of its symbols can be encoded by a word of length $O(\log(k))$ for
$\mathbf{a}$ and $\overrightarrow{add}_{a,b}$ and $O(k.\log(k))$ for
$relab_{h}$. Hence, its size $\Vert t\Vert$ is $O(k.\log(k).|t|).$ Clearly,
$|V_{G(t)}|\leq|t|\leq\Vert t\Vert$ but $\Vert t\Vert$ is not bounded by a
function of $|V_{G(t)}|$ because a graph can be denoted by arbitrary large
terms, in particular because $\overrightarrow{add}_{a,b}$ is idempotent.\ To
avoid this, we define a term $t\in T(F_{\infty})$ as \emph{good} if, for some
$k,$ we have $t\in T(F_{k})$, $k\leq|V_{G(t)}|$ and $\ |t|\leq(k+1)^{2}%
.|V_{G(t)}|+1$. We denote by $T_{\mathrm{good}}(F_{\infty})$ the set of good
terms. In Proposition 35 of the appendix, we give an algorithm that transforms
a term $t\in T(F_{k})$ into an equivalent good term in $T(F_{k^{\prime}})$ for
some $k^{\prime}\leq k$. Its proof constructs a kind of normal form that
justifies the bound $(k+1)^{2}.|V_{G(t)}|+1$ in the definition. For example
the term $relab_{5\rightarrow1}(add_{1,9}(add_{1,8}(\boldsymbol{1}%
\oplus\boldsymbol{5}\oplus\boldsymbol{8})))$ is not good and can be replaced
by the good term $relab_{2\rightarrow1}(add_{1,3}(\boldsymbol{1}%
\oplus\boldsymbol{2}\oplus\boldsymbol{3}))$. This preprocessing takes time
$\mathrm{poly}(\Vert t\Vert).$

If $t$ is good, we have $\Vert t\Vert=O(k.\log(k).|t|)=O(k^{3}.\log
(k).|V_{G(t)}|)$ where $k=\max\mu(t).$ A computation time is this case bounded
by a polynomial in $\Vert t\Vert$\ if and only if it is by a polynomial in
$|V_{G(t)}|+\max\mu(t)$.

A term $t$ is \emph{irredundant} if, for each of its subterms of the form
$\overrightarrow{add}_{a,b}(t^{\prime})$ (or $add_{a,b}(t^{\prime})$), there
is in $G(t^{\prime})$ no edge from an $a$-port to a $b$-port (or between an
$a$-port and a $b$-port). This means that none of its operations
$\overrightarrow{add}_{a,b}$ tries to add an edge, say from $x$ to $y,$ when
there exists already one.\ The construction of several automata in Section 4
will be based on the assumption that the input terms are irredundant. The
corresponding preprocessing is considered in Proposition 35.

We do not investigate the \emph{parsing problem}, that consists, for fixed
$k$, in finding a term in $T(F_{k})$ that denotes a given graph. See however
Section 1.5.

\subsection{Sets of positions of terms and sets of vertices}

Let $E$ be a set, $X\subseteq E$ and $u\in E$. Then $[u\in X]$ denotes the
Boolean value 1 (i.e., $True$) if $u\in X$ and 0 otherwise. An $s$-tuple
$\overline{X}=(X_{1},...,X_{s})$ of subsets of $E$ can be described by the
function $\widetilde{X}:E\rightarrow\{0,1\}^{s}$ such that, for $u\in E$,
$\widetilde{X}(u)$ is the word $[u\in X_{1}]...[u\in X_{s}]$. If $\overline
{X}$\ is a partition of $E$ (a typical case is when it represents a vertex
coloring with $s$ colors of a graph $G$ and $E=V_{G}$), then $\widetilde{X}$
can be replaced by $\widehat{X}:E\rightarrow\lbrack s]$ such that
\ $\widehat{X}(u)=i$ if and only if $u\in X_{i}$. We now consider in more
detail the two cases where $E$ is the set of positions of a term and the set
of vertices of a graph defined by a term.

\bigskip

\textit{Sets of positions of terms}.

Let $F$ be a signature and $s$ be a positive integer.\ Our objective is to
encode a pair $(t,\overline{X})$ such that $t\in T(F)$ and\ $\overline{X}%
\in\mathcal{P}(Pos(t))^{s}$ by a term $t\ast\overline{X}$ $\in$ $T(F^{(s)})$
where $F^{(s)}$ is the new signature $F\times\{0,1\}^{s}$ with arity mapping
$\rho((f,w)):=\rho(f)$. We \ let $pr_{s}:F^{(s)}\rightarrow F$ \ be the
relabelling that deletes the second component of a symbol \ $(f,w)$. We denote
it by $pr$ if $s$ need not be specified.

If $t\in T(F)$ and $\overline{X}\in\mathcal{P}(Pos(t))^{s}$, then the term
$t\ast\overline{X}\in T(F^{(s)})$ is obtained from $t$ by replacing, at each
position $u$ of $t$, the symbol $f$ occurring there by $(f,\widetilde
{X}(u))\in F^{(s)}$. It is clear that $t\ast\overline{X}\in T(F^{(s)})$ and
$pr_{s}(t\ast\overline{X})=t$; we define $\nu(t\ast\overline{X}):=\overline
{X}$. Every term in $T(F^{(s)})$ is of the form $t\ast\overline{X}$ and
encodes a term $t$ in $T(F)$ and the $s$-tuple $\nu(t\ast\overline{X}%
)\in\mathcal{P}(Pos(t))^{s}$.

A property\footnote{$\overline{X}$\ abbreviates $(X_{1},...,X_{s})$\ and
$P(\overline{X})$\ stands for $P(X_{1},...,X_{s})$.} $P(X_{1},...,X_{s})$ of
sets of positions of terms over a signature $F$ is thus characterized by the
language $T_{P(\overline{X})}:=\{t\ast\overline{X}\mid t\models P(\overline
{X})\}\subseteq T(F^{(s)}).$ It can also be considered as the property
$\overline{P}$ of the terms in $T(F^{(s)})$ such that $t\ast\overline
{X}\models\overline{P}$ if and only if $\ t\models P(\overline{X})$.
Conversely, every subset of $T(F^{(s)})$ is $T_{P(\overline{X})}$ for some
property $P(\overline{X}).$ A key fact about the relabelling $pr_{s}$ is that
$T_{\exists\overline{X}.P(\overline{X})}=pr_{s}(T_{P(\overline{X})})$.

More generally (because every property is a Boolean-valued function) a
function $\alpha$ whose arguments are $t\in T(F)$ and $s$-tuples $\overline
{X}$ of positions of $t,$ and whose values are in a set $\mathcal{D}$,
corresponds to the function $\overline{\alpha}:T(F^{(s)})\rightarrow
\mathcal{D}$ such that $\overline{\alpha}(t\ast\overline{X}):=\alpha
(t,\overline{X})$.

In a situation where the tuples $\overline{X}$ are partitions of $Pos(t)$, we
can use $\widehat{X}$ instead of $\widetilde{X}$, and the signature
$F\times\lbrack s]$\ denoted by $F_{\operatorname{col}}^{(s)}$ (because of the
applications to coloring problems) instead of $F^{(s)}=F\times\{0,1\}^{s}$.

\bigskip

\textit{Sets of vertices}.

A similar technique applies to sets of vertices of graphs defined by terms in
$T(F_{\infty}).$ We first recall that the vertices are the occurrences of the
nullary symbols $\mathbf{a}$.\ We define $F_{\infty}^{(s)}$ from $F_{\infty}$
by replacing each symbol $\mathbf{a}$ by the nullary symbols\footnote{We need
not modify the operations $\overrightarrow{add}_{a,b}$ and $relab_{h}$ because
they do not create vertices.\ Hence, the notation $F_{\infty}^{(s)}$ is not an
instance of the notation $F^{(s)}$ of the previous case where $F$ is an
arbitrary signature and we want to encode sets of positions of terms in
$T(F)$. We do not set a specific notation, the context will make things
clear.} $(\mathbf{a},w)$ for all $w\in\{0,1\}^{s}.$ We define $pr:F_{\infty
}^{(s)}\rightarrow F_{\infty}$ as the mapping that deletes the sequences $w$
from nullary symbols. It extends into a relabelling $pr:T(F_{\infty}%
^{(s)})\rightarrow T(F_{\infty})$. A term $t^{\prime}$ in $T(F_{\infty}%
^{(s)})$ defines the graph $G(pr(t^{\prime}))$ and the $s$-tuple $\overline
{X}\in\mathcal{P}(V_{G(pr(t^{\prime}))})^{s}$ such that $\widetilde{X}(u)=w$
if and only if $u$ is an occurrence of $(\mathbf{a},w)$ for some $\mathbf{a}$.
The nullary symbol $(\mathbf{a},w)$ defines an isolated $a$-port together with
the information about the components of $\overline{X}$\ to which it belongs,
hence it does not define an $(a,w)$-port. The edge additions and relabellings
do not depend on the components $w$.\ They act in a term $t\in T(F_{\infty
}^{(s)})$ exactly as in the term $pr(t)\in T(F_{\infty}).$

As for sets of positions in terms, we use the notation $t\ast\overline{X}$
(where$\ t=pr(t^{\prime})$). Hence, a property \ $P(X_{1},...,X_{s})$ of sets
of vertices of $G(t)$ is characterized by the language $L_{P(X_{1},...,X_{s}%
)}:=\{t\ast\overline{X}\in T(F_{\infty}^{(s)})\mid G(t)\models P(\overline
{X})\}$. It can also be considered as the property $\overline{P}$ of terms in
$T(F_{\infty}^{(s)})$ such that $t\ast\overline{X}\models\overline{P}$ if and
only if \ $G(t)\models P(\overline{X})$.

As for terms, this definition extends to functions on graphs taking sets of
vertices as auxiliary arguments.\ For example, let $e(X_{1},X_{2})$ be the
number of undirected edges between sets $X_{1}$ and $X_{2}$\ if these sets are
disjoint and $\bot$, a special symbol that means "undefined", if $X_{1}$ and
$X_{2}$\ are not disjoint.\ It can be handled as a mapping \ $\overline
{e}:T(F_{\infty}^{\mathrm{u}(2)})\rightarrow\{\bot\}\cup\mathbb{N}$,
cf.\ Section 4.2.2.

For handling coloring problems, hence, partitions of vertex sets, we can also
use $\widehat{X}$ instead of $\widetilde{X}$, as for positions of terms (cf.
\cite{BCID}, Section 7.3.3). Hence, we can use $F_{\infty\operatorname{col}%
}^{(s)}$, where each unary symbol $\mathbf{a}$ is replaced by the symbols
$(\mathbf{a},i)$ for all $i\in\lbrack s].$ \ 

\bigskip

\emph{Set terms and substitutions of variables.}

We consider set variables $X_{1},...,$ $X_{s}$ denoting subsets of $E$, the
set of positions of a term $t\in T(F)$. A \emph{set term} over $X_{1}%
,...,X_{s}$ is a term $S$ written with them, the constant symbol
$\boldsymbol{\varnothing}$ for denoting the empty set and the operations
$\cap$, $\cup$ and $^{c}$ (for complementation).\ Hence,
$\boldsymbol{\varnothing}^{c}$ denotes $E$. An example is $S_{0}=(X_{1}\cup
X_{3}^{c})\cap(X_{2}\cup X_{5})^{c}$.

To each set term $S$\ over $X_{1},...,X_{s}$ corresponds a mapping
$\widetilde{S}:\{0,1\}^{s}\rightarrow\{0,1\}$ such that, for each $u\in E$,
$[u\in S(\overline{X})]=\widetilde{S}(\widetilde{X}(u))$ where $\overline
{X}=(X_{1},...,X_{s})$. For $S_{0}$ as above, $\widetilde{S_{0}}(w_{1}%
...w_{5})=(w_{1}\vee\lnot w_{3})\wedge\lnot(w_{2}\vee w_{5}).$ The general
definition is clear from this example.

If now $\overline{Y}=(Y_{1},...,Y_{m})$ is defined from $\overline{X}%
=(X_{1},...,X_{s})$ by $Y_{i}:=S_{i}(\overline{X})$\ for set terms
$S_{1},...,S_{m}$ over $X_{1},...,X_{s}$.\ Let $\overline{X}\in\mathcal{P}%
(Pos(t))^{s}$. \ Then $t\ast\overline{Y}=h(t\ast\overline{X})$ \ where $h$ is
the relabelling $h$:$F^{(s)}\rightarrow F^{(m)}$ that replaces, in each symbol
$(f,w),$ the word $w\in\{0,1\}^{s}$ by the word $\widetilde{S_{1}%
}(w)...\widetilde{S_{m}}(w)\in\{0,1\}^{m}$.

Let now $\alpha(Y_{1},...,Y_{m})$ be a function on terms in $T(F)$ with set
arguments $Y_{1},...,Y_{m}$ and values in a set $\mathcal{D}$. Let
$S_{1},...,S_{m}$ be set terms over $X_{1},...,X_{s}$ and $\beta(\overline
{X}):=\alpha(S_{1}(\overline{X}),...,S_{m}(\overline{X}))$.\ Hence
$\overline{\alpha}$ maps $T(F^{(m)})$ into $\mathcal{D}$ \ and $\overline
{\beta}$ maps $T(F^{(s)})$\ into $\mathcal{D}.$ We have $\overline{\beta
}=\overline{\alpha}\circ h$ where $h$:$T(F^{(s)})\rightarrow T(F^{(m)})$ is
the relabelling that encodes the tuple ($S_{1},...,S_{m}$). For an example, we
take $s:=4$, $m:=3$, $S_{1}:=X_{1}\cup X_{3}^{c}$, $S_{2}%
:=\boldsymbol{\varnothing}$, $S_{3}:=\boldsymbol{\varnothing}^{c}$. Then
$\beta(X_{1},X_{2},X_{3},X_{4})\ $defined as $\alpha(X_{1}\cup X_{3}%
^{c},\boldsymbol{\varnothing},\boldsymbol{\varnothing}^{c})$ \ satisfies the
equality $\overline{\beta}=\overline{\alpha}\circ h$ with $h$ defined by:

\begin{quote}
$h((f,x_{1}x_{2}x_{3}x_{4})):=(f,(x_{1}\vee\lnot x_{3})01)$, that is, for all
$x,y\in\{0,1\}$ and \ $f\in F$:

$h((f,1x0y)):=h((f,1x1y)):=h((f,0x0y)):=(f,101)$ and \newline%
$h((f,0x1y)):=(f,001)$.
\end{quote}

This shows that from an automaton that computes $\overline{\alpha}$, we get by
composition with the relabelling $h$ an automaton having the same states that
computes $\overline{\beta}$ (cf. Definition 4(5) in Section 2.1 below). This
technique can also be used if the terms $S_{1},...,S_{m}$ are just variables,
say $X_{i_{1}},...,X_{i_{m}}$, hence for handling a substitution of variables.
We have stated these facts for an arbitrary signature $F$.\ They hold with
obvious adaptations for the signature $F_{\infty}$.\ In this case, $t\in
T(F_{\infty})$, $E=Pos_{0}(t)=V_{G(t)}$.

\bigskip

\emph{Induced subgraphs and relativization}

Let $\alpha(X_{1},\cdots,X_{s-1})$ be a function with (vertex) set arguments
in graphs $G$ to be defined by terms.\ We define $\beta(X_{1},\cdots,X_{s})$
as\ $\alpha(X_{1}\cap X_{s},\cdots,X_{s-1}\cap X_{s})$ computed in the induced
subgraph $G[X_{s}]$. We define $h$ as the relabelling: $F_{\infty}%
^{(s)}\rightarrow F_{\infty}^{(s-1)}$ such that, for every $\mathbf{a}%
\in\mathbf{C}$ and $w\in\{0,1\}^{s-1}$, we have $h((\mathbf{a}%
,w0)):=\boldsymbol{\varnothing}$, $h((\mathbf{a},w1)):=(\mathbf{a},w)$ and
$h(f):=f$ for all other operations of $F_{\infty}$.\ With these hypotheses and
notation, we have $\overline{\beta}=\overline{\alpha}\circ h$ \ and a
corresponding transformation of automata as in the case of set terms.\ This
fact motivates the introduction of the nullary symbol $\boldsymbol{\varnothing
}$\ to denote the empty graph.

If $\alpha$\ is a property $P$ and $s=1$, we obtain a property denoted by
$P[X_{1}]$\ called the \emph{relativization} of $P$\ to $X_{1}$.\ 

\bigskip

\textit{First-order variables}

If $P(X,Y,Z)$ is a property of subsets of a set $E$, we denote by $P(X,y,Z)$
the property $P(X,\{y\},Z)$\ where $y\in E$.\ Accordingly, $\exists
y.P(X,y,Z)$\ abbreviates $\exists Y.(P(X,Y,Z)\wedge Sgl(Y))$ where $Sgl(Y)$
means that $Y$ is singleton.\ If $\alpha$ is a ternary function on
$\mathcal{P}(E)$, we let similarly $\alpha(X,y,Z)$ abbreviate $\alpha
(X,\{y\},Z)$.

\subsection{Effectively given sets}

\bigskip

A set $\mathcal{D}$ is \emph{effectively given} if it is a decidable subset of
$Z^{\ast}$ for some finite alphabet $Z$ and, furthermore, the list of its
elements is computable if it is finite.\ More precisely, such a set can be
specified either by a list of words (if not too long) or by a triple
$(Z,\mathcal{M},k)$ such that $\mathcal{M}$ is an algorithm that decides the
membership in $\mathcal{D}$ of a word in $Z^{\ast}$, $k=\omega$ if
$\mathcal{D}$ is infinite and $k\in\mathbb{N}$, $k\geq\left\vert w\right\vert
$ for every $w$ in $\mathcal{D}$ if it is finite. From $\mathcal{M}$ and $k$,
one can compute $\mathcal{D}$ whenever it is finite\footnote{In \cite{CouEng},
Definition 2.8, we take for $k$ the cardinality of $\mathcal{D}$.\ This gives
an equivalent notion but in our applications, it is easier to bound the length
of a word in $\mathcal{D}$ than to determine its exact cardinality.}. Examples
of effectively given sets are $\mathbb{B}:=\{False,True\},\mathbb{N}^{k},$
$Pos(t)$ (for a term $t$, it is a set of Dewey sequences, cf.\ Section 1.2).
The set of finite graphs up to isomorphism is effectively given (the proof is
left to the reader).\ 

We get immediately the notion of a \emph{computable mapping} from an
effectively given set to another one. If $\mathcal{D}$ is effectively given,
then so are $\mathcal{D}^{s}$, $\mathcal{P}_{f}(\mathcal{D})$ and
$\mathcal{M}_{f}(\mathcal{D})$.

In many cases, an effectively given set $\mathcal{D}$ has a special element
that we call a \emph{zero}, denoted by $zero_{\mathcal{D}}$. It can be a
special symbol $\bot$ standing for an undefined value, it can be 0\ if
$\mathcal{D}=\mathbb{N}$, the empty set if $\mathcal{D}=\mathcal{P}_{f}(E)$ or
the neutral element $\mathbf{0}_{\mathcal{D}}$ if $\mathcal{D}$ is a
distributive algebra.\ A mapping $f:\mathcal{D}^{\prime}\rightarrow
\mathcal{D}$ is \emph{finite} if the set of elements $d$ of $\mathcal{D}%
^{\prime}$ such that $f(d)\neq zero_{\mathcal{D}}$ is finite.\ Then, $f$ can
be identified with the finite set $\{(d,f(d))\mid f(d)\neq zero_{\mathcal{D}%
}\}.$ If $\mathcal{D}^{\prime}$ is also effectively given, the set
$[\mathcal{D}^{\prime}\rightarrow\mathcal{D}]_{f}$ of finite mappings:
$\mathcal{D}^{\prime}\rightarrow\mathcal{D}$ is effectively given.

\bigskip

We will consider terms over \emph{finite} or\ \emph{countable signatures}
$F$\ that satisfy the following conditions:

(a) \ the set $F$ is effectively given,

(b) the arity of a symbol can be computed in constant time,

(c) its symbols have bounded arity and $\rho(F)$ denotes the maximal arity.

We will simply say that $F$ is an \emph{effectively given signature}. To
insure (b), we can begin the word that encodes a symbol by its arity. It
follows that one can check in linear time whether a labelled tree is actually
the syntactic tree of a "well-formed" term in $T(F)$.\ We will only use
relabellings: $F\rightarrow F^{\prime}$ that are computable in linear time.
Their extensions: $T(F)\rightarrow T(F^{\prime})$ are also computable in
linear time by our definition of the size of a term (cf. Section 1.1).

An $F$-algebra $\mathcal{D}$ is \emph{effectively given} if its signature and
its domain are effectively given and its operations are computable.\ The
mapping $val_{\mathcal{D}}$ is then computable.

\subsection{Parameterization}

\bigskip

We give definitions relative to \emph{parameterized complexity}
\cite{DF,DF2,FG}.

Let $F$ be a signature, for which the notion of size of a term is fixed. A
function $h:T(F)\rightarrow\mathbb{N}$ is P-\emph{bounded} if there exists a
constant $a$ such that $h(t)\leq\Vert t\Vert^{a}$ for every term $t$ in
$T(F)$.\ It is FPT-\emph{bounded} if $h(t)\leq f(Sig(t)).\Vert t\Vert^{a}$ and
XP-\emph{bounded} if $h(t)\leq f(Sig(t)).\Vert t\Vert^{g(Sig(t))}$ \ for some
fixed functions $f$ and $g$ and constant $a$. Since $\left\vert t\right\vert
\leq\Vert t\Vert\leq\left\vert t\right\vert .\ell(Sig(t))$ for some function
$\ell$, $\Vert t\Vert$\ can be replaced by $\left\vert t\right\vert $ in the
last two cases.

A function $\alpha:T(F)\rightarrow\mathcal{D}$ is \textbf{P}-\emph{computable}
(resp. \textbf{FPT}-\emph{computable}, \textbf{XP}-\emph{computable}) \ if it
has an algorithm whose computation time is P-bounded (resp. FPT-bounded,
XP-bounded).\ We use $Sig(t)$ as a parameter in the sense of parameterized
complexity. If $F$ is finite, these three notions are equivalent. If $\alpha$
is a property, we say that it is, respectively, \textbf{P}-, \textbf{FPT}- or
\textbf{XP}-\emph{decidable}.

\bigskip

We will consider graph algorithms whose inputs are given by terms $t$ over
$F_{\infty}$.\ By constructing automata, we will obtain algorithms that are
polynomial-time, FPT or XP\ for $Sig(t)$ as parameter. The size of the input
is $\Vert t\Vert.$\ If the graph is given without any defining term $t$, we
must construct such a term and we get algorithms with same parameterized time
complexity for the following reasons.

First we observe that every graph with $n$ vertices is defined by a good term
in $T(F_{n})$ where each vertex has a distinct label and no relabelling is
made.\ Such a term has size $O(n^{2}.\log(n))$ (cf. Section 1.2) and can be
constructed in polynomial time in $n$. Hence, if a function $\alpha$ on graphs
whose input is a term in $T(F_{\infty})$ is P-computable, then it is also
P-computable if the graph of interest is given without any defining term.

The situation is more complicated for FPT- and XP-computability.\ The parsing
problem, i.e., the problem of deciding if a graph has clique-width at most $k$
is \textbf{NP}-complete where $k$ part of the input \cite{Fell}. However,
finding an optimal term is not necessary. There is an algorithm that computes,
for every directed or undirected graph $G,$ a good term in $T(F_{h(cwd(G))})$
that defines this graph without being necessarily optimal (\cite{CouEng},
Proposition 6.8).\ This algorithm takes time $g(cwd(G)).|V_{G}|^{3}$ where $g$
and $h$ are fixed functions.\ It follows that an FPT or XP graph algorithm
taking as input a term in $T(F_{\infty})$ yields an equivalent FPT or XP graph
algorithm for clique-width as parameter that takes a graph as input.\ 

\section{Fly-automata}

\bigskip

\subsection{Fly-automata: definitions}

\bigskip

We review definitions from \cite{BCID} and we extend\ them by equipping
automata with output functions.

\bigskip

\textbf{Definitions 1}: \textit{Fly-automata that recognize languages.}

(a) Let $F$ be an effectively given signature. A \emph{fly-automaton}
\emph{over} $F$ (in short, an \emph{FA\ over} $F$) is a $4$-tuple
$\mathcal{A}=\langle F,Q_{\mathcal{A}},\delta_{\mathcal{A}},\mathit{Acc}%
_{\mathcal{A}}\rangle$ such that $Q_{\mathcal{A}}$ is an effectively given set
called the set of \emph{states}, $\mathit{Acc}_{\mathcal{A}}$ is a decidable
subset of $Q_{\mathcal{A}}$ called the set of \emph{accepting states,
}(equivalently, $\mathit{Acc}_{\mathcal{A}}=\alpha^{-1}(True)$ for some
computable mapping\ $\alpha:Q_{\mathcal{A}}\rightarrow\{True,False\})$, and
$\delta_{\mathcal{A}}$ is a computable function such that, for each tuple
$(f,q_{1},\dots,q_{m})$ such that $q_{1},\dots,q_{m}\in Q_{\mathcal{A}}$,
$f\in F$ and $\rho(f)=m$, $\delta_{\mathcal{A}}(f,q_{1},\dots,q_{m})$\ is a
finite (enumerated) set of states.\ The \emph{transitions} are $f[q_{1}%
,\dots,q_{m}]\rightarrow_{\mathcal{A}}q$ if and only if $q\in\delta
_{\mathcal{A}}(f,q_{1},\dots,q_{m})$. We say that $f[q_{1},\dots
,q_{m}]\rightarrow_{\mathcal{A}}q$ \ is a transition that \emph{yields} $q$.

\bigskip

Each state is a word over a finite alphabet $Z$ hence has a size defined as
the length of that word.\ Each set $\delta_{\mathcal{A}}(f,q_{1},\dots,q_{m})$
is ordered by some linear order on $Z^{\ast}$.\ We say that $\mathcal{A}$ is
\emph{finite} if $F$ and $Q_{\mathcal{A}}$ are finite. If furthermore,
$Q_{\mathcal{A}}$, $Acc_{\mathcal{A}}$ and its transitions are listed in
tables, we call $\mathcal{A}$ a \emph{table-automaton}.

\bigskip

\emph{Remark}: An infinite FA $\mathcal{A}$ is specified by a finite tuple
$\underline{\mathcal{A}}$ of programs, or in an abstract setting, of Turing
machines, that decide membership in $F$, $Q_{\mathcal{A}}$ and $\mathit{Acc}%
_{\mathcal{A}}$\ and compute $\delta_{\mathcal{A}}$ and the arity function of
$F$.\ But since one cannot decide if the function defined by a program or a
Turing machine is total on its domain, the set of such tuples $\underline
{\mathcal{A}}$ is not recursive.\ We could strengthen the definition (and make
it heavier) by requiring that each program of $\underline{\mathcal{A}}$ \ is
accompanied with a proof that it is terminating.\ This requirement will hold
for the FA we will construct because their "termination properties" will be
straightforward to prove.\ Furthermore, all transformations and combinations
of fly-automata will preserve these termination properties.

\bigskip

(b) A \emph{run} of an FA $\mathcal{A}$ on a term $t\in T(F)$ is a mapping
$r:\mathit{Pos}(t)\rightarrow Q_{\mathcal{A}}$ such that:

\begin{quote}
if $u$ is an occurrence of a function symbol $f\in F$ and $u_{1}%
,...,u_{\rho(f)}$ is the sequence of its sons, then $f[r(u_{1}),\dots
,r(u_{\rho(f)})]\rightarrow_{\mathcal{A}}r(u)$; if $\rho(f)=0$, the condition
reads $f\rightarrow_{\mathcal{A}}r(u)$.
\end{quote}

Automata are bottom-up without $\varepsilon$-transition. For state $q$,
$L(\mathcal{A},q)$ is the set of terms $t$ in $T(F)$ on which there is a run
$r$ of $\mathcal{A}$ such that $r(\mathit{root}_{t})=q$. A run $r$ on $t$ is
\emph{accepting} if $r(\mathit{root}_{t})$ is accepting. \ The language
\emph{recognized (}or\emph{ accepted} by $\mathcal{A}$) is $L(\mathcal{A}%
):=\bigcup\{L(\mathcal{A},q)\mid q\in\mathit{Acc}_{\mathcal{A}}\}\subseteq
T(F)$. A state $q$ is \emph{accessible} if $L(\mathcal{A},q)\neq\emptyset$. We
denote by $Q_{\mathcal{A}}\upharpoonright t$ the\ set of states that occur in
the runs on $t$ and on its subterms, and by $Q_{\mathcal{A}}\upharpoonright L$
the\ union of the sets $Q_{\mathcal{A}}\upharpoonright t$ for $t$ in
$L\subseteq T(F)$.

A \emph{sink} is a state $s$ such that, for every transition $f[q_{1}%
,\dots,q_{\rho(f)}]\rightarrow_{\mathcal{A}}q,$ we have $q=s$ if $q_{i}=s$ for
some $i$. If $F$ has at least one symbol of arity at least 2, an automaton has
at most one sink. A state named $Success$ (resp. $Error$) will always be an
accepting (resp. a nonaccepting) sink, but accepting (resp.\ nonaccepting)
states may be different from $Success$ (resp. from $Error$).

Unless $\mathcal{A}$ is finite, we cannot decide if a state is accessible,
hence we cannot perform on FA the classical trimming operation that removes
the inaccessible states.\ This fact raises no problem as we will see next.

\medskip

(c) \textit{Deterministic automata}. An FA $\mathcal{A}$ is
\emph{deterministic} if all sets $\delta_{\mathcal{A}}(f,q_{1},\dots,$
$q_{\rho(f)})$\ have cardinality 1, hence, "deterministic" means
\emph{deterministic} \emph{and complete}. A deterministic FA $\mathcal{A}$
has, on each term $t$, a unique run denoted by $run_{\mathcal{A},t}$ and
$q_{\mathcal{A}}(t):=run_{\mathcal{A},t}(root_{t})$.\ The mapping
$q_{\mathcal{A}}$\ is computable and the membership in $L(\mathcal{A})$ of a
term $t$ is decidable.

Every FA $\mathcal{A}$ over $F$\ can be \emph{determinized} as follows.\ For
every term $t\in T(F),$ we denote by $\mathit{run}_{\mathcal{A},t}^{\ast}$ the
mapping: $Pos(t)\rightarrow\mathcal{P}_{f}(Q_{\mathcal{A}})$ that associates
with every position $u$ the finite set of states of the form $r(root_{t/u})$
for some run $r$ on the subterm $t/u$ of $t$. If $\mathcal{A}$ is finite, then
$\mathit{run}_{\mathcal{A},t}^{\ast}=\mathit{run}_{\mathcal{B},t}$ where
$\mathcal{B}$ is its classical determinized automaton, denoted by
$\det(\mathcal{A}),$ with set of states included in $\mathcal{P}%
_{f}(Q_{\mathcal{A}})$. If $\mathcal{A}$ is infinite, we have the same
equality where $\mathcal{B}$\ is a deterministic FA with set of states
$\mathcal{P}_{f}(Q_{\mathcal{A}})$ that we denote also by $\det(\mathcal{A})$
(cf. \cite{BCID}, Proposition 45(2)).\ In both cases, the run of
$\det(\mathcal{A})$ on a term is called \emph{the determinized run} of
$\mathcal{A}$ on this term. The mapping $\mathit{run}_{\mathcal{A},t}^{\ast}$
is computable and the membership\ in $L(\mathcal{A})$ of a term in $T(F)$ is
decidable because $t\in L(\mathcal{A})$ if and only if the set\ $\mathit{run}%
_{\mathcal{A},t}^{\ast}(root_{t})$\ contains an accepting state. We define
$ndeg_{\mathcal{A}}(t)$, the \emph{nondeterminism degree of} $\mathcal{A}$
\emph{on} $t,$ as the maximal cardinality of $\mathit{run}_{\mathcal{A}%
,t}^{\ast}(u)$ for $u$ in $Pos(t)$. We have $ndeg_{\mathcal{A}}(t)\leq
|Q_{\mathcal{A}}\upharpoonright t|.$\ 

\bigskip

If $\mathcal{A}$\ is deterministic, then $\det(\mathcal{A})$ is not identical
to $\mathcal{A}$ because its accessible states are singletons $\{q\}$ such
that $q\in Q_{\mathcal{A}}.$ However, the determinized run of $\mathcal{A}$ is
isomorphic to the run $\det(\mathcal{A})$ and the two automata recognize the
same languages. It is not decidable whether an FA $\mathcal{A}$ given by a
tuple $\underline{\mathcal{A}}$ is deterministic.\ However, when we construct
an FA, we know whether it is deterministic.

\bigskip

Whether all states of an FA\ are accessible or not, does not affect the
membership algorithm: the inaccessible states never appear in any run.\ There
is no need to remove them as for table-automata, in order to get small
transition tables. The emptiness of $L(\mathcal{A})$ is semi-decidable (one
can enumerate all terms and, for each of them, check its membership in
$L(\mathcal{A})$) but undecidable (\cite{CouEng}; Proposition 3.95).

\bigskip

\textbf{Definition 2}: \textit{Fly-automata that compute functions.}

An FA \emph{with output }is a $4$-tuple $\mathcal{A}=\langle F,Q_{\mathcal{A}%
},\delta_{\mathcal{A}},\mathit{Out}_{\mathcal{A}}\rangle$\ as in Definition 1
except that the set $Acc_{\mathcal{A}}$ is replaced by a computable
\emph{output function} $Out_{\mathcal{A}}$: $Q_{\mathcal{A}}\rightarrow
\mathcal{D}$ where $\mathcal{D}$ is effectively given. If $\mathcal{A}$ is
deterministic, the \emph{function computed by} $\mathcal{A}$ is
$Comp(\mathcal{A}):T(F)\rightarrow\mathcal{D}$ such that $Comp(\mathcal{A}%
)(t):=Out_{\mathcal{A}}(\mathit{q}_{\mathcal{A}}(t))$.\ \ In the general case,
the computed function is $Comp_{nd}(\mathcal{A}):T(F)\rightarrow
\mathcal{P}_{f}(\mathcal{D})$ such that\ $Comp_{nd}(\mathcal{A}%
)(t):=\{Out_{\mathcal{A}}(q)\mid q\in\mathit{run}_{\mathcal{A},t}^{\ast
}(root_{t})\}.\ $The latter set is equal to $Comp(\mathcal{B})(t)$ where
$\mathcal{B}$ is $\det(\mathcal{A})$ equipped with the output function
$Out_{\mathcal{B}}:\mathcal{P}_{f}(Q_{\mathcal{A}})\rightarrow\mathcal{P}%
_{f}(\mathcal{D})$ such that $Out_{\mathcal{B}}(\alpha):=\{Out_{\mathcal{A}%
}(q)\mid q\in\alpha\}$. If $\mathcal{A}$ is deterministic, then $Comp_{nd}%
(\mathcal{A})(t):=\{Comp_{\mathcal{A}}(t)\}.$

\bigskip

\textbf{Examples 3}: (a) The height $ht(t)$ of a term $t$ is computable by a
deterministic FA. More generally, if $\mathcal{M}$ is an effectively given
$F$-algebra, then $val_{\mathcal{M}}$ is computable by a deterministic FA over
$F$ with set of states $M$, the identity as output function \ and transitions
$f[m_{1},...,m_{\rho(f)}]\rightarrow f_{\mathcal{M}}(m_{1},...,m_{\rho(f)}).$

(b) Let $F$ be an effectively given signature, $r:=\rho(F)$ and $f\in F$. If
$t$ $\in T(F)$, $Pos_{f}(t)$ is the set of occurrences of $f$ in $t$.\ The
function $Pos_{f}$ is computed by the following deterministic FA\ $\mathcal{A}%
_{f}$: its states are the finite sets of words over $[r]$ (denoting positions
of terms in $T(F)$). The transitions are as follows, for $q_{1},...,q_{r}%
\in\mathcal{P}_{f}([r]^{\ast}):$

\begin{quote}
$f[q_{1},...,q_{s}]\rightarrow\{\varepsilon\}\cup1.q_{1}\cup...\cup s.q_{s}$,

$f^{\prime}[q_{1},...,q_{s^{\prime}}]\rightarrow1.q_{1}\cup...\cup s^{\prime
}.q_{s^{\prime}}$ \ if $f^{\prime}\neq f$.
\end{quote}

At each position $u$ of $t$, $run_{\mathcal{A}_{f},t}(u)=Pos_{f}(t/u)$, hence
$Comp(\mathcal{A}_{f})=Pos_{f}$ if we take the identity as output function.

\bigskip

\textbf{Definitions 4}: \textit{Subautomata; products and other
transformations of automata .}

(1) \emph{Subautomata. }We say that a signature $H$ is a \emph{subsignature}
of $F$, written $H\subseteq F$, if every operation of $H$ is an operation of
$F$ with same arity. We say that an FA $\mathcal{B}$\ over $H$ is a
\emph{subautomaton} of an FA $\mathcal{A}$\ over $F$, which we denote by
$\mathcal{B}\subseteq\mathcal{A}$, if:

\begin{quote}
$H\subseteq F$, $Q_{\mathcal{B}}\subseteq Q_{\mathcal{A}}$,

$\delta_{\mathcal{B}}(f,q_{1},\dots,q_{\rho(f)})=\delta_{\mathcal{A}}%
(f,q_{1},\dots,q_{\rho(f)})\subseteq Q_{\mathcal{B}}$ if $f\in H$ and

$q_{1},\dots,q_{\rho(f)}\in Q_{\mathcal{B}}$,

and $\mathit{Acc}_{\mathcal{B}}=\mathit{Acc}_{\mathcal{A}}\cap Q_{\mathcal{B}%
}$ or $\mathit{Out}_{\mathcal{B}}=\mathit{Out}_{\mathcal{A}}\upharpoonright
Q_{\mathcal{B}}$\ $.$
\end{quote}

If $\mathcal{A}$ is deterministic then $\mathcal{B}$ is so. If $\mathcal{A}$
recognizes a language, then $L(\mathcal{B})=L(\mathcal{A})\cap T(H)$.\ If it
computes a function and is deterministic, then $Comp(\mathcal{B}%
)=Comp(\mathcal{A})\upharpoonright T(H)$; in the general case, $Comp_{nd}%
(\mathcal{B})=Comp_{nd}(\mathcal{A})\upharpoonright T(H)$. If $\mathcal{A}$ is
an FA over $F$ and $H\subseteq F,$ then $\mathcal{A}\upharpoonright H:=\langle
H,Q_{\mathcal{A}},\delta_{\mathcal{A}\upharpoonright H},\mathit{Out}%
_{\mathcal{A}}\rangle$\ where $\delta_{\mathcal{A}\upharpoonright H}$ is the
restriction of $\delta_{\mathcal{A}}$\ to the tuples $(f,q_{1},\dots
,q_{\rho(f)})$ such that $f\in H,$ is a subautomaton of $\mathcal{A}$. Its set
of states is $Q_{\mathcal{A}}$ (some states may not be accessible).

The \emph{Weak Recognizability Theorem} (\cite{CouEng}, Chapters\ 5 and 6
and\ \cite{FREC2014}) states that, for each MS\ sentence $\varphi$ expressing
a graph property and each integer $k$, one can construct a deterministic
finite automaton\ $\mathcal{A}_{\varphi,k}$ over $F_{k}$ that recognizes the
set of terms $t\in T(F_{k})$ such that $G(t)\models\varphi$.\ In \cite{BCID},
Section 7.3.1 we prove more: we construct a deterministic FA $\mathcal{A}%
_{\varphi,\infty}$ on $F_{\infty}$ that recognizes the terms $t\in
T(F_{\infty})$ such that $G(t)\models\varphi$.\ The automata $\mathcal{A}%
_{\varphi,k}$ are subautomata of $\mathcal{A}_{\varphi,\infty}$.

\bigskip

(2) \emph{Products of fly-automata}.\ Let $\mathcal{A}_{1},...,\mathcal{A}%
_{k}$ be FA over a signature $F$, and $g$ be a computable mapping from
$Q_{\mathcal{A}_{1}}\times...\times Q_{\mathcal{A}_{k}}$ to some effectively
given domain $\mathcal{D}$. We define $\mathcal{A}:=\mathcal{A}_{1}\times
_{g}...\times_{g}\mathcal{A}_{k}$ as the FA\ with set of states
$Q_{\mathcal{A}_{1}}\times...\times Q_{\mathcal{A}_{k}}$, transitions defined by:

\begin{quote}
$\delta_{\mathcal{A}}(f,\overline{q_{1}},\dots,\overline{q_{\rho(f)}}):=$

$\qquad\{(p_{1},\dots,p_{\rho(f)})\mid p_{i}\in\delta_{\mathcal{A}_{i}%
}(f,\overline{q_{1}}[i],\dots,\overline{q_{\rho(f)}}[i])$\ for each $i\}$

\qquad where $\overline{q}[i]$ is the $i$-th component of a $\rho(f)$-tuple of
states $\overline{q},$
\end{quote}

and output function defined by:

\begin{quote}
$Out_{\mathcal{A}}((p_{1},\dots,p_{k})):=g(p_{1},\dots,p_{k})$.
\end{quote}

Depending on $g$, $\mathcal{A}$ recognizes a language or defines a function.

\bigskip

(3) \emph{Output composition}. Let $\mathcal{A}$ be an FA\ with output
mapping: $Q_{\mathcal{A}}\rightarrow\mathcal{D}$ and $g$ be computable:
$\mathcal{D}\rightarrow\mathcal{D}^{\prime}$. We let $g\circ\mathcal{A}$ be
the automaton obtained from $\mathcal{A}$ by replacing $Out_{\mathcal{A}}$ by
$g\circ Out_{\mathcal{A}}$. If\ $\mathcal{A}$ \ is deterministic, then
$Comp(g\circ\mathcal{A})=g\circ Comp(\mathcal{A}).$ In the general case,
$Comp_{nd}(g\circ\mathcal{A})=\widehat{g}\circ Comp_{nd}(\mathcal{A})$ where
$\widehat{g}(\alpha):=\{g(d)\mid d\in\alpha\}.$

\bigskip

(4) \emph{Image}.\ Let $h:T(H)\rightarrow T(F)$ be a relabelling having a
\emph{computable inverse} $h^{-1}$ such that $h^{-1}(f)$ is finite for each
$f\in F$. If $L\subseteq T(H)$, then $h(L):=\{h(t)\mid t\in L\}$. If
$\mathcal{A}$ is an FA over $H$, we let $h(\mathcal{A})$ be the automaton over
$F$ obtained from $\mathcal{A}$ by replacing each transition $f[q_{1}%
,\cdots,q_{\rho(f)}]\rightarrow_{\mathcal{A}}q$ by $h(f)[q_{1},\cdots
,q_{\rho(f)}]\rightarrow q$. It is an FA\ by Proposition 45 of \cite{BCID}. We
say that $h(\mathcal{A})$ is the \emph{image of} $\mathcal{A}$ under $h$. It
is not deterministic in general, even if $\mathcal{A}$ is. We have
$h(L(\mathcal{A},q))=L(h(\mathcal{A}),q)$ for every state $q$ and, if
$\mathcal{A}$ defines a language, then $h(L(\mathcal{A}))=L(h(\mathcal{A}))$
because $h(\mathcal{A})$ has the same accepting states as $\mathcal{A}$. If
$\mathcal{A}$ computes a function, then $Comp_{nd}(h(\mathcal{A}))(t)$=
$\bigcup\{Comp_{nd}(\mathcal{A})(t^{\prime})\mid t^{\prime}\in h^{-1}(t)\}$.

\bigskip

(5) \emph{Inverse image}. Let $h:T(H)\rightarrow T(F)$ be a computable
relabelling. If $K\subseteq T(F)$, then $h^{-1}(K):=\{t\in T(H)\,\mid h(t)\in
K\}$. If $\mathcal{A}$ is an FA over $F$, we define $h^{-1}(\mathcal{A})$ as
the FA over $H$ with transitions of the form $f[q_{1},\cdots,q_{\rho
(f)}]\rightarrow q$ such that $h(f)[q_{1},\cdots,q_{\rho(f)}]\rightarrow
_{\mathcal{A}}q$. We call $h^{-1}(\mathcal{A})$ the \emph{inverse image} of
$\mathcal{A}$ under $h$ (\cite{BCID}, Definition 17(h)); it is deterministic
if $\mathcal{A}$ is so. We have $L(h^{-1}(\mathcal{A}),q)=h^{-1}%
(L(\mathcal{A},q))$ for every state $q$.\ If $\mathcal{A}$ defines a language,
then $L(h^{-1}(\mathcal{A}))=h^{-1}(L(\mathcal{A}))$. If $\mathcal{A}$
computes a function $\alpha$: $T(F)\rightarrow\mathcal{D}$, then
$h^{-1}(\mathcal{A})$ \ defines $\alpha\circ h:T(H)\rightarrow\mathcal{D}$. In
Section 1.3 we have noted that if $\alpha(Y_{1},...,Y_{m})$ is a function on
terms in $T(F)$, $S_{1},...,S_{m}$ are set terms over $X_{1},...,X_{s}$ and
$\beta(\overline{X}):=\alpha(S_{1}(\overline{X}),...,S_{m}(\overline{X}%
))$\ (with $\overline{X}=(X_{1},...,X_{s})$) \ then $\overline{\beta
}=\overline{\alpha}\circ h$ where $h$:$T(F^{(s)})\rightarrow$ $T(F^{(m)})$ is
the relabelling that encodes the tuple $(S_{1},...,S_{m})$. If $\overline
{\alpha}$ is computed by an FA\ $\mathcal{A}$, then $\overline{\beta}$ is
computed by \ $h^{-1}(\mathcal{A})$.

\bigskip

\textbf{Example 5}: \emph{The number of runs of a nondeterministic FA}.

Let $\mathcal{A}$\ be a nondeterministic FA over $F$ without output. For each
$t\in T(F),$ we define $\#_{AccRun}(t)$ as the number of accepting runs of
$\mathcal{A}$ on $t$. We will construct a deterministic FA\ $\mathcal{B}%
$\ that computes $\#_{AccRun}.$ We define it from $\det(\mathcal{A})$ in such
a way that, for each term $t$, if $q_{\det(\mathcal{A})}(t)=\{q_{1}%
,...,q_{p}\},$ \ then $q_{\mathcal{B}}(t)=\{(q_{1},m_{1}),...,(q_{p},m_{p})\}$
where $m_{i}$ is the number of runs of $\mathcal{A}$ that yield $q_{i}$ at the
root of $t$. It is convenient to consider such a state as the finite mapping
$\mu$:$Q_{\mathcal{A}}\rightarrow\mathbb{N}$ \ such that $\mu(q_{i})=m_{i}$
and $\mu(q)=0$ if $q\notin\{q_{1},...,q_{p}\}$. \ As output function, we take
$Out_{\mathcal{B}}(\mu):=\Sigma\llbracket \mu(q)\mid q\in Acc_{\mathcal{A}%
}\rrbracket $. Some typical transitions are as follows, with states handled as
finite mappings:

\begin{quote}
$a\rightarrow\mu\ $such that $\mu(q):=$ \texttt{if }$a\rightarrow
_{\mathcal{A}}q$\ \texttt{then }1\ \texttt{else }0, for each $q\in
Q_{\mathcal{A}},$

$f[\mu_{1},\mu_{2}]\rightarrow\mu$ such that $\mu(q):=\Sigma\llbracket\mu
_{1}(q_{1}).\mu_{2}(q_{2})\mid f[q_{1},q_{2}]\rightarrow_{\mathcal{A}%
}q\rrbracket,$ for each $q\in Q_{\mathcal{A}}$.
\end{quote}

The summations are over multisets and do not give the infinite value $\omega$.
If $\mathcal{A}$ has nondeterminism degree $d$ on a term $t$, then it has at
most $|t|^{d}$ runs on this term; the size of a state of $\mathcal{B}$\ is
thus $O(d^{2}.\log(|t|))$ where numbers of runs are written in binary.

In this example, we can consider that a state $q$ of $\mathcal{A}$\ at a
position $u$ is enriched with an \emph{attribute} that records information
about all the runs of $\mathcal{A}$ on the subterm issued from $u$ that reach
state $q$ at $u$.\ This information is the number of such runs. We get a
nondeterministic FA $\mathcal{A}^{\prime}$\ whose states are pairs $(q,m)$ in
$Q_{\mathcal{A}}\times\mathbb{N}_{+}.$ The FA\ $\mathcal{B}$ is then obtained
from $\det(\mathcal{A}^{\prime}).$ This observation will be developped and
formalized in Section 3.2.

\subsection{Polynomial-time fly-automata}

\bigskip

We now classify fly-automata according to their computation times.

\bigskip

\textbf{Definitions 6}: \textit{Polynomial-time fly-automata and related
notions}

A deterministic FA over a signature $F$, possibly with output, is a
\emph{polynomial-time FA} (a \emph{P-FA}) if its computation time on any term
$t\in T(F)$ is P-bounded (cf.\ Section 1.5).\ It is an \emph{FPT-FA} or an
\emph{XP-FA} if its computation time is, respectively, FPT-bounded or
XP-bounded. It is a \emph{linear FPT-FA}\ if the computation time is bounded
by $f(Sig(t)).\Vert t\Vert$ (equivalently by $f^{\prime}(Sig(t)).\left\vert
t\right\vert $) for some fixed function $f$ (or $f^{\prime}$). The first three
notions coincide if $F$ is finite.\ A deterministic FA $\mathcal{A}$ over $F$
is an XP-FA if and only if $\mathcal{A}\upharpoonright F^{\prime}$ is a P-FA
for each finite subsignature $F^{\prime}$ of $F$.\ 

\bigskip

\textbf{Lemma 7}: Let $\mathcal{A}$ be an FA over a signature $F.$

(1) If $\mathcal{A}$ is deterministic, it is a P-FA, an FPT-FA or an XP-FA if
and only if there are functions $p_{1},p_{2},p_{3}$ such that, in the run of
$\mathcal{A}$\ \ on any term $t\in T(F)$:

\begin{quote}
$p_{1}(\Vert t\Vert)$ bounds the time for computing a transition,

$p_{2}(\| t\| )$ bounds the size of a state,

$p_{3}(\Vert t\Vert)$ bounds the time for checking if a state is accepting or
for computing the output\footnote{By using $Out_{\mathcal{A}}$; it bounds also
the size of the output.},
\end{quote}

and these functions are respectively polynomials, FPT-bounded or XP-bound-ed.

\bigskip

(2) In the general case, $\det(\mathcal{A})$ is a P-FA,\ an FPT-FA or\ an
XP-FA if and only if there are functions $p_{1},...,p_{4}$ such that, in the
determinized run of $\mathcal{A}$\ \ on any term $t\in T(F)$:

\begin{quote}
$p_{1}(\Vert t\Vert)$ bounds the time for computing the next
transition\footnote{We recall from Definition 1 that the sets $\delta
_{\mathcal{A}}(f,q_{1},\dots,q_{\rho(f)})$ are linearly ordered; firing the
next transition includes recognizing that there is no next transition.},

$p_{2}(\Vert t\Vert)$ and $p_{3}(\Vert t\Vert)$ are as in (1),

$p_{4}(\Vert t\Vert)$ bounds the nondeterminism degree of $\mathcal{A}$ \ on
$t$,
\end{quote}

and these functions are respectively polynomials, FPT-bounded or XP-bound-ed.

\bigskip

\textbf{Proof}: We prove (2) that yields (1).

"Only if". If $\det(\mathcal{A})$ is a P-FA with bounding polynomial $p$
(i.e., the computation time is bounded by $p(\Vert t\Vert)$), then, one can
take $p_{i}=p$ for $i=1,...,4$.

"If". Let us conversely assume that $\mathcal{A}$ has bounding polynomials
$p_{1},...,p_{4}$.\ Let $t$ be a term of size $\Vert t\Vert=n$.\ The states of
$\det(\mathcal{A})$ on $t$ are sets of at most $p_{4}(n)$ words of length at
most $p_{2}(n)$, that we organize as trees with at most $p_{4}(n)$ branches.
Firing a transition of \ $\det(\mathcal{A})$ at an occurrence $u$ in $t$ of a
binary symbol $f$ with sons $u_{1}$ and $u_{2}$ uses the following operations:

\begin{quote}
for all states $q_{1}$ at $u_{1}$ and $q_{2}$ at $u_{2}$, we compute in time
bounded by $p_{4}(n)^{2}.p_{1}(n)$ the states of $\delta_{\mathcal{A}}%
(f,q_{1},q_{2})$ and we insert them in the already constructed tree intended
to encode the state of $\det(\mathcal{A})$ at\ $u$.\ In this way we eliminate
duplicates.\ Each insertion takes time at most $p_{2}(n)$, hence the total
time is bounded by $p_{4}(n)^{3}.(p_{1}(n)+$ $p_{2}(n)).$
\end{quote}

In time bounded by $p_{3}(n).p_{4}(n)$ we can check if the state at the root
is accepting, and in this case, we can compute the output. The case of symbols
of different arities is similar.\ As $\left\vert t\right\vert \leq n$, we can
take the polynomial $p(n):=n.(p_{1}(n)+p_{2}(n)).p_{4}(n)^{\rho(F)+1}%
+p_{3}(n).p_{4}(n)$ to bound the global computation time.

The proof yields the result for the two other types of bound. $\square$\ 

\bigskip

\textbf{Remarks and examples 8}: (1) For finding if a deterministic FA is a
P-FA, an FPT-FA or an XP-FA, the main value to examine is the maximal size of
a state, to be bounded by $p_{2}$, because in most cases, computing the output
or the state yielded by a transition is doable in polynomial time (with a
small constant exponent) in the size of the considered states.\ For an FA that
is not deterministic, we must also examine the degree of nondeterminism to be
bounded by $p_{4}$.

(2) For every MS\ formula $\varphi(\overline{X})$ with $\overline{X}%
=(X_{1},...,X_{s})$ that expresses a graph property, we can construct a linear
FPT-FA $\mathcal{A}_{\infty}$ over $F_{\infty}^{(s)}$ that recognizes the set
of terms $t\ast\overline{X}$ such that $G(t)\models\varphi(\overline{X})$. The
functions $p_{1},p_{2},p_{3}$ of Lemma 7(1) depend only on the minimum $k$
such that $t\in T(F_{k}^{(s)})$. The recognition time is thus $f(k).\left\vert
t\right\vert $ and even $f^{\prime}(k).\left\vert V_{G(t)}\right\vert $ if $t$
is a good term (cf.\ the end of Section 1.2). The function $f(k)$ may be a
polynomial or a hyper-exponential function.\ (Concrete cases are shown in
Table 20\ of \cite{BCID}.) For each $k$, $\mathcal{A}_{\infty}$ has a finite
subautomaton $\mathcal{A}_{k}$ over $F_{k}^{(s)}$ that recognizes the set
$\{t\ast\overline{X}\in T(F_{k}^{(s)})\mid G(t)\models\varphi(\overline{X}%
)\}$. We have $\mathcal{A}_{k}\subseteq\mathcal{A}_{k^{\prime}}$ if
$k<k^{\prime}$ (\cite{BCID}, Section 7.3.1).

(3) In our applications to graphs, $\rho(F)=2.$ Furthermore, the only
nondeterministic transitions are those from the nullary symbols.\ It follows
that the bound $p_{4}(n)^{3}.(p_{1}(n)+$ $p_{2}(n))$ in the proof of Lemma 7
can be replaced by $p_{4}(n)^{2}.(p_{1}(n)+$ $p_{2}(n))$.\ As global time
complexity, we get $n.(p_{1}(n)+p_{2}(n)).p_{4}(n)^{2}$ \ $+p_{3}(n).p_{4}(n)$
and, in most cases, $O(n.p_{1}(n).p_{4}(n)^{2})$.

(4) If $t\in T(F_{\infty}),$ its height, the number of vertices of $G(t)$ (it
is the number of occurrences of the nullary symbols in \textbf{C}) and the
finite sets of port labels $\pi(t)$\ and $\mu(t)$ (cf.\ Section 1.2) can be
computed by P-FA. The set of good terms is thus P-FA recognizable.

The states of the P-FA\ $\mathcal{A}_{ht}$ that computes the height are
positive integers and its transitions are such that $q_{\mathcal{A}_{ht}%
}(t)=ht(t)$. A term $t$ is \emph{uniform} if and only if any two leaves of its
syntactic tree are at the same distance to the root.\ This property is not MS
expressible.\ It is equivalent to the condition that, for every position $u$
with sons $u^{\prime}$ and $u^{\prime\prime},$ the subterms $t/u^{\prime}$ and
$t/u^{\prime\prime}$ have same height. The automaton $\mathcal{A}_{ht}$ can
thus be modified into a P-FA\ $\mathcal{A}_{\mathit{Unif}}$\ that decides
uniformity.\ Its set of states is $\mathbb{N}_{+}\cup\{Error\}$ and its
transitions are such that $q_{\mathcal{A}_{\mathit{Unif}}}(t)$ is $ht(t)$\ if
$t$ is uniform and $Error$ otherwise.

(5) The mapping $\mathrm{Sat}X.P(X)$ that associates with a term $t$ the set
of sets $X\subseteq Pos(t)$ that satisfy $P(X)$ is not P-FA computable, and
even not XP-FA computable in general for the obvious reason that its output is
not always of polynomial size (take $P(X)$ always true)\footnote{Unless
$\mathrm{Sat}X.P(X)$ is encoded in a particular compact way; here we take it
as a straight list of sets.}.

\bigskip

\textbf{Proposition 9}:\ Let $F$ be a signature.\ Every \textbf{P}-computable
(resp.\ \textbf{FPT}- computable or \textbf{XP}-computable) function $\alpha
$\ on $T(F)$ is computable by a P-FA (resp.\ by an FPT-FA or an XP-FA).

\bigskip

\textbf{Proof}: Consider the deterministic FA\ $\mathcal{A}$ over $F$ with set
of states $T(F)$ that associates with each position $u$ of the input term $t$
the state $t/u,$ i.e., the subterm of $t$ issued from $u$.\ The state at the
root is $t$ itself, and is obtained in linear time.\ We take $\alpha$\ as
output function.\ Then $\mathcal{A}$ is a P-FA (resp.\ an FPT-FA or an
XP-FA).$\ \square$

\bigskip

Hence, our three notions of FA\ may look uninteresting.\ Actually, we will be
interested by giving effective constructions of P-FA, FPT-FA and XP-FA from
logical expressions of functions and properties (possibly \emph{not
MS\ expressible}) that are computable or decidable\ in polynomial time on
graphs of bounded tree-width or clique-width. Our motivation is to obtain
uniform, flexible and implementable constructions.

All our existence proofs are effective. When we say that a function is P-FA
computable, we mean that it is computable by a P-FA that we have constructed
or that we know how to construct by an algorithm, and for which the polynomial
bound on the computation time can be proved. The same remark applies to FPT-FA
and XP-FA\ computability.\ 

\subsection{Transformations and compositions of automata}

In view of building algorithms by combining previously constructed automata,
we define and analyze several operations on automata.

\bigskip

\textbf{Proposition 10}: Let $\mathcal{A}_{1},...,\mathcal{A}_{r}$ be P-FA
that compute functions $\alpha_{1},...,\alpha_{r}:T(F)\rightarrow\mathcal{D}%
$.\ There exists a P-FA $\mathcal{A}$ that computes the function
$\mathcal{\alpha}:T(F)\rightarrow\mathcal{D}^{r}$ such that $\alpha
(t):=(\alpha_{1}(t),...,\alpha_{r}(t)).$ If $\mathcal{A}_{1},...,\mathcal{A}%
_{r}$ are FPT-FA or XP-FA, then $\mathcal{A}$ is of same type.

\bigskip

\textbf{Proof}: The product automaton $\mathcal{A}=\mathcal{A}_{1}\times
_{g}...\times_{g}\mathcal{A}_{r}$ where $g(q_{1},...,q_{r}):=(Out_{\mathcal{A}%
_{1}}(q_{1}),...,Out_{\mathcal{A}_{r}}(q_{r}))$ \ is a deterministic FA
(cf.\ Definition 4(2)) that computes $\alpha$. The computation time of
$\mathcal{A}$ on a term is the sum of the computation times of $\mathcal{A}%
_{1},...,\mathcal{A}_{r}$ on this term.\ The claimed results follow.$\square$

\bigskip

Next we consider operations defined in Definition 4 that transform single automata.

\bigskip

\textbf{Proposition 11}: Let $\mathcal{A}$ be a P-FA that computes
$\alpha:T(F)\rightarrow\mathcal{D}.$

(1) If $g$ is a \textbf{P}-computable function $\mathcal{D}\rightarrow
\mathcal{D}^{\prime}$, then, there is a P-FA\ over $F$ \ that computes
$g\circ\alpha$.

(2) Let $h:F^{\prime}\rightarrow F$ be a relabelling. There exists a P-FA over
$F^{\prime}$ that computes the mapping $\alpha\circ h:T(F^{\prime}%
)\rightarrow\mathcal{D}.$

The same implications hold for FPT-FA and XP-FA.

\bigskip

\textbf{Proof}: (1)\ The deterministic FA $g\circ\mathcal{A}$ defined from
$\mathcal{A}$ (output composition) by replacing $Out_{\mathcal{A}}$ by $g\circ
Out_{\mathcal{A}}$ computes $g\circ\alpha$. The size of an output is
polynomially bounded, hence, we get a P-FA.

(2) Immediate by the inverse image construction. Recall that $h$ is computable
in linear time (cf.\ Section 1.4).

Each class P-FA, FPT-FA and XP-FA is preserved in both cases. $\square$

\bigskip

\textbf{Proposition 12: }Let $h:F\rightarrow F^{\prime}$ be a relabelling with
a computable inverse. Let $\mathcal{A}$ be a P-FA (resp. an FPT-FA or an
XP-FA) that computes $\alpha:T(F)\rightarrow\mathcal{D}.$ The fly-automaton
$\det(h(\mathcal{A}))$ over $F^{\prime}$ is a P-FA (resp.\ an FPT-FA or an
XP-FA) if and only if the nondeterminism degree of $h(\mathcal{A})$ is
P-bounded (resp.\ FPT-bounded or\ XP-bounded) in the size of terms over
$F^{\prime}$.

\bigskip

\textbf{Proof}: Immediate consequence of the definitions and Lemma 7.
$\square$

\bigskip

In the sufficient conditions, the bounds on $ndeg_{h(\mathcal{A)}}(t)$ can be
replaced by bounds on $\left\vert Q_{\mathcal{A}}\upharpoonright
h^{-1}(t)\right\vert $, the number of states of $\mathcal{A}$ used on input
terms $t^{\prime}$ such that $h(t^{\prime})=t$, that are frequently easier to evaluate.

\bigskip

The following counter-example shows that unless \textbf{P}$\mathbf{=}%
$\textbf{NP}, there is no alternative image construction that preserves the
polynomial-time property.

\bigskip

\textbf{Counter-example 13}: There exist a finite signature $F$ and a
P-FA\ decidable property $P(X)$ of terms in $T(F^{(1)})$ such that $\exists
X.P(X)$ is not P-FA decidable unless \textbf{P}$\mathbf{=}$\textbf{NP.}

We give a sketch of proof that uses a reduction from SAT, the satisfiability
problem for propositional formulas.\ There exists a finite signature $F$ and a
\textbf{P}-decidable property $P(X)$ of terms in $T(F^{(1)})$ such that each
instance of SAT is encoded by a term $t\in T(F^{(1)})$ and each solution of
this problem corresponds to a set $X$\ of positions of $t$ that satisfies
$P(X)$. Hence $P(X)$ is P-FA\ decidable by Proposition 9.\ Since $\exists
X.P(X)$ is not \textbf{P}-decidable unless \textbf{P}$\mathbf{=}$\textbf{NP,
}it is not P-FA decidable, again by Proposition 9. $\square$

\bigskip

\textbf{Examples 14}: \textit{P-FA for cardinality and identity.}

(a) We consider the function $Card$ that associates with a set $X$ of
positions of a term $t\in T(F)$ its cardinality $|X|$. Hence, the
corresponding mapping $:T(F^{(1)})\rightarrow\mathbb{N}$ is computed by a P-FA
$\mathcal{A}_{Card(X)}$\ whose states are the natural numbers.\ The
computation time is $O(n.\log(n))$. It is $O(n)$ if we admit that the addition
of two numbers can be done in constant time.

From $\mathcal{A}_{Card(X)}$ we can construct, for each integer $p,$ a P-FA
$\mathcal{A}_{Card(X)\leq p}$ to check that $X$ has at most $p$
elements.\ However, the automata $\mathcal{A}_{Card(X)\leq p}$ can be handled
as instanciations of a unique P-FA that takes as input a term $t$, a set of
positions $X$ of this term and an integer $p$ as auxiliary input.

(b) We consider the function $\mathit{Id}$ that associates with a set $X$ of
positions of a term this set itself. The construction of a FA denoted by
$\mathcal{A}_{\mathit{Id}(X)}$\ for the function $\mathit{Id}$ is
straightforward (cf.\ Example 3(b)). Its states are sets of positions of the
input term, hence have size $O(\left\Vert t\right\Vert ^{2})$ (cf.\ Section
1.1).\ The automaton $\mathcal{A}_{\mathit{Id}(X)}$\ is a P-FA.\ It\ may look
trivial, but it will be useful for Corollary 18\ or when combined with others,
by means of Proposition 15 (see Section 4.1.1 for an example).

\section{Fly-automata for logically defined properties and functions}

We now examine if and when the transformations of automata representing
certain logical constructions preserve the classes P-FA, FPT-FA and XP-FA.
From Counter-example 13, we know that this is not the case for existential set
quantifications. We also examine in the same perspective the logic based
functions defined in the Introduction. We consider automata on general
effectively given signatures, that check properties or compute functions on
terms.\ Applications to graphs will be considered in Section 4.

\bigskip

Two functions (or properties) $\alpha$ and $\beta$ are of \emph{same type} if
they have the same number of set arguments.\ 

\bigskip

\textbf{Proposition 15}: (1) If $\mathcal{\alpha}_{1},...,\mathcal{\alpha}%
_{r}$ are P-FA computable functions of same type and $g$ is a \textbf{P}%
-computable function (or relation) of appropriate type, the function (or the
property) $g\circ(\mathcal{\alpha}_{1},...,\mathcal{\alpha}_{r})$ is P-FA
computable (or P-FA decidable).

(2) If $\mathcal{\alpha}_{1},\mathcal{\alpha}_{2}$ and $P$ are P-FA computable
functions of same type and $P$ is Boolean-valued, then the function \texttt{if
}$P$\texttt{ then} $\alpha_{1}$\texttt{ else }$\alpha_{2}$ \ is P-FA computable.\ 

(3) If $P$ and $Q$ are P-FA decidable properties of same type, then, so are
\ $\lnot P$, $P\vee Q$ and $P\wedge Q.$

(4) The same three properties hold with FPT-FA and XP-FA.\ 

\bigskip

\textbf{Proof: }Straightforward consequences of Propositions 10 and 11(1).
$\square$

\bigskip

We denote by $\alpha\upharpoonright P\wedge...\wedge Q$ the function
\texttt{if }$P\wedge...\wedge Q$\texttt{ then} $\alpha$\texttt{ else }$\bot$:
it is the restriction of $\alpha$ to its arguments that satisfy $P\wedge
...\wedge Q$ and could be written $(...(\alpha\upharpoonright
P)\upharpoonright...)\upharpoonright Q$. (The symbol $\bot$ stands for an
undefined value). We now consider substitutions of set terms and variables
(cf. Section 1.3).\ 

\bigskip

\textbf{Proposition 16}: Let $\alpha(Y_{1},...,Y_{m})$ denote a P-FA\ function
on terms in $T(F)$ with set arguments $Y_{1},...,Y_{m}$.\ Let $S_{1}%
,...,S_{m}$ be set terms over $X_{1},...,X_{s}.$ The function $\beta
(X_{1},...,X_{s}):=\alpha(S_{1},...,S_{m})$ is P-FA computable. The same holds
with FPT-FA and XP-FA.\ 

\bigskip

\textbf{Proof}: We recall from Section 1.3 that $\overline{\beta}%
=\overline{\alpha}\circ h$ \ where $h$ is a relabelling: $T(F^{(s)}%
)\rightarrow T(F^{(m)})$ that modifies only the Boolean part of each symbol.
If $\mathcal{A}$ is a P-FA\ that computes $\overline{\alpha}$, then
$\mathcal{B}:=h^{-1}(\mathcal{A})$ is a P-FA\ by Proposition 11(2) that
computes $\overline{\beta}$. The same proof works for FPT-FA and
XP-FA.\ $\square$

\bigskip

In Proposition 15, we combine functions and properties of same type.\ With the
previous proposition, we can extend it to properties and functions that are
\emph{not of same type}. For example if we need $P(X_{1},X_{2})\wedge
Q(X_{1},X_{2},X_{3}),$ we redefine $P(X_{1},X_{2})$ into $P^{\prime}%
(X_{1},X_{2},X_{3})$ that is true if and only if $P(X_{1},X_{2})$ is,
independently of $X_{3}$.\ Proposition 16\ shows how to transform an automaton
for $P(X_{1},X_{2})$ into one for $P^{\prime}(X_{1},X_{2},X_{3})$.\ Then
$P(X_{1},X_{2})\wedge Q(X_{1},X_{2},X_{3})$ is equivalent to $P^{\prime}%
(X_{1},X_{2},X_{3})\wedge Q(X_{1},X_{2},X_{3})$ and we can apply Proposition 15.

\subsection{First-order constructions}

\bigskip

Let $P$ be a property of terms $t$ taking also as argument an $s$-tuple of
sets of positions $\overline{X}=(X_{1},...,X_{s}).$\ We recall that $\exists
x_{1},...,x_{s}.P(x_{1},...,x_{s})$\ (also written $\exists\overline
{x}.P(\overline{x})$) abbreviates $\exists\overline{X}.(P(\overline{X})\wedge
Sgl(X_{1})\wedge...\wedge Sgl(X_{s})).$

We define $\mathrm{Sat}\overline{x}.P(\overline{x})(t)$ as $\{(u_{1}%
,...,u_{s})\in(Pos(t))^{s}\mid P(\{u_{1}\},...,\{u_{s}\})$ holds in term
$t\}$. This set is in bijection with $\mathrm{Sat}\overline{X}.(P(\overline
{X})\wedge Sgl(X_{1})\wedge...\wedge Sgl(X_{s}))(t).$ (The function
$\mathrm{Sat}\overline{X}.(.)$ is defined in the introduction).

If\ $\alpha(\overline{X})$\ is a function, we define $\mathrm{SetVal}%
\overline{X}.\alpha(\overline{X})(t)$ as the set of values $\alpha
(\overline{X})$\ different from $\bot$ and $\mathrm{SetVal}\overline{x}%
.\alpha(\overline{x})(t)$\ as $\mathrm{SetVal}\overline{X}.(\alpha
(\overline{X})\upharpoonright Sgl(X_{1})\wedge...\wedge Sgl(X_{s}))(t)$.

\bigskip

\textbf{Theorem 17}: (1) If $P(\overline{X})$ is a P-FA\ decidable property,
then the properties $\exists\overline{x}.P(\overline{x})$\ and $\forall
\overline{x}.P(\overline{x})$\ are P-FA decidable.

(2) If $\alpha(\overline{X})$ is a P-FA\ computable function, then the
function $\ \mathrm{SetVal}\overline{x}.\alpha(\overline{x})$ \ is P-FA computable.

(3) The same implications hold for the classes FPT-FA and XP-FA.

\bigskip

\textbf{Proof}: (1) and (3). We let $\mathcal{A}$ be a deterministic FA\ over
$F^{(s)}$ that decides $P(\overline{X}).$\ We let $\mathcal{B}_{i}$ be the
deterministic FA\ over $F^{(s)}$ \ for $Sgl(X_{i})$ with states 0,1 and
$Error_{\mathcal{B}_{i}}$ such that:

\begin{quote}
$run_{\mathcal{B}_{i},t\ast\overline{X}}(u)=0$ if $X_{i}/u=\emptyset$,

$run_{\mathcal{B}_{i},t\ast\overline{X}}(u)=1$ if $|X_{i}/u|=1$ and

$run_{\mathcal{B}_{i},t\ast\overline{X}}(u)=Error_{\mathcal{B}_{i}}$ if
$|X_{i}/u|\geq2.$
\end{quote}

There is, by Proposition 15, a deterministic FA that decides property
$P(X_{1},$\ \ \ $...,X_{s})\wedge Sgl(X_{1})\wedge...\wedge Sgl(X_{s})$. Its
set of states is $\ Q_{\mathcal{A}}\times Q_{\mathcal{B}_{1}}\times...\times
Q_{\mathcal{B}_{s}}$ and its set of accepting states is $Acc_{\mathcal{A}%
}\times\{1\}\times...\times\{1\}.$\ We build a smaller deterministic FA
$\mathcal{C}$ with set of states $\{Error_{\mathcal{C}}\}\cup((Q_{\mathcal{A}%
}-\{Error_{\mathcal{A}}\})\times\{0,1\}^{s})$ and same set of accepting states
by merging into a unique error state $Error_{\mathcal{C}}$ all tuples of
$Q_{\mathcal{A}}\times Q_{\mathcal{B}_{1}}\times...\times Q_{\mathcal{B}_{s}}%
$, one component of which is an error state.

The nondeterministic automaton $pr_{s}(\mathcal{C})$ decides the
property$\ \exists\overline{x}.P(\overline{x})$.\ Its states at a position $u$
in a term $t\in T(F)$ are $Error_{\mathcal{C}}$ or the tuples of the form
$(run_{\mathcal{A},t\ast\overline{X}}(u),|X_{1}/u|,...,|X_{s}/u|)$\ such that
$|X_{1}/u|,...,|X_{s}/u|\leq1.$ Since $\mathcal{A}$ is deterministic, there
are at most $1+(|t|+1)^{s}$ different such states and the nondeterminism
degree of $pr_{s}(\mathcal{C})$ is bounded by the polynomial $p(n)=1+(n+1)^{s}%
$ that does not depend on $Sig(t)$. \ Hence $\det(pr_{s}(\mathcal{C}))$ is a
P-FA, an FPT-FA or an XP-FA by Lemma 7 if $\mathcal{A}$ is so.

Property $\forall\overline{x}.P(\overline{x})$ can be written $\lnot
\exists\overline{x}.\lnot P(\overline{x}).$\ The results follow since, by by
Proposition 15(3,4) the classes of P-FA, FPT-FA and XP-FA that check
properties are closed under the transformation implementing negation.

(2) and (3). We apply the same construction to an FA\ $\mathcal{A}$ over
$F^{(s)}$ that computes $\alpha(\overline{X}).$ As output function for
$\mathcal{C}$, we take:

\begin{quote}
$Out_{\mathcal{C}}((q,1,...,1)):=Out_{\mathcal{A}}(q)$, for $q\in
Q_{\mathcal{A}},$

$Out_{\mathcal{C}}(p):=\bot$, \ for all other states $p$ of $\mathcal{C}$.
\end{quote}

By the definitions, $Comp(\det(pr_{s}(\mathcal{C})))$ is equal to
$\mathrm{SetVal}\overline{x}.\alpha(\overline{x})$ hence, is P-FA, or FPT-FA
or XP-FA computable by Lemma 7, depending on $\mathcal{A}$ as above.
$\ \square$

\bigskip

The construction of this proof is \emph{generic} in that it applies to
\emph{any} deterministic FA $\mathcal{A}$\ over $F^{(s)}$, even that is not of
type XP.\ The hypotheses on the type, P, FPT or XP of $\mathcal{A}$ are only
used to determine the type of the resulting automaton.

\begin{quote}

\end{quote}

\textbf{Corollary 18}: If $P(\overline{X})$ is a P-FA\ decidable property,
then the functions $\mathrm{Sat}\overline{x}.P(\overline{x})$ and
$\ \#\overline{x}.P(\overline{x})$ \ are P-FA computable. The same implication
holds with FPT-FA and XP-FA.

\bigskip

\textbf{Proof:} We observe that $\mathrm{Sat}\overline{x}.P(\overline
{x})=\mathrm{SetVal}\overline{x}.\alpha(\overline{x})$ where $\alpha
(\overline{x}):=$\ \texttt{if} $P(\overline{x})$ \texttt{then} $\overline{x}%
$\ \texttt{else} $\bot.$ The result follows then from Propositions 15(2),
Theorem 17 and a variant of $\mathcal{A}_{Id(X)}$ of Example
14(b).\ \ However, we can give a direct construction that modifies the one of
the proof of Theorem 17. We replace each $\mathcal{B}_{i}$ by $\mathcal{B}%
_{i}^{\prime}$\ such that:

\begin{quote}
$run_{\mathcal{B}_{i}^{\prime},t\ast\overline{X}}(u)=\emptyset$\ if
$X_{i}/u=\emptyset$,

$run_{\mathcal{B}_{i}^{\prime},t\ast\overline{X}}(u)=\{w\}$ \ if
$X_{i}/u=\{u.w\}$ (positions are Dewey words) and

$run_{\mathcal{B}_{i}^{\prime},t\ast\overline{X}}(u)=Error_{\mathcal{B}%
_{i}^{\prime}}$ if $|X_{i}/u|\geq2.$
\end{quote}

Then, we make the product $\mathcal{A}\times\mathcal{B}_{1}^{\prime}%
\times...\times\mathcal{B}_{s}^{\prime}$ into a deterministic automaton
$\mathcal{C}^{\prime}$ with set of states $\{Error_{\mathcal{C}^{\prime}%
}\}\cup((Q_{\mathcal{A}}-\{Error_{\mathcal{A}}\})\times\mathcal{P}_{\leq
1}([\rho(F)]^{\ast})^{s})$ similarly as in the proof of Theorem 17.\ The
deterministic automaton $\mathcal{C}^{\prime\prime}$, defined as
$\det(pr(\mathcal{C}^{\prime}))$ equipped with the output function such that
for $Z\subseteq Q_{\mathcal{C}^{\prime}}=Q_{pr(\mathcal{C}^{\prime})}$:

\begin{quote}
$Out_{\mathcal{C}^{\prime\prime}}(Z):=\{(x_{1},...,x_{s})\mid(q,\{x_{1}%
\},...,\{x_{s}\})\in Z,Acc_{\mathcal{A}}(q)=True\},\,\,(1)$
\end{quote}

defines $\mathrm{Sat}\overline{x}.P(\overline{x})$. The states of
$pr(\mathcal{C}^{\prime})$ at a position $u$ of $t$ are $Error_{\mathcal{C}%
^{\prime}}$ and tuples $(run_{\mathcal{A},t\ast(X_{1},...,X_{s})}%
(u),X_{1},...,X_{s})$ such that $|X_{1}|,...,|X_{s}|\leq1$ and $X_{1}\cup
...$\ \ $\cup X_{s}\subseteq\lbrack r]^{\ast}.$ Since $\mathcal{A}$ is
deterministic, there are at most $1+(|t|+1)^{s}$ different such states at each
position $u$.\ The nondeterminism degree of $\det(pr(\mathcal{C}^{\prime}))$
is bounded as in the proof of Theorem 17. The conclusions follow from Lemma 7.

Since the value $\#\overline{x}.P(\overline{x})$ on a term $t$ is computable
in linear time from that of $\mathrm{Sat}\overline{x}.P(\overline{x})$, we get
the corresponding assertions (by using Proposition 11(1)). However there is a
more direct construction that does not use $\mathrm{Sat}\overline
{x}.P(\overline{x})$ as an intermediate step (see below (3.2,2.4)). It is
related to (but does not coincide with) counting the number of accepting runs
of $pr(\mathcal{C}^{\prime})$, which we did in Example 5.\ $\square$

\bigskip

\textbf{Remarks\ 19}: (1) From $\mathrm{SetVal}\overline{x}.\alpha
(\overline{x})$, we can obtain in polynomial time the maximum or the minimum
value of $\alpha(\{x_{1}\},...,\{x_{s}\})$ \ if the range of $\alpha$\ \ is
linearly ordered and two values can be compared in polynomial time. The
corresponding functions are thus P-FA (or FPT-FA, or XP-FA) \ computable.
Alternative constructions will be given below.

(2) The results of Theorem 17 and Corollary 18\ remain valid if each condition
$Sgl(X_{i})$ is replaced by $Card(X_{i})=c_{i}$ or $Card(X_{i})\leq c_{i}$ for
fixed integers $c_{i}$.\ In particular, we can compute:

\begin{quote}
$\#\overline{X}.(P(\overline{X})\wedge Card(X_{1})\leq c_{1}\wedge...\wedge
Card(X_{s})\leq c_{s}).$ \ 
\end{quote}

The exponents in the bounding polynomial become larger, but they still depend
only on the numbers $c_{1},...,c_{s}.$ (The polynomial $p(n)=1+(n+1)^{s}$ in
the proof of Theorem 17(1) is replaced by $1+(n+1)^{c_{1}+...+c_{s}}$). By
Counter-example 20 below, this fact does not hold with $Card(X_{i})\geq c_{i}%
$: just take $c_{i}=0$. $\square$

\bigskip

In Theorem 17, we only handle first-order quantifications. Counter-example 13
has shown that we cannot replace them by arbitrary set quantifications. We now
give a counter-example that does not use any complexity hypothesis. \ 

\bigskip

\textbf{Counter-example 20}: We sketch a proof that the image construction for
FA\ that corresponds to an existential set quantification does not preserve
the polynomial-time property.

We consider terms over $F=\{f,g,a\}$ where $f$ is binary, $g$ is unary and $a$
is nullary. For every position $u$ of $t\in T(F)$, we let $s(u):=|Pos(t)/u|.$
For a set $X$ of positions of $t$, we define $m(X)$ as the multiset of numbers
$\llbracket s(u)\mid u\in X\rrbracket $. We let $P(X)$ mean the following:

\begin{quote}
(i) $X\neq\emptyset$, its elements are first sons of occurrences of $f$ \ and

(ii) the multiset $m(X)$ contains exactly two occurrences of each of its elements.
\end{quote}

There is a P-FA $\mathcal{A}$ over $F^{(1)}$ that decides $P(X)$. The state
$run_{\mathcal{A},t\ast X}(u)$ is $Error$ if $X/u$ contains a position
different from $u$ that is not the first son of an occurrence of $f$ or if
$m(X/u)$ contains at least three occurrences of some integer. Otherwise,
$run_{\mathcal{A},t\ast X}(u)=(\alpha,m(X/u))$ with $\alpha:=$ \texttt{if}
$u\in X$ \texttt{then}\ $1$ \texttt{else} $0$. The accepting states are
$(0,m)$ where $m$ is not empty and contains exactly two occurrences of each of
its elements.

The nondeterministic FA $pr_{1}(\mathcal{A})$ decides $\exists X.P(X).$ The
second components of any state belonging to $run_{pr(\mathcal{A}),t}^{\ast
}(u)$ are the multisets\ $m(X/u)$ that do not contain three occurrences of a
same integer and are associated with a set $X$ of positions containing only
first sons of occurrences of $f$. The maximum cardinality of the
set\ $run_{pr(\mathcal{A}),t}^{\ast}(u)$ is the nondeterminism degree of
$pr(\mathcal{A})$ on $t$.\ It is not polynomially bounded in $|t|$ hence
$pr(\mathcal{A})$ is not a P-FA.

For a comparison with Counter-example 13, note that we can easily build a
P-FA\ that decides $\exists X.P(X)$ without using $pr(\mathcal{A)}$ as an
intermediate step. $\square$

\subsection{Monadic Second-order constructions}

\bigskip

Although Theorem 17 does not extend to arbitrary existential set
quantifications, we can get some results for them and more generally, for the
computation of multispectra and the derived functions such as $\#\overline
{X}.P(\overline{X})$, $\mathrm{Sp}\overline{X}.P(\overline{X})$,
$\mathrm{MinCard}X.P(X)$ defined in the introduction and some others.\ In
particular, we will consider $\mathrm{Sat}\overline{X}.P(\overline{X})(t)$
(the set of tuples $\overline{X}$ that satisfy $P$ in $t$). This function
generalizes $\mathrm{Sat}\overline{x}.P(\overline{x})(t)$ considered in
Section 3.1.\ We first present some general constructions relative to FA. To
simplify notation, we write definitions, conditions and transitions of
automata for operation symbols of arity 0 or 2.\ The generalization to other
arities is immediate.

\subsubsection{Attributed automata}

Let $H$ be an effectively given signature and $\mathcal{A}$ be a deterministic
FA\ over $H$ without output.\ Let $\mathcal{D}$ be an effectively given
$H$-algebra. The mapping $val_{\mathcal{D}}\upharpoonright L(\mathcal{A})$ (a
partial function: $T(H)\rightarrow D)$ is computed by the deterministic
FA\ $\mathcal{A}\times_{g}\mathcal{D}$ with set of states $Q_{\mathcal{A}%
}\times D$ and output function $g$ (cf.\ Definition 4(2) and Example 3(a))
such that $g(q,d):=$ \texttt{if}\ $q\in Acc_{\mathcal{A}}$ \texttt{then} $d$
\texttt{else} $\bot$ (with $\bot\notin D$, standing for "undefined").

We will denote this FA by $\mathcal{A}\ltimes\mathcal{D}$ and call it an
\emph{attributed fly-automaton}. We consider $d$ in a state $(q,d)$ as an
\emph{attribute} of $q$ (cf.\ Example 5). We will give a slightly more general
notion of attributed FA at the end of this section.

Assume that we also have a signature $F$ and a computable relabelling
$h:H\rightarrow F$ (cf. Definition 4(4); in particular $h^{-1}(f)$ is finite
for each $f$) extended into $h:T(H)\rightarrow T(F)$. We want to compute, for
every term $t\in T(F),$ the following objects:

\begin{quote}
(a) $\gamma(t):=\{val_{\mathcal{D}}(t^{\prime})\mid t^{\prime}\in
L(\mathcal{A})\cap h^{-1}(t)\}\in\mathcal{P}_{f}(D),$

(b) $\xi(t):=\llbracket val_{\mathcal{D}}(t^{\prime})\mid t^{\prime}\in
L(\mathcal{A})\cap h^{-1}(t)\rrbracket \in\mathcal{M}_{f}(D)$ ($\xi(t)$ is a
finite multiset over $D$).
\end{quote}

In the next case, $\mathcal{D}$ is a distributive $H$-algebra and we want to compute:

\begin{quote}
(c) $\theta(t):=\squplus\xi(t),$
\end{quote}

where $\squplus$ is applied to finite multisets over $D$, that is\ a
commutative monoid with neutral element $\boldsymbol{0}_{\mathcal{D}}$.\ We
recall from\ Section 1.1 that a multiset over $D$\ is finite if the total
number of occurrences of its elements different from $\boldsymbol{0}%
_{\mathcal{D}}$ is finite. Then, $\squplus\alpha$ is well-defined if $\alpha
$\ is finite.\ We have $\squplus\alpha:=\squplus\beta$ where $\beta$\ is
obtained from $\alpha$ by removing all occurrences of $\boldsymbol{0}%
_{\mathcal{D}}$ and we evaluate $\squplus\beta$ with the rules $\squplus(\beta
_{1}\sqcup\beta_{2}):=(\squplus\beta_{1})\squplus(\squplus\beta_{2})$ and
$\squplus\emptyset:=\boldsymbol{0}_{\mathcal{D}}$.

We will prove that $\mathrm{Sp}\overline{X}.P(\overline{X})$ and
$\mathrm{Sat}\overline{X}.P(\overline{X})$ are instances of Case (a),
$\mathrm{MSp}\overline{X}.P(\overline{X})$ of Case (b) and $\#\overline
{X}.P(\overline{X})$ of Case (c) with $\mathcal{A}=\mathcal{A}_{P(\overline
{X})}$, $H=F^{(s)}$ and $h=pr:F^{(s)}\rightarrow F$.\ Case (c) will also be
useful for computing optimizing functions (see Section (3.2.3)).

\bigskip

\textbf{Proposition 21}: Let $F,H,h$,$\mathcal{A}$ and $\mathcal{D}$ be as
above.\ The functions $\gamma$, $\xi$ and $\theta$ are computable by
deterministic FA's.

\bigskip

\textbf{Proof}: In all cases we will use $\mathcal{B}:=h(\mathcal{A}%
\ltimes\mathcal{D})$, the image of the deterministic FA $\mathcal{A}%
\ltimes\mathcal{D}$\ under $h$, that is not deterministic in general. Cases
(a) and (b) are particular instances of Case (c), but we think useful to
present Case (a) first.

\bigskip

\emph{Case} (a) The function computed by $\det(\mathcal{B})$ is $\gamma$, up
to the value $\bot$ that is not in $D.$\ More precisely $\gamma(t)=Comp_{nd}%
(\mathcal{B})(t)-\{\bot\}.$

To prove this claim, we consider an element of $\gamma(t)$ of the form
$val_{\mathcal{D}}(t^{\prime})$ for $t^{\prime}\in L(\mathcal{A})\cap
h^{-1}(t).\ $We have $q_{\mathcal{A}\ltimes\mathcal{D}}(t^{\prime
})=(q,val_{\mathcal{D}}(t^{\prime}))$ for some $q$ in $Acc_{\mathcal{A}}$.
Then $\mathcal{B}$ has a run on $t=h(t^{\prime})$ that yields state
$(q,val_{\mathcal{D}}(t^{\prime}))$ at the root.\ Since $val_{\mathcal{D}%
}(t^{\prime})\neq\bot$, we have $val_{\mathcal{D}}(t^{\prime}%
)=g((q,val_{\mathcal{D}}(t^{\prime})))\in Comp_{nd}(\mathcal{B})(t)-\{\bot\}.$
For the other direction, let $d\in Comp_{nd}(\mathcal{B})(t)-\{\bot\}.$ Then
$(q,d)\in run_{\mathcal{B},t}^{\ast}(root_{t})$ for some accepting state $q$.
There is $t^{\prime}\in L(\mathcal{A},q)$ such that $h(t^{\prime})=t$ and
$d=val_{\mathcal{D}}(t^{\prime})$. Hence, $d\in\gamma(t)$\ and we have the
claimed equality.

The set $\gamma(t)$ can thus be computed by running $\mathcal{B}$
deterministically (i.e., by running$\ \det(\mathcal{B})$, cf.\ Section 2.1) or
by using an enumeration algorithm that outputs one by one its elements
\cite{Dur}.

\bigskip

\emph{Remarks}\textbf{:} (1) When defining $\det(\mathcal{B})$ or running
$\mathcal{B}$ deterministically, we can eliminate the pairs $(Error,d)$ as the
values $d$ arising from the corresponding runs will not contribute to
$\gamma(t)$ (but they can occur in alternative accepting runs).

(2) The states of $\det(\mathcal{B})$ are finite subsets of $Q_{\mathcal{A}%
}\times D$ (or rather $(Q_{\mathcal{A}}-\{Error\})\times D$).\ It is
convenient to identify such a set $\alpha$ with the mapping $\overline{\alpha
}:Q_{\mathcal{A}}\rightarrow\mathcal{P}_{f}(D)$ such that $\overline{\alpha
}(q):=\{d\in D\mid(q,d)\in\alpha\}$. This mapping is finite in the sense of
Section (1.4) if the empty set is the "zero element" of $\mathcal{P}_{f}(D).$
That is, $\overline{\alpha}^{-1}(\mathcal{P}_{f}(D)-\{\emptyset\})$\ is
finite. It can also be identified with the finite set of pairs $(q,\overline
{\alpha}(q))$ such that $\overline{\alpha}(q)\neq\emptyset.$ In further
constructions, the sets $\overline{\alpha}(q)$ will be \emph{aggregated} into
combinations of values by the associative and commutative operation $\squplus$
of a distributive algebra with domain of $D$.

(3) For clarity, we spell out the transitions of $\det(\mathcal{B})$ by using
the latter presentation of its states. For a nullary symbol $a\in F$, we
have$\ a\rightarrow_{\det(\mathcal{B})}\overline{\beta}$ \ where
$\overline{\beta}$\ is the set of pairs $(q,\{b_{\mathcal{D}}\mid b\in
h^{-1}(a)\cap\delta_{\mathcal{A}}^{-1}(q)\})$ \ such that $h^{-1}(a)\cap
\delta_{\mathcal{A}}^{-1}(q)\neq\emptyset.$ \ For a binary symbol $f\in F,$ we
have $\ f[\overline{\alpha}_{1},\overline{\alpha}_{2}]\rightarrow
_{\det(\mathcal{B})}\overline{\beta}$ where $\overline{\beta}$\ is the set of
pairs of the form \ $(q,\bigcup\{g_{\mathcal{D}}(\overline{\alpha_{1}}%
(q_{1}),\overline{\alpha_{2}}(q_{2}))\mid g\in h^{-1}(f),g[q_{1}%
,q_{2}]\rightarrow_{\mathcal{A}}q\})$ such that the second component of this
pair is not empty. This formulation shows that $\overline{\beta}$\ can be
computed with the following operations on $\mathcal{P}_{f}(D)$: set union and
the extensions to sets of the operations $g_{\mathcal{D}}$. $\square$

\bigskip

\emph{Case} (c) Here $\mathcal{D=}\langle D,\squplus ,\boldsymbol{0}%
,(g_{\mathcal{D}})_{g\in H}\rangle$ is a distributive $H$-algebra, and we want
to compute:

\begin{quote}
$\theta(t):=\squplus \llbracket val_{\mathcal{D}}(t^{\prime})\mid t^{\prime
}\in L(\mathcal{A})\cap h^{-1}(t)\rrbracket $.
\end{quote}

First we extend the mapping $val_{\mathcal{D}}$ to finite sets of terms
$T\subseteq T(H)$ by:

\begin{quote}
$val_{\mathcal{D}}(T):=\squplus \llbracket val_{\mathcal{D}}(t)\mid t\in
T\rrbracket $.
\end{quote}

Note that $\llbracket val_{\mathcal{D}}(t)\mid t\in T\rrbracket$ is a finite
multiset.\ The associativity and commutativity of $\squplus$ and the
distributivity of $g_{\mathcal{D}}$ over $\squplus$ yield:

\begin{quote}
$val_{\mathcal{D}}(T\uplus T^{\prime})=val_{\mathcal{D}}(T)\squplus
val_{\mathcal{D}}(T^{\prime})\qquad$and$\qquad\qquad\qquad\ \ \ \ \ \ \ (2)$

$val_{\mathcal{D}}(g(T,T^{\prime}))=g_{\mathcal{D}}(val_{\mathcal{D}%
}(T),val_{\mathcal{D}}(T^{\prime})).\qquad\qquad\qquad\qquad\qquad(3)$
\end{quote}

Recall that we only write such equalities for binary symbols $g$ because their
extensions to other positive arities are obvious.

For $q\in Q_{\mathcal{A}}$, we define $\theta(t,q):=val_{\mathcal{D}%
}(L(\mathcal{A},q)\cap h^{-1}(t))$ and we get $\theta
(t)=\squplus\llbracket\theta(t,q)\mid q\in Acc_{\mathcal{A}}\rrbracket.$ \ The
righthand side of this equality is well-defined because $\theta(t,q)\neq
\boldsymbol{0}$ for finitely many states $q$, since $h^{-1}(t)$ is finite. The
sets $L(\mathcal{A},q)\cap h^{-1}(t)$ for $q\in Q_{\mathcal{A}}$ are pairwise
disjoint because $\mathcal{A}$ is deterministic, which ensures the equality.

\bigskip

We define as follows a deterministic FA\ $\mathcal{C}$ over $F$:

\begin{quote}
its states are functions $\sigma:Q_{\mathcal{A}}\rightarrow D$ such that
$\sigma^{-1}(D-\{\boldsymbol{0}\})$ is finite (they can be seen as finite
subsets of $Q_{\mathcal{A}}\times(D-\{\boldsymbol{0}\}))$;

its transitions are defined in such a way that $q_{\mathcal{C}}(t)$, the state
reached by $\mathcal{C}$ at the root of any term $t\in T(F),$ is the mapping

$\qquad\lambda q\in Q_{\mathcal{A}}.\theta(t,q)$ (that can be seen as the
finite set of pairs

$\qquad\qquad(q,\theta(t,q))\in Q_{\mathcal{A}}\times D$ such that
$\theta(t,q)\neq\boldsymbol{0}$);

its output function is $Out_{\mathcal{C}}(\sigma):=\squplus \llbracket \sigma
(q)\mid q\in Acc_{\mathcal{A}}\rrbracket .$
\end{quote}

We now define the transitions. For a nullary symbol $a\in F$, we define:

\begin{quote}
$a\rightarrow_{\mathcal{C}}\lambda q\in Q_{\mathcal{A}}.val_{\mathcal{D}%
}(h^{-1}(a)\cap\delta_{\mathcal{A}}^{-1}(q)).$
\end{quote}

It is well-defined because $h^{-1}(a)\cap\delta_{\mathcal{A}}^{-1}(q)$ is
finite. For a binary symbol $f\in F,$ we define:

\begin{quote}
$f[\sigma_{1},\sigma_{2}]\rightarrow_{\mathcal{C}}\lambda q.\squplus\llbracket
g_{\mathcal{D}}(\sigma_{1}(q_{1}),\sigma_{2}(q_{2}))\mid h(g)=f$,
$g[q_{1},q_{2}]\rightarrow_{\mathcal{A}}q\rrbracket.$\ (4)
\end{quote}

The operation $\squplus$ is applied to a finite multiset (having finitely many
elements different from $\boldsymbol{0}$) because $h^{-1}(f)$ is finite,
$\sigma_{1}(q_{1})\neq\boldsymbol{0}$ for finitely many states $q_{1}$,
similarly for $\sigma_{2}(q_{2})$ and $g_{\mathcal{D}}(\boldsymbol{0}%
,d)=g_{\mathcal{D}}(d,\boldsymbol{0})=\boldsymbol{0}.$

\bigskip

Before proving the validity of this construction, we compare $\mathcal{C}$
with $\det(\mathcal{B})$.\ The state $q_{\det(\mathcal{B})}(t)$ is a finite
subset, say $\alpha$, of $Q_{\mathcal{A}}\times D$ (we use the same notation
as in the remark after Case (a)).\ The state $q_{\mathcal{C}}(t)$ can be seen
as the finite subset \ of $Q_{\mathcal{A}}\times D$ obtained by replacing the
pairs $(q,d)$ of $\alpha$ having the same first component $q$ by the single
pair $(q,\squplus\overline{\beta}(q))$ where $\overline{\beta}(q)$ is a finite
multiset whose underlying set is $\overline{\alpha}(q)$. The multiplicity of
an element $d$ of $\overline{\beta}(q)$ counts the number of ways it can be
produced with state $q$.

\bigskip

\emph{Claim}: For every $t\in T(F)$, we have $q_{\mathcal{C}}(t)=\lambda q\in
Q_{\mathcal{A}}.\theta(t,q).$

\emph{Proof}: By induction on the structure of $t$.

If $t=a\in F$, the equality follows from the definitions.

Let $t=f(t_{1},t_{2})$ and $q\in Q_{\mathcal{A}}$. By definition, we have
$\theta(t,q)=val_{\mathcal{D}}(L(\mathcal{A},q)\cap h^{-1}(t)).$ For each term
$t^{\prime}$\ in $L(\mathcal{A},q)\cap h^{-1}(t),$\ there is a unique 5-tuple
$(g,t_{1}^{\prime},t_{2}^{\prime},q_{1},$ $q_{2})$ \ such that $t^{\prime
}=g(t_{1}^{\prime},t_{2}^{\prime})$ and:

\begin{quote}
$h(g)=f$, $t_{1}^{\prime}\in L(\mathcal{A},q_{1})\cap h^{-1}(t_{1}),$

$t_{2}^{\prime}\in L(\mathcal{A},q_{2})\cap h^{-1}(t_{2})$ and $g[q_{1}%
,q_{2}]\rightarrow_{\mathcal{A}}q.\qquad\qquad\qquad\qquad(5)$
\end{quote}

The existence and unicity of $(g,t_{1}^{\prime},t_{2}^{\prime})$ follows from
the equality $h(t^{\prime})=t$.\ The pair $(q_{1},q_{2})$ such that
$t_{1}^{\prime}\in L(\mathcal{A},q_{1}),$ $t_{2}^{\prime}\in L(\mathcal{A}%
,q_{2})$ is unique because $\mathcal{A}$ is deterministic.\ Then we have
$g[q_{1},q_{2}]\rightarrow_{\mathcal{A}}q$\ because $t\in L(\mathcal{A},q).$

Conversely, every such 5-tuple satisfying (5) yields a term $t^{\prime
}=g(t_{1}^{\prime},t_{2}^{\prime})\in L(\mathcal{A},q)\cap h^{-1}(t).$\ It
follows that $L(\mathcal{A},q)\cap h^{-1}(t)$\ is the disjoint union of the
sets $g(T_{1}(q_{1}),T_{2}(q_{2}))$ for all triples $(g,q_{1},q_{2})$ such
that $h(g)=f$ and $g[q_{1},q_{2}]\rightarrow_{\mathcal{A}}q$ where, for every
state $p\in Q_{\mathcal{A}}$, $T_{1}(p):=L(\mathcal{A},p)\cap h^{-1}(t_{1})$
and $T_{2}(p):=L(\mathcal{A},p)\cap h^{-1}(t_{2}).$ \ For each such triple:

\begin{quote}
$val_{\mathcal{D}}(g(T_{1}(q_{1}),T_{2}(q_{2})))=g_{\mathcal{D}}%
(val_{\mathcal{D}}(T_{1}(q_{1})),val_{\mathcal{D}}(T_{2}(q_{2})))$ \ \ \ (6)
\end{quote}

by (3). Hence, by (2) and the definitions:

\begin{quote}
$\theta(t,q)=\squplus\llbracket g_{\mathcal{D}}(val_{\mathcal{D}}(T_{1}%
(q_{1})),val_{\mathcal{D}}(T_{2}(q_{2})))\mid h(g)=f$ 

\qquad\qquad\qquad\qquad\qquad\qquad\qquad\qquad\qquad and $g[q_{1}%
,q_{2}]\rightarrow_{\mathcal{A}}q\rrbracket$

$=\squplus\llbracket g_{\mathcal{D}}(\theta(t_{1},q_{1}),\theta(t_{2}%
,q_{2}))\mid h(g)=f$ \ and $g[q_{1},q_{2}]\rightarrow_{\mathcal{A}%
}q\rrbracket$. \ (7)
\end{quote}

This equality is true for all states $q\in Q_{\mathcal{A}}$. By induction, we
have $\theta(t_{1},p)=q_{\mathcal{C}}(t_{1})(p)$ \ and $\theta(t_{2}%
,p)=q_{\mathcal{C}}(t_{2})(p)$ for all $p\in Q_{\mathcal{A}}$. Hence,

\begin{quote}
$\theta(t,q)=\squplus\llbracket g_{\mathcal{D}}(q_{\mathcal{C}}(t_{1}%
)(q_{1}),q_{\mathcal{C}}(t_{2})(q_{2}))\mid h(g)=f$, $g[q_{1},q_{2}%
]\rightarrow_{\mathcal{A}}q\rrbracket$ (8)
\end{quote}

and $\lambda q\in Q_{\mathcal{A}}.\theta(t,q)=q_{\mathcal{C}}(t)$\ by the
definition of $\mathcal{C}$, which completes the proof of the claim.$\square$

\bigskip

Hence, the deterministic FA\ $\mathcal{C}$\ computes $\theta$, as desired.

As noted above in Case (a), we can delete the $Error$ state of $\mathcal{A}$
and define $\mathcal{C}$ so that $q_{\mathcal{C}}(t)=\lambda q\in
(Q_{\mathcal{A}}-\{Error\}).\theta(t,q).$

\bigskip

Case (b) is a special case of (c): we replace the effectively given
$H$-algebra $\mathcal{D}$ by the distributive $H$-algebra $\mathcal{E}%
:=\mathcal{M}_{f}(\mathcal{D})$ \ (cf.\ Section 2.1). For $t^{\prime}\in
T(H)$, $val_{\mathcal{E}}(t^{\prime})=\{val_{\mathcal{D}}(t^{\prime})\},$ as
observed in Section 2.1. It follows that

\begin{quote}
$\xi(t):=\llbracket val_{\mathcal{D}}(t^{\prime})\mid t^{\prime}\in
L(\mathcal{A})\cap h^{-1}(t)\rrbracket =val_{\mathcal{E}}(L(\mathcal{A})\cap
h^{-1}(t))$.
\end{quote}

The states of $\mathcal{C}$\ are finite mappings $\sigma:Q_{\mathcal{A}%
}\rightarrow\mathcal{M}_{f}(D)$ such that we have $q_{\mathcal{C}}(t)=\lambda
q\in Q_{\mathcal{A}}.\llbracket val_{\mathcal{D}}(t^{\prime})\mid t^{\prime
}\in L(\mathcal{A},q)\cap h^{-1}(t)\rrbracket.$

\bigskip

Case (a) is the instance of Case (c) where we take similarly $\mathcal{E}%
:=\mathcal{P}_{f}(\mathcal{D})$. $\square$

\bigskip

\textbf{Remark 22}: \emph{More general attributed automata}.

Let $\mathcal{A}$\ be a deterministic FA\ over a signature $H$. We define
$H\ast Q_{\mathcal{A}}$ as the signature of $\rho(f)$-ary symbols
$(f,q_{1},...,q_{\rho(f)})$ for all $f\in H$ and $q_{1},...,q_{\rho(f)}\in
Q_{\mathcal{A}}$. Let $\mathcal{D}$ be an effectively given $H\ast
Q_{\mathcal{A}}$-algebra. Extending the notation of Section 1.1, we define
$val_{\mathcal{D}}:T(H)\rightarrow D$ by using the run of $\mathcal{A}$ on the
considered term:

\begin{quote}
$val_{\mathcal{D}}(f(t_{1},...,t_{\rho(f)}):=(f,q_{1},...,q_{\rho
(f)})_{\mathcal{D}}(d_{1},...,d_{\rho(f)})$

where $q_{i}=q_{\mathcal{A}}(t_{i})$ and $d_{i}=val_{\mathcal{D}}(t_{i})$ for
$i=1,...,\rho(f)$.
\end{quote}

Hence $val_{\mathcal{D}}(t)$ is computed by a deterministic FA\ with set of
states $Q_{\mathcal{A}}\times D$.\ We denote this FA by $\mathcal{A}%
\ltimes\mathcal{D}$ and call it also an \emph{attributed fly-automaton}. (As
we do not exclude to extend in future articles the notion of an attributed FA,
we leave "open" the definition).\ 

As in Proposition 21, we let $h$ be a computable relabelling: $T(H)\rightarrow
T(F)$ and we are interested in computing the functions $\gamma,\xi$ and
$\theta$ defined as above in terms of $val_{\mathcal{D}}$, now based on the
$H\ast Q_{\mathcal{A}}$-algebra $\mathcal{D}$. For $\theta$, we also assume
that $\mathcal{D}$ is distributive. The construction for Case (c) (that yields
the two other cases) works with the following adaptations:\ Equality (3) is
replaced by:

\begin{quote}
$val_{\mathcal{D}}(g(T,T^{\prime}))=$

$\squplus\llbracket(g,q_{1},q_{2})_{\mathcal{D}}(val_{\mathcal{D}}(T\cap
L(\mathcal{A},q_{1})),val_{\mathcal{D}}(T^{\prime}\cap L(\mathcal{A}%
,q_{2})))\mid q_{1},q_{2}\in Q_{\mathcal{A}}\rrbracket,$
\end{quote}

and, in Equalities (4),(6), (7) and (8), $g_{\mathcal{D}}$ is replaced by
$(g,q_{1},q_{2})_{\mathcal{D}}.\square$

\subsubsection{Sets of satisfying tuples and counting functions.}

We now compute the functions $\mathrm{Sat}\overline{X}.P(\overline{X})$,
$\#\overline{X}.P(\overline{X})$, $\mathrm{Sp}\overline{X}.P(\overline{X})$,
$\mathrm{MSp}\overline{X}.P(\overline{X})$, $\mathrm{MinCard}X_{1}%
.P(\overline{X})$ and a\ few others by FA derived in uniform ways from a
deterministic FA $\mathcal{A}$ that recognizes the language $T_{P(\overline
{X})}$ representing $P$.\ (This language is defined Section 1.3).

As before, $F$ is an effectively given signature, $\overline{X}=(X_{1}%
,...,X_{s})$ and $P(\overline{X})$ is a property of terms in $T(F)$ with $s$
set arguments. We will use Proposition 21 with $H:=F^{(s)}$ and $h=pr:F^{(s)}%
\rightarrow F$. We first consider the computation of the function
$\mathrm{Sat}\overline{X}.P(\overline{X})$.\ All other functions (they are
called \emph{aggregate functions} in the context of databases \cite{AHV}) can
be computed from it, but we will give direct constructions yielding XP
algorithms whereas $\mathrm{Sat}\overline{X}.P(\overline{X})$ is not
XP-computable in general.

\bigskip

(3.2.2.1) \emph{Computation of} $\mathrm{Sat}\overline{X}.P(\overline{X})$.\ 

In order to apply Case (a) of Proposition 21, we define an $F^{(s)}$-algebra
$\mathcal{D}$ such that:

\begin{quote}
$val_{\mathcal{D}}(t\ast\overline{X})=\overline{X}$ for all $t\ast\overline
{X}\in T(F^{(s)}).\qquad\qquad\qquad\qquad\qquad(9)$
\end{quote}

Each tuple $\overline{X}$ is an $s$-tuple of finite sets positions of $t$
(they are Dewey words). We let $r:=\rho(F)$ and we take $D:=\mathcal{P}%
_{f}([r]^{\ast})^{s}.$ If $\overline{X}\in\mathcal{D}$ and $i\in\lbrack r]$,
we define $i.\overline{X}$ by replacing in $\overline{X}$ each word
$u\in\lbrack r]^{\ast}$ by $i.u$.

If $(a,w)$ is a nullary symbol in $F^{(s)}$, ($a\in F$ and $w\in\{0,1\}^{s}$)
we define:

\begin{quote}
$(a,w)_{\mathcal{D}}:=\overline{w}\ $where \ $\overline{w}:=\{(X_{1}%
,...,X_{s})\}$ \ such that:

$X_{i}:=$ \texttt{if }$w[i]=1$ \texttt{then} $\{\varepsilon\}$ \texttt{else}
\ $\emptyset.$
\end{quote}

If $(f,w)$ is binary, and with \ $\overline{w}$\ as above, we define:

\begin{quote}
$(f,w)_{\mathcal{D}}(\overline{X},\overline{Y}):=\overline{w}\cup
1.\overline{X}\cup2.\overline{Y}$
\end{quote}

where the union of sets is extended to tuples by: $\overline{X}\cup
\overline{Y}:=(X_{1}\cup Y_{1},...,X_{s}\cup Y_{s}).$ The validity of (9) is
easy to check. We will denote by $\mathcal{A}^{\mathrm{Sat}}$\ the
deterministic FA\ $\det(\mathcal{B})$ obtained by Proposition 21 to compute
$\gamma(t):=\{\overline{X}\mid t\ast\overline{X}\in L(\mathcal{A}%
)\}=\mathrm{Sat}\overline{X}.P(\overline{X})(t)$.\ 

For later use of $\mathcal{D}$, we will denote it by $\mathcal{D}_{F,s}$.

\bigskip

\emph{Remarks}: (1) The definitions are similar if $\overline{X}$ is a
partition of $Pos(t)$ encoded by a finite subset $\widehat{X}\ $of
$\ [r]^{\ast}\times\lbrack s]$ (cf.\ Section 1.3).

(2) To make things (hopefully) clear we work out the construction of
$\mathcal{A}^{\mathrm{Sat}}$.\ Our description is based on the construction of
Proposition 21 and the remark about Case (a). Each state of $\det
(\mathcal{B})$ is handled as a finite function $\sigma$: $Q_{\mathcal{A}%
}\rightarrow\mathcal{P}_{f}(D)=\mathcal{P}_{f}(\mathcal{P}_{f}([r]^{\ast}%
)^{s}).$ We fix $t\in T(F)$.\ For each state $q$ of $\mathcal{A}$, we define
$\sigma(q)$ as the finite set of $s$-tuples $\overline{X}\in\mathcal{P}%
(Pos(t))^{s}$ such that $q_{\mathcal{A}}(t\ast\overline{X})=q$. Since
$\mathcal{A}$ is deterministic, $\sigma(q)\cap\sigma(q^{\prime})=\emptyset$
\ if $q\neq q^{\prime}$ and clearly, $\mathrm{Sat}\overline{X}.P(\overline
{X})(t)=\bigcup\nolimits_{q\in Acc_{\mathcal{A}}}\sigma(q).$ The transitions
of $\det(\mathcal{B})$ are thus, for a nullary symbol $a$ in $F$:

\begin{quote}
$a\rightarrow\lambda q\in Q_{\mathcal{A}}.\{\overline{w}\mid(a,w)\rightarrow
_{\mathcal{A}}q\}.$
\end{quote}

For definining in a compact way the transitions on binary symbols, we define
for disjoint sets $E$ and $E^{\prime}$, $Z\subseteq\mathcal{P}(E)^{s}$ and
$Z^{\prime}\subseteq\mathcal{P}(E^{\prime})^{s}$:

\begin{quote}
$Z\circledast Z^{\prime}:=\{(X_{1}\cup Y_{1},\ldots,X_{s}\cup Y_{s}%
)\mid\overline{X}\in Z,\overline{Y}\in Z^{\prime}\}.$
\end{quote}

This operation is nothing but the extension to sets of the union of tuples of
sets. Then, for a binary symbol $f$:

\begin{quote}
$(f,w)[\sigma_{1},\sigma_{2}]\rightarrow\lambda q\in Q_{\mathcal{A}}%
.\bigcup\nolimits_{(f,w)[q_{1},q_{2}]\rightarrow_{\mathcal{A}}q}\overline
{w}\circledast1.\sigma_{1}(q_{1})\circledast2.\sigma_{2}(q_{2}),$

$Out_{\mathcal{A}^{\mathrm{Sat}}}(\sigma):=\bigcup\nolimits_{q\in
Acc_{\mathcal{A}}}\sigma(q).$\ $\qquad\qquad\qquad\qquad\qquad\qquad
\qquad\square$\ 
\end{quote}

\bigskip

(3.2.2.2) \emph{Computation of} $\mathrm{Sp}\overline{X}.P(\overline{X})$.\ 

We use again Case (a) of Proposition 21 with $\mathcal{D}$ such that:

\begin{quote}
$val_{\mathcal{D}}(t\ast\overline{X})=(\left\vert X_{1}\right\vert
,...,\left\vert X_{s}\right\vert )$ for all $t\ast\overline{X}\in
T(F^{(s)}).\qquad(10)$
\end{quote}

We take $D:=\mathbb{N}^{s}$ and we define (with $\boldsymbol{+}$ denoting the
addition of vectors):

\begin{quote}
$(a,w)_{\mathcal{D}}:=(w[1],...,w[s])$,

$(f,w)_{\mathcal{D}}(\overline{m},\overline{p}):=(w[1],...,w[s])\boldsymbol{+}%
\overline{m}\boldsymbol{+}\overline{p}.$
\end{quote}

The verification that (10) is true is straightforward. We will denote by
$\mathcal{A}^{\mathrm{Sp}}$\ the deterministic FA\ $\det(\mathcal{B})$
obtained in this way to compute

\begin{quote}
$\gamma(t):=\{(\left\vert X_{1}\right\vert ,...,\left\vert X_{s}\right\vert
)\mid t\ast\overline{X}\in L(\mathcal{A})\}=\mathrm{Sp}\overline
{X}.P(\overline{X})(t)$.\ 
\end{quote}

\bigskip

(3.2.2.3) \emph{Computation of} $\mathrm{MSp}\overline{X}.P(\overline{X})$.\ 

We use Case (b) of Proposition 21 with the same $F^{(s)}$-algebra
$\mathcal{D}$ as in the previous case. We will denote by $\mathcal{A}%
^{\mathrm{MSp}}$\ the obtained deterministic FA that computes $\xi
(t):=\llbracket(\left\vert X_{1}\right\vert ,...,\left\vert X_{s}\right\vert
)\mid t\ast\overline{X}\in L(\mathcal{A})\rrbracket=\mathrm{MSp}\overline
{X}.P(\overline{X})(t)$.\ 

We now detail the transitions of $\mathcal{A}^{\mathrm{MSp}}$. A finite
multiset over $\mathbb{N}^{s}$ is a function $m:\mathbb{N}^{s}\rightarrow
\mathbb{N}$ such that $m^{-1}(\mathbb{N}_{+})$ \ is finite.\ We have the
following transitions:

\begin{quote}
$a\rightarrow\lambda q\in Q_{\mathcal{A}}.(\lambda\overline{x}\in
\mathbb{N}^{s}.$\texttt{if} $\overline{x}\in\{0,1\}^{s}\wedge(a,\overline
{x})\rightarrow_{\mathcal{A}}q$\ \texttt{then }$1$\texttt{ else }$0)$
\end{quote}

and

\begin{quote}
$f[\sigma_{1},\sigma_{2}]\rightarrow\lambda q\in Q_{\mathcal{A}}%
.(\lambda\overline{x}\in\mathbb{N}^{s}.\Sigma\llbracket \sigma_{1}%
(q_{1})(\overline{y}).\sigma_{2}(q_{2})(\overline{z})\mid w\in\{0,1\}^{s},$

$\qquad\qquad\qquad\qquad\qquad(f,w)[q_{1},q_{2}]\rightarrow_{\mathcal{A}}q$
$\ $and $\ \overline{x}=w\boldsymbol{+}\overline{y}\boldsymbol{+}\overline
{z}\rrbracket )$.
\end{quote}

In the second transition, the multiset is indexed by the 5-tuples
$(w,q_{1},q_{2},\overline{y},\overline{z})$ that satisfy $\overline
{x}=w+\overline{y}+\overline{z}\wedge(f,w)[q_{1},q_{2}]\rightarrow
_{\mathcal{A}}q.$

Hence, $q_{\mathcal{A}^{\mathrm{MSp}}}(t)$ is a finite mapping, say $\sigma$,
from $Q_{\mathcal{A}}$ to $\mathcal{M}_{f}(\mathbb{N}^{s})$ such that, for
every state $q$ of $\mathcal{A}$, $\sigma(q)$ is the finite multiset of tuples
$(\left\vert X_{1}\right\vert ,...,\left\vert X_{s}\right\vert )$ such that
$q_{\mathcal{A}}(t\ast\overline{X})=q$.\ Here, $\sigma(q)$ is a particular
aggregation of the values in $\gamma(q)$ relative to the FA\ $\mathcal{A}%
^{\mathrm{Sat}}$.

\bigskip

(3.2.2.4) \emph{Computation of} $\#\overline{X}.P(\overline{X})$.

We want to compute $\#\overline{X}.P(\overline{X})(t)=\theta(t)$ defined as
the cardinality of the set $\{\overline{X}\mid t\ast\overline{X}\in
L(\mathcal{A})\}.$ As the multiset $\llbracket\overline{X}\mid t\ast
\overline{X}\in L(\mathcal{A})\rrbracket$ has only one occurrence of each
element, $\theta(t)$ is its cardinality. In order to apply Case (c) of
Proposition 21, we define a distributive $F^{(s)}$-algebra $\mathcal{D}%
:=\langle\mathbb{N},+,0,(g_{\mathcal{D}})_{g\in F^{(s)}}\rangle$ with
$(a,w)_{\mathcal{D}}:=1$ and $(f,w)_{\mathcal{D}}(m,p):=m.p.$ Clearly,
$val_{\mathcal{D}}(t\ast\overline{X})=1$ for all $t\ast\overline{X}\in
T(F^{(s)}).$ We will denote by $\mathcal{A}^{\#}$\ the deterministic FA
$\mathcal{C}$ obtained in this way by Case (c) of Proposition 21.

\bigskip

\textbf{Theorem 23}: Let $\mathcal{A}$ be a deterministic FA\ over $F^{(s)}%
$\ that decides a property $P(\overline{X})$. The functions $\mathrm{Sat}%
\overline{X}.P(\overline{X}),\mathrm{MSp}\overline{X}.P(\overline{X}),$
$\mathrm{Sp}\overline{X}.P(\overline{X})$ and $\#\overline{X}.P(\overline{X})$
are computable by deterministic FA's constructed from the tuple $\underline
{\mathcal{A}}$\ that defines $\mathcal{A}$.

\bigskip

Complexity issues will be discussed in Section (3.2.4).

\subsubsection{Optimizing functions}

We now consider how to compute certain values defined by \emph{optimizing
functions} that minimize or maximize values defined from the set
$\mathrm{Sat}\overline{X}.P(\overline{X})(t)$ without using it as intermediate
value for efficiency purposes. We will only discuss minimizations because
maximizations are fully similar.

\bigskip

(3.2.3.1) \emph{Minimizing cardinalities or other values}.

In order to compute:

\begin{quote}
$\mathrm{MinCard}X_{1}.P(\overline{X})(t):=$

\texttt{if} $t\models\exists\overline{X}.P(\overline{X})$ \texttt{then }%
$\min\{\left\vert X_{1}\right\vert \mid t\models\exists X_{2},...,X_{s}%
.P(\overline{X})\}$ \ \texttt{else} $\infty,$
\end{quote}

we use Case (c) of Proposition 21. We take $\mathcal{D}:=\langle\mathbb{N\cup
}\{\infty\},\min,\infty,(g_{\mathcal{D}})_{g\in F^{(s)}}\rangle$ with
$(a,w)_{\mathcal{D}}:=w[1]$ and $(f,w)_{\mathcal{D}}(m,p):=w[1]+m+p$. Clearly,
$\min\emptyset=\infty$. We will denote by $\mathcal{A}^{\mathrm{MinCard}}$ the
obtained deterministic FA.

\bigskip

More generally, in order to compute:

\begin{quote}
$\mathrm{Min\_\alpha}\overline{X}.P(\overline{X})(t):=$

\texttt{if} $t\models\exists\overline{X}.P(\overline{X})$ \texttt{then }%
$\min\{\mathrm{\alpha}(\overline{X})\mid t\models P(\overline{X})\}$
\ \texttt{else} $\infty$
\end{quote}

where $\mathrm{\alpha}(\overline{X}):=c_{1}.\left\vert X_{1}\right\vert
+...+c_{s}.\left\vert X_{s}\right\vert $ \ for fixed integers $c_{1}%
,...,c_{s}$ in $\mathbb{Z}$, we take $\mathcal{D}:=\langle\mathbb{Z\cup
}\{\infty\},\min,\infty,(g_{\mathcal{D}})_{g\in F^{(s)}}\rangle$ with:

\begin{quote}
$(a,w)_{\mathcal{D}}:=c_{1}.w[1]+...+c_{s}.w[s],$

$(f,w)_{\mathcal{D}}(m,p):=c_{1}.w[1]+...+c_{s}.w[s]+m+p.$
\end{quote}

This construction works because $\alpha(X_{1}\uplus Y_{1},...,X_{s}\uplus
Y_{s})=\alpha(\overline{X})+\alpha(\overline{Y}).$ We will denote by
$\mathcal{A}^{\mathrm{Min\_\alpha}}$ the obtained deterministic FA.

\bigskip

(3.2.3.2) \emph{Minimal satisfying sets}.

We describe, in a uniform way, several FA that extract particular "minimal"
sets from $\mathrm{Sat}X.P(X)(t)$.\ (The extension to $\mathrm{Sat}%
\overline{X}.P(\overline{X})(t)$ is easy.)

Let $\leq$ be a partial order on $\mathcal{P}_{f}([r]^{\ast})$.\ For each
$Z\subseteq\mathcal{P}_{f}([r]^{\ast})$, we define $\mathrm{Min}(Z)$ as the
subset of $Z$ consisting of its minimal elements with respect to $\leq$. We
want to compute, for each term $t\in T(F),$ the set $\mathrm{Min}_{\leq
}X.P(X)(t):=\mathrm{Min}(\mathrm{Sat}X.P(X)(t)).$ Some interesting orders
$X\leq Y$ on $\mathcal{P}_{f}([r]^{\ast})$ are:

\begin{quote}
(i) $X\subseteq Y,$

(ii) $X\subseteq\mathit{Pref}(Y)$, ($\mathit{Pref}(Y)$ is the set of prefixes
of the words in $Y$),

(iii) $\left\vert X\right\vert \leq\left\vert Y\right\vert ,$

(iv) $\left\vert X\right\vert <\left\vert Y\right\vert $ or, $\left\vert
X\right\vert =\left\vert Y\right\vert $ and $X\leq_{lex}Y,$

(v) $X\leq_{lex}Y,$

where $\leq_{lex}$ is a lexicographic order on $\mathcal{P}_{f}([r]^{\ast})$
defined below.
\end{quote}

In the last two cases, $\leq$ is a linear order so that $\mathrm{Min}(Z)$ is
empty or singleton. Our method is also applicable to the quasi-order
$\min\{\left\vert u\right\vert \mid u\in X\}\leq\min\{\left\vert u\right\vert
\mid u\in Y\}$, but we will not discuss this extension.

\bigskip

The following notion will be useful in several cases.

\bigskip

\textbf{Definition 24}: \emph{Minimizing algebras.}

Let $\mathcal{D}$ be an effectively given $H$-algebra whose domain $D$ has a
decidable partial order $\leq$ and whose functions $g_{\mathcal{D}}$ are
increasing, i.e., \ $g_{\mathcal{D}}(...,d,...)\leq g_{\mathcal{D}%
}(...,d^{\prime},...)$ if $d\leq d^{\prime}$. For $Z\in\mathcal{P}_{f}(D)$,
the subset $\mathrm{Min}(Z)$ of $Z$ consists of its minimal elements with
respect to $\leq$. Hence it is empty if and only if $Z$ is empty; it is
computable if $Z$ is finite.

We let $\mathcal{M}in(D)\subseteq\mathcal{P}_{f}(D)$ be the set of finite
subsets $Z$ of $D$ such that $\mathrm{Min}(Z)=Z$. It is effectively given as
the property $\mathrm{Min}(Z)=Z$ is decidable. We define a distributive $H$-algebra:

\begin{quote}
$\mathcal{M}in(\mathcal{D})=\left\langle \mathcal{M}in(D),\squplus,\emptyset
,(g_{\mathcal{D}})_{g\in H}\right\rangle $ such that:

$Z\squplus Z^{\prime}:=\mathrm{Min}(Z\cup Z^{\prime})$,

$a_{\mathcal{M}in(\mathcal{D})}:=\{a_{\mathcal{D}}\}$ \ if $a$ is nullary,

$g_{\mathcal{M}in(\mathcal{D})}(Z,Z^{\prime}):=\mathrm{Min}(g_{\mathcal{D}%
}(Z,Z^{\prime}))$ ($=\mathrm{Min}(\{g_{\mathcal{D}}(d,d^{\prime})\mid d\in
Z,d^{\prime}\in Z^{\prime}\})$ if $g$ is binary.
\end{quote}

It is clear that $\squplus$\ is associative and commutative with neutral
element $\emptyset$. We need only verify the distributivity property of
$g_{\mathcal{D}}$ over $\squplus$. We check that, for $Z,Z^{\prime}%
,Z^{\prime\prime}$ in $\mathcal{M}in(D)$:

\begin{quote}
$g_{\mathcal{M}in(\mathcal{D})}(Z\squplus Z^{\prime},Z^{\prime\prime
})=g_{\mathcal{M}in(\mathcal{D})}(Z,Z^{\prime\prime})\squplus g_{\mathcal{M}%
in(\mathcal{D})}(Z^{\prime},Z^{\prime\prime}),$
\end{quote}

i.e., by the definitions:

\begin{quote}
$\mathrm{Min}(g_{\mathcal{D}}(\mathrm{Min}(Z\cup Z^{\prime}),Z^{\prime\prime
}))=\mathrm{Min}(\mathrm{Min}(g_{\mathcal{D}}(Z,Z^{\prime\prime}%
))\cup\mathrm{Min}(g_{\mathcal{D}}(Z^{\prime},Z^{\prime\prime}))).$
\end{quote}

The righthand side is $\mathrm{Min}(g_{\mathcal{D}}(Z,Z^{\prime\prime})\cup
g_{\mathcal{D}}(Z^{\prime},Z^{\prime\prime})).$ Clearly:

\begin{quote}
$g_{\mathcal{D}}(\mathrm{Min}(Z\cup Z^{\prime}),Z^{\prime\prime})\subseteq
g_{\mathcal{D}}(Z,Z^{\prime\prime})\cup g_{\mathcal{D}}(Z^{\prime}%
,Z^{\prime\prime}),$
\end{quote}

but, since $g_{\mathcal{D}}$ is increasing, for every $d\in g_{\mathcal{D}%
}(Z,Z^{\prime\prime})\cup g_{\mathcal{D}}(Z^{\prime},Z^{\prime\prime}),$ there
is $d^{\prime}\in g_{\mathcal{D}}(\mathrm{Min}(Z\cup Z^{\prime}),Z^{\prime
\prime})$ such that $d^{\prime}\leq d$. It follows that:

\begin{quote}
$\mathrm{Min}(g_{\mathcal{D}}(\mathrm{Min}(Z\cup Z^{\prime}),Z^{\prime\prime
}))=\mathrm{Min}(g_{\mathcal{D}}(Z,Z^{\prime\prime})\cup g_{\mathcal{D}%
}(Z^{\prime},Z^{\prime\prime})).$
\end{quote}

Hence, $\mathcal{M}in(\mathcal{D})$ is a distributive $H$-algebra.\ We call it
a \emph{minimizing H-algebra}.

\bigskip

In order to compute minimizing functions by FA, we will use the $F^{(1)}%
$-algebra $\mathcal{D}=\mathcal{D}_{F,1}:=\left\langle \mathcal{P}%
_{f}([r]^{\ast}),(g_{\mathcal{D}})_{g\in F^{(1)}}\right\rangle $ defined for
computing $\mathrm{Sat}X.P(X)$ (cf.\ Section 3.2.2.1).\ For each partial order
$\leq$ on $\mathcal{P}_{f}([r]^{\ast})$ such that the functions
$(f,w)_{\mathcal{D}}$ with $f$ of positive arity and $w\in\{0,1\}$ are
increasing, we make $\mathcal{D}$\ into a minimizing $F^{(1)}$-algebra. We
recall the definition of $(f,w)_{\mathcal{D}}$ for a binary function
$f(X,X^{\prime})$ where $X,X^{\prime}$\ are finite subsets of $[r]^{\ast}$):

\begin{quote}
$(f,w)_{\mathcal{D}}(X,X^{\prime}):=\overline{w}\cup1.X\cup2.X^{\prime}%
,\qquad\qquad\qquad\qquad(11)$

where $\overline{w}:=$ \texttt{if }$w=1$ \texttt{then} $\{\varepsilon\}$
\texttt{else} \ $\emptyset.$
\end{quote}

\bigskip

\textbf{Proposition 25}: Let $F$ be an effectively given signature and
$P(X)$\ be a property of terms over it defined by a deterministic
FA\ $\mathcal{A}$ over $F^{(1)}$. Let $\leq$\ be partial order making
$\mathcal{D}_{F,1}$\ into a minimizing algebra.\ The function $\mathrm{Min}%
_{\leq}X.P(X)$ is computable by a deterministic FA\ constructed from
$\mathcal{A}$ (defined by a tuple of programs $\underline{\mathcal{A}}$) and
the algorithm that decides $\leq$.

\bigskip

\textbf{Proof}: We apply Case (c) of Proposition 21 to the distributive and
minimizing $F^{(1)}$-algebra $\mathcal{M}in(\mathcal{D})$ defined from $\leq$.
$\square$

\bigskip

We now examine the first four partial orders on $\mathcal{P}_{f}([r]^{\ast})$
defined above.\ In each case we use Equality (11) to verify that
$(f,w)_{\mathcal{D}}$ is increasing.

\bigskip

(i) \emph{Case of} $\subseteq$.

Each function $(f,w)_{\mathcal{D}}$ is increasing, hence, we can compute for
$t\in T(F)$ the set $\mathrm{Min}_{\subseteq}X.P(X)(t):=\mathrm{Min}%
(\mathrm{Sat}X.P(X)(t))$ of inclusion minimal sets $X$ such that $t\models
P(X)$.

\bigskip

(ii) \emph{Case of }$X\leq_{anc}Y:\Leftrightarrow X\subseteq\mathit{Pref}(Y)$.

Each function $(f,w)_{\mathcal{D}}$ is increasing, in particular because
$X\leq_{anc}Y\Rightarrow i.X\leq_{anc}i.Y$. \ Hence, $\mathrm{Min}%
(\mathrm{Sat}X.P(X)(t))$ is the set of minimal sets $X$ such that $t\models
P(X)$ where minimality means that one cannot reduce a satisfying set by
removing a node $u$ or replacing it by one of its ancestors in $\mathit{Pref}%
(\{u\}).$ We denote by $\mathrm{Min}_{anc}X.P(X)$ the corresponding function.

\bigskip

(iii) \emph{Case of }$X\leq_{card}Y:\Leftrightarrow$ $\left\vert X\right\vert
\leq\left\vert Y\right\vert :$

Equality (7) shows that $\left\vert (f,w)_{\mathcal{D}}(X,X^{\prime
})\right\vert =w+\left\vert X\right\vert +\left\vert X^{\prime}\right\vert $.
Hence, $\left\vert X\right\vert \leq\left\vert Y\right\vert $ implies
$\left\vert (f,w)_{\mathcal{D}}(X,X^{\prime})\right\vert \leq\left\vert
(f,w)_{\mathcal{D}}(Y,X^{\prime})\right\vert $. We can thus compute the set
$\mathrm{Min}(\mathrm{Sat}X.P(X)(t))$ of sets $X$ of minimal cardinality such
that $t\models P(X)$. Their common cardinality is $\mathrm{MinCard}X.P(X)(t)$
that we already know how to compute. We denote by $\mathrm{Min}_{card}X.P(X)$
the corresponding function.

\bigskip

(iv) \emph{Case of} $X\leq_{clex}Y:\Leftrightarrow\left\vert X\right\vert
<\left\vert Y\right\vert $ or, $\left\vert X\right\vert =\left\vert
Y\right\vert $ and $X\leq_{lex}Y.$

We denote by $\preceq_{lex}$the lexicographic order on $[r]^{\ast}$.\ Hence,
every\ finite subset $X$ of $[r]^{\ast}$ can be written in a unique way as a
sequence of words $Seq(X):=(w_{1},...,w_{p})$ such that $X=\{w_{1}%
,...,w_{p}\}$ and $w_{1}\prec_{lex}...\prec_{lex}w_{p}$; we have
$Seq(\emptyset)=()$ not to be confused with $(\varepsilon)$. The set
$\mathcal{P}_{f}([r]^{\ast})$ can thus be ordered lexicographically ; we
denote this order by $\leq_{lex}.$ Its least element is the empty set.\ If
$X=\{1,2,11,\varepsilon,222\}$ and $Y=\{1,2,\varepsilon,111\}$\ then
$Seq(X)=(\varepsilon,1,11,2,222)$ and $Seq(Y)=(\varepsilon,1,111,2)$ so that
$X<_{lex}Y$. The order $\leq_{clex}$ is lexicographic with priority on
cardinality. We\ will denote the corresponding function by $\mathrm{Min}%
_{clex}X.P(X)$. To verify that the functions $(f,w)_{\mathcal{D}}$ are
increasing for $\leq_{clex}$, we have by (11):

\begin{quote}
$Seq((f,w)_{\mathcal{D}}(X,Y))=Seq(\overline{w})\circ1.Seq(X)\circ
2.Seq(Y),\qquad\qquad\qquad(12)$
\end{quote}

where $\circ$\ denotes the concatenation of sequences and $i.(w_{1}%
,...,w_{p}):=(i.w_{1},...,$ $i.w_{p}).$\ We have $Seq(\overline{w}):=$
\texttt{if} $w=0$\ \texttt{then} $()$\ \texttt{else} $(\varepsilon)$. It is
then clear that $X\leq_{clex}Y$ and $X^{\prime}\leq_{clex}Y^{\prime}$
imply$\ (f,w)_{\mathcal{D}}(X,X^{\prime})\leq_{clex}(f,w)_{\mathcal{D}%
}(Y,Y^{\prime}).$

\bigskip

This technique does not apply to $\leq_{lex}$\ because the functions
$(f,w)_{\mathcal{D}}$ are not increasing.

\bigskip

\emph{Example}: Let $F=\{f,g,a,b,c\}$ with $a,b,c$ nullary, $g$ unary and $f$
binary. We let $P(X)$ mean that, either each occurrence of $a$ and no
occurrence of $b$ or $c$ is below a position in $X$\ or, that each occurrence
of $b$ and no occurrence of $a$ or $c$ is below a position in $X.$\ One can
construct terms showing that the five minimization functions based on property
$P(X)$ and the orders (i)-(v) are pairwise different.$\square$

\bigskip

The constructions of this section establish the following theorem, where $F$
is an effectively given signature and $\overline{X}$ is an $s$-tuple of set variables.

\bigskip

\textbf{Theorem 26}: Let $\mathcal{A}$ be a deterministic FA\ over $F^{(s)}%
$\ that decides a property $P(\overline{X})$ and $\alpha(\overline{X})$ be a
linear function of the cardinalities of the sets forming its argument. The
functions$\ \mathrm{MinCard}X_{1}.P(\overline{X})$, $\mathrm{Min\_\alpha
}\overline{X}.P(\overline{X})$, $\mathrm{Min}_{\subseteq}X.P(X)$,
$\mathrm{Min}_{anc}X.P(X)$, $\mathrm{Min}_{card}X.P(X)$ and $\mathrm{Min}%
_{clex}X.P(X)$ are computable by deterministic FA\ constructed from $\alpha$
and the tuple $\underline{\mathcal{A}}$\ that defines $\mathcal{A}$.

\bigskip

\textbf{Proof}: The corresponding constructions are done in Section (3.2.3.1)
and Proposition 25.\ $\square$

\bigskip

This theorem does not exhaust the possibilities of building FA by general
methods, see Section 4.2.1.

\subsubsection{Parameterized complexity}

We now consider conditions ensuring that the automata constructed by Theorems
23 and 26 are P-FA, FPT-FA or XP-FA. We recall that if the signature $F$ is
finite, the notions of P-FA, FPT-FA and XP-FA\ coincide. Lemma 7 shows the
importance of the nondeterminism degree for analyzing the computation time of
determinized automata.\ 

\bigskip

\textbf{Theorem 27}: Let $F,s$, $\mathcal{A}$, $P(\overline{X})$ and $\alpha$
be as in Theorems 23 and 26.

(1) If $\mathcal{A}$ is a P-FA such that the mapping $ndeg_{pr(\mathcal{A}%
)}\ $is P-bounded, then, the properties $\exists\overline{X}.P(\overline{X})$
and $\forall\overline{X}.P(\overline{X})$ are P-FA decidable and the functions
$\mathrm{MSp}\overline{X}.P(\overline{X}),$ $\mathrm{Sp}\overline
{X}.P(\overline{X})$, $\#\overline{X}.P(\overline{X})$,$\ \mathrm{MinCard}%
X_{1}.P(\overline{X})$, $\mathrm{Min\_\alpha}\overline{X}.P(\overline{X})$ and
$\mathrm{Min}_{clex}X.P(X)$ are P-FA computable.

(2) If $\beta$: $T(F^{(s)})\rightarrow\mathcal{D}$ is computed by a P-FA
$\mathcal{A}$ such that $ndeg_{pr(\mathcal{A})}$ is P-bounded, then the
function $\mathrm{SetVal}\overline{X}.\beta(\overline{X})$ is P-FA computable.

(3) These implications hold if we replace P- by FPT- or \ XP-.

\bigskip

Since we have $ndeg_{pr(\mathcal{A})}(t)\leq|Q_{\mathcal{A}}\upharpoonright
pr^{-1}(t)|$ (cf.\ Proposition 12), we can replace in these statements, the
P-, FPT- or XP-boundings of $ndeg_{pr(\mathcal{A})}$ by the corresponding ones
for the mapping $\ t\longmapsto|Q_{\mathcal{A}}\upharpoonright pr^{-1}(t)|.$

\bigskip

\textbf{Proof}: We use Lemma 7 for all proofs.

(1) As $\mathcal{A}$ is a P-FA, $p_{1},p_{2},p_{3}$\ are polynomials.\ Then,
$pr(\mathcal{A})$ satisfies the hypotheses of Lemma 7(2).\ Hence,
$\exists\overline{X}.P(\overline{X})$ and $\forall\overline{X}.P(\overline
{X})$ are checked by $\det(pr(\mathcal{A}))$ that is a P-FA.

Next we consider the deterministic FA $\mathcal{A}^{\mathrm{MSp}}$ that
computes $\mathrm{MSp}\overline{X}.P(\overline{X})$ (Section 3.2.2.3).\ At
position $u$ in a term $t$, the state of $\mathcal{A}^{\mathrm{MSp}}$ is the
set $\{(q,m)\mid m\neq\emptyset\}$ where $m$ is a multiset of $s$-tuples of
integers that are cardinalities of subsets of $Pos(t)$. The cardinality of
this set is bounded by $ndeg_{pr(\mathcal{A)}}(t)$. Each multiset $m$\ is a
function: $\mathbb{N}^{s}\mathbb{\rightarrow N}$\ that maps $[0,n]^{s}%
\ $to$\ [0,2^{s.n}]$ where $n:=|Pos(t)|$. It is finite and can be encoded by a
word of length at most $(n+1)^{s}.\log(2^{s.n})=O(n^{s+1}).$ (The numbers
$m(\overline{x})$ for $\overline{x}\in\lbrack0,n]^{s}\ $are written in
binary). Hence the size of a state is $O(ndeg_{pr(\mathcal{A)}}(t).\Vert
t\Vert^{s+1}).$

We must also bound the time for computing the transitions and the output.

The time for computing a transition of $\mathcal{A}$ is bounded by
$p_{1}(\Vert t\Vert)$. Computing the transition of $\mathcal{A}^{\mathrm{MSp}%
}$ at a nullary symbol of $t$ takes time at most $2^{s}.p_{1}(\Vert t\Vert).$
We now examine the computation of a transition $f[\sigma_{1},\sigma
_{2}]\rightarrow\sigma$ by using the description made in Section
(3.2.2.3).\ Given $\sigma_{1},\sigma_{2}$ we build $\sigma$, defined as a
finite subset of $Q_{\mathcal{A}}\times\lbrack0,n]^{s}\times\lbrack2^{s.n}%
]$.\ The last component is a number written in binary and a tuple in $\sigma$
is denoted by $(q,\overline{x},\sigma(q)(\overline{x}))$. We omit the tuples
$(q,\overline{x},\sigma(q)(\overline{x}))$ such that $\sigma(q)(\overline
{x})=0.$ We initialize $\sigma$ with the empty set.\ There are at most
$2^{s}.(ndeg_{pr(\mathcal{A})}(t))^{2}.(n+1)^{2s}$ tuples of the form
$(w,q_{1},q_{2},\overline{y},\overline{z})$ such that $\sigma_{1}%
(q_{1})(\overline{y})\neq0,\sigma_{2}(q_{2})(\overline{z})\neq0.$ For each of
them, we compute $q$ such that $(f,w)[q_{1},q_{2}]\rightarrow_{\mathcal{A}}q$,
$\overline{x}=w\boldsymbol{+}\overline{y}\boldsymbol{+}\overline{z}$, and we
add $\sigma_{1}(q_{1})(\overline{y}).\sigma_{2}(q_{2})(\overline{z})$ to the
current value of $\sigma(q)(\overline{x})$. The computation time is
$O((ndeg_{pr(\mathcal{A})}(t))^{2}.n^{2s}.(p_{1}(\Vert t\Vert)+n^{2}))$.\ (The
term $n^{2}$ represents the computation time for the arithmetic operations on
integers in $[2^{s.n}]$.)  For $f$ of arity $r$, we get
$O((ndeg_{pr(\mathcal{A})}(t))^{r}.n^{r.s}.(p_{1}(\Vert t\Vert)+n^{2}))$,
which is P-bounded.

Similarly, for computing the output, we need at most $ndeg_{pr(\mathcal{A)}%
}(t)$ checks that a state is accepting, with cost at most $p_{3}(\Vert
t\Vert)$ for each and the same number of unions of multisets defined as
functions: $[0,n]^{s}\rightarrow\lbrack0,2^{s.n}]$.\ This gives a computation
time bounded by $ndeg_{pr(\mathcal{A)}}(t).(p_{3}(\Vert t\Vert)+O(n^{s+1})).$
Again, as $ndeg_{pr(\mathcal{A)}}(t)$ is P-bounded, the bound on the
computation time of the output is of same type. \ 

We get the announced result for $\mathrm{MSp}\overline{X}.P(\overline{X})$.
For $\mathrm{Sp}\overline{X}.P(\overline{X})$, $\#\overline{X}.P(\overline
{X})$, $\mathrm{MinCard}X_{1}.P(\overline{X})$ and $\mathrm{Min\_\alpha
}\overline{X}.P(\overline{X})$, the functions used to compute transitions are
simpler than those for $\mathrm{MSp}\overline{X}.P(\overline{X})$. The size of
$m$ in a state $(q,m)$ is smaller, and so are the computation times of the
transitions and the output.\ Hence, the above argument applies as well. For
$\mathrm{Min}_{clex}X.P(X)$ we observe that a state is a pair $(q,m)$ where
$m$ is the empty set or a single ($\leq_{clex}$-minimal) \ set ($s=1$).

\bigskip

(2) We now consider $\mathrm{SetVal}\overline{X}.\beta(\overline{X})$ where
$\beta$ is computed by a deterministic FA $\mathcal{A}$ over $F^{(s)}$.\ For
each term $t$ and $\overline{X}\in\mathcal{P}(Pos(t))^{s}$ we have
$\beta(t\ast\overline{X})=$ \ $Out_{\mathcal{A}}(q_{\mathcal{A}}%
(t\ast\overline{X}))$.\ Hence, $\mathrm{SetVal}\overline{X}.\beta(\overline
{X})(t)$ is the set of values $Out_{\mathcal{A}}(q)$ for $q\in q_{\det
(pr(\mathcal{A}))}(t).$ The time taken to compute $\mathrm{SetVal}\overline
{X}.\beta(\overline{X})(t)$ is that for computing the set $q_{\det
(pr(\mathcal{A}))}(t)$ of cardinality at most $ndeg_{pr(\mathcal{A)}}(t)$ plus
that for computing the final output, bounded by $ndeg_{pr(\mathcal{A)}%
}(t).p_{3}(\Vert t\Vert)$.\ Hence, we conclude as in the cases considered in (1).

(3) The proofs are similar if $p_{1},p_{2},p_{3}$ and $ndeg_{pr(\mathcal{A)}}$
are FPT- or XP-bounded.$\square$

\bigskip

\textbf{Remarks 28:} (1) Even if $F$ is finite, we cannot omit in Theorem
27\ the hypothesis that $pr(\mathcal{A})$\ has a nondeterminism degree bounded
in some way, because the validity of $\exists\overline{X}.P(\overline{X})$ can
be determined in polynomial time from either $\mathrm{MSp}\overline
{X}.P(\overline{X})$, $\mathrm{Sp}\overline{X}.P(\overline{X})$,
$\#\overline{X}.P(\overline{X})$ or $\mathrm{MinCard}X_{1}.P(\overline{X})$.
Otherwise, by Counter-example 13, we would have \textbf{P}=\textbf{NP.}

(2) Theorem 27 does not apply to $\mathrm{Min}_{\subseteq}X.P(X)$,
$\mathrm{Min}_{anc}X.P(X)$ and

$\mathrm{Min}_{card}X.P(X)$ because their outputs may be of exponential size
in the size of the input tree.\ 

\subsection{Summary of results}

\bigskip

The following table summarizes the \emph{preservation results} of this
section: we mean by this that the classes of functions and properties that are
P-FA, FPT-FA or XP-FA computable (or decidable) are preserved under
constructions of three types: composition, first-order and monadic
second-order constructions.%

\begin{tabular}
[c]{|l|c|c|}\hline
& Construction \  & Conditions and proofs\\\hline\hline
Composition & \multicolumn{1}{|l|}{$g\circ(\mathcal{\alpha}_{1}%
,...,\mathcal{\alpha}_{r})${\small , }} & \multicolumn{1}{|l|}{$g$ is
\textbf{P}-computable, \ }\\
& \multicolumn{1}{|l|}{{\small \texttt{if} }$P${\small \texttt{ then} }%
$\alpha_{1}${\small \texttt{else} }$\alpha_{2},$} &
\multicolumn{1}{|l|}{$S_{1},...,S_{m}$ are set terms;}\\
& \multicolumn{1}{|l|}{$\lnot P$, $P\vee Q$, $P\wedge Q$, $\alpha
\upharpoonright P,$} & \multicolumn{1}{|l|}{by Proposition 15}\\
& \multicolumn{1}{|l|}{$\mathcal{\alpha(}S_{1},...,S_{m})$, $P(S_{1}%
,...,S_{m}).$} & \multicolumn{1}{|l|}{and Theorem 17.}\\\hline
FO const. & \multicolumn{1}{|l|}{$\exists\overline{x}.P(\overline{x})$,
$\forall\overline{x}.P(\overline{x})$, $\mathrm{SetVal}\overline{x}%
.\alpha(\overline{x}),$} & \multicolumn{1}{|l|}{by Theorem 17, Corollary
18.}\\
& \multicolumn{1}{|l|}{$\mathrm{Sat}\overline{x}.P(\overline{x})$,
$\#\overline{x}.P(\overline{x}).$} & \multicolumn{1}{|l|}{}\\\hline
MS const. & \multicolumn{1}{|l|}{$\exists\overline{X}.P(\overline{X})$,
$\forall\overline{X}.P(\overline{X})$,} & \multicolumn{1}{|l|}{$P$ or $\alpha$
is defined by}\\
& \multicolumn{1}{|l|}{$\mathrm{SetVal}\overline{X}.\alpha(\overline{X}),$} &
\multicolumn{1}{|l|}{a P-FA $\mathcal{A}$}\\
& \multicolumn{1}{|l|}{$\#\overline{X}.P(\overline{X})$, $\mathrm{MSp}%
\overline{X}.P(\overline{X})$,} & \multicolumn{1}{|l|}{such that
$ndeg_{pr(\mathcal{A})}$}\\
& \multicolumn{1}{|l|}{$\mathrm{Sp}\overline{X}.P(\overline{X})$,
$\mathrm{MinCard}X.P(X),$} & \multicolumn{1}{|l|}{is P-, FPT- or
XP-bounded;}\\
& \multicolumn{1}{|l|}{$\mathrm{Min}_{clex}X.P(X)$} & \multicolumn{1}{|l|}{by
Theorem 27.}\\\hline
\end{tabular}

\begin{center}
Table 1: Preservation results.
\end{center}

In the next section, we develop constructions specific to graphs.

\section{Application to graphs}

\bigskip

We wish to check $\exists\overline{X}.P(\overline{X})$, $\forall\overline
{X}.P(\overline{X})$ and to compute $\mathrm{MSp}\overline{X}.P(\overline
{X}),\mathrm{Sp}\overline{X}.P(\overline{X})$ etc. in graphs $G(t)$ defined by
terms $t$ in $T(F_{\infty})$. We recall that if $P(\overline{X})$ is a graph
property with $s$ sets of vertices as auxilliary arguments, then
$L_{P(\overline{X})}:=\{t\ast\overline{X}\in T(F_{\infty}^{(s)})\mid
G(t)\models P(\overline{X})\}$.\ The following fundamental result is proved in
\cite{BCID}, Section 7.3.1\ and in \cite{FREC2014}.

\bigskip

\textbf{Theorem 29}: If $P(\overline{X})$ is MS expressible, then the language
$L_{P(\overline{X})}$ is recognized by a linear FPT-FA.

\bigskip

The proof uses an induction on the structure of the formula $\varphi$ that
expresses $P(\overline{X})$.\ Fly-automata are built\ for the atomic
formulas\footnote{For formulas of \emph{counting monadic second-order logic},
we also need FA for the atomic formulas $Card_{p,q}(X_{1})$ expressing that
$X_{1}$ has cardinality $p$ modulo $q$, see the appendix.}\ $X_{1}\subseteq
X_{2}$ and $edg(X_{1},X_{2})$.\ The constructions of Proposition 15(3,4) and
Theorem 27 are then used\ for handling logical connectives. The inductive
construction shows that for each automaton built in this way, the number of
states it reaches by runs on a term $t$ depends only on $\varphi$ and $\max
\mu(t)$ (this number bounds the clique-width of the graph $G(t)$). It follows
from Theorem 27 that the functions $\mathrm{MSp}\overline{X}.P(\overline{X}),$
$\mathrm{Sp}\overline{X}.P(\overline{X})$, $\#\overline{X}.P(\overline{X})$,
$\mathrm{MinCard}X_{1}.P(\overline{X})$, $\mathrm{Min\_\alpha}\overline
{X}.P(\overline{X})$ and $\mathrm{Min}_{clex}X.P(X)$ are computable by
FPT-FA\footnote{We will also apply this theorem to properties $P(\overline
{X})$ that are defined by FA\ without being MS\ expressible.}.

\bigskip

\textbf{Remark 30}: To simplify the discussion, we let $P$ be an
MS\ expressible graph property without set arguments. A consequence of Theorem
29\ (called in \cite{CouEng} the \emph{Weak Recognizability Theorem}) is that
for every integer $k$, the language $L_{P}\cap T(F_{k})$ is recognized by a
finite automaton $\mathcal{A}_{P,k}$. A quick proof of this fact follows from
the observation that the mapping $t\mapsto G(t)$ is a \emph{monadic
second-order transduction} from $T(F_{k})$ to the class of graphs of
clique-width at most $k$ and the \emph{Backwards Translation Theorem}%
\footnote{It says that if $\tau$ is a monadic second-order transduction and
$L$ is a monadic second-order definable class of structures, then $\tau
^{-1}(L)$ is monadic second-order definable (\cite{CouEng}, Theorem
7.10).}.\ However, this technique is not applicable to $L_{P}\subseteq
T(F_{\infty})$ because the signature $F_{\infty}$ is infinite so that the
mapping $t\mapsto G(t)$ is not a monadic second-order transduction on
$T(F_{\infty})$. From the practical view point, an FA $\mathcal{A}_{P,k}$
constructed from this observation would be anyway very complicated and hard to
implement.$\ \square$

\bigskip

Graphs are always given by terms over $F_{\infty}$ or $F_{\infty}^{\mathrm{u}%
}$ (and not by adjacency lists).\ The constructions of Section 3 that are done
for FA\ over an arbitrary effectively given signature have immediate
applications to graphs via the signature $F_{\infty}$.\ One adaptation to make
is due to the fact that the set arguments $X_{1},...,X_{s}$ denote sets of
vertices of the defined graphs, hence sets of positions in the input terms of
the nullary symbols in $\mathbf{C}$. For example, the algebra $\mathcal{D}%
_{F,s}$ used in Section (3.2.2.1) for computing $\mathrm{Sat}\overline
{X}.P(\overline{X})$ must be modified into $\mathcal{D}^{\prime}$ such that,
for the binary symbol $\oplus$, $\oplus_{\mathcal{D}^{\prime}}(\overline
{X},\overline{Y}):=1.\overline{X}\cup2.\overline{Y}$. Similarily, for
computing $\mathrm{Sp}\overline{X}.P(\overline{X})$, we take $\mathcal{D}%
^{\prime\prime}$ such that $\oplus_{\mathcal{D}^{\prime\prime}}(\overline
{m},\overline{p}):=\overline{m}+\overline{p}.$ For the unary symbols $f$ of
$F_{\infty}$ (they are $relab_{h}$ or $\overrightarrow{add}_{a,b}$), we take
$f_{\mathcal{D}^{\prime}}(\overline{X}):=1.\overline{X}$ \ and $f_{\mathcal{D}%
^{\prime\prime}}(\overline{m}):=\overline{m}.$

\bigskip

Although the FA\ for the atomic formulas $X_{1}\subseteq X_{2}$,
$edg(X_{1},X_{2})$\ and $Card_{p,q}(X_{1})$\ suffice for proving Theorem 29,
it is useful to "precompute" FA for other frequently used MS\ properties.
Table 2 lists bounds the sizes of the states in their runs on terms in
$T(F_{k})$.\ We will define FA for some other basic properties and functions.
By combining these automata as explained in the previous section, we can
easily build automata for checking properties and computing functions
expressed by formulas written with the basic ones and the logical connectives
of MS\ logic. The FA\ of Table 2\ concern the following
properties:\ $Partition(X_{1},...,X_{s})$ meaning that $(X_{1},...,X_{s})$ is
a partition of the vertex set, $St$ that the considered graph is
\emph{stable,} i.e., has no edge, \ $Link(X_{1},X_{2})$ that it has at least
one edge from some vertex of $X_{1}$ to some vertex of $X_{2}$, \ $Path(X_{1}%
,X_{2})$ that $X_{1}$ consists of two vertices linked by an undirected path
with vertices in $X_{2}$ ($X_{2}$\ must contain $X_{1}$), $Clique$ that the
graph is a clique, $Conn$ that it is connected, $Cycle$ that it has an
undirected cycle and $DirCycle$ that it has a directed cycle. Finally,
$edg(X_{1},X_{2})$ is equivalent to $Link(X_{1},X_{2})\wedge Sgl(X_{1})\wedge
Sgl(X_{2}).$ The automata are constructed in \cite{BCID} and the bounds on
sizes of states are clear by inspecting the constructions.

\bigskip%

\begin{tabular}
[c]{|c|c|}\hline
Property & Size of a state\\\hline\hline
\multicolumn{1}{|l|}{$Sgl$, $X_{1}\subseteq X_{2}$, $X_{1}=\emptyset$,
$Card_{p,q}(X_{1})$} & \multicolumn{1}{|l|}{independent of $k$}\\\hline
\multicolumn{1}{|l|}{$Partition(X_{1},...,X_{s})$} &
\multicolumn{1}{|l|}{independent of $k$}\\\hline
\multicolumn{1}{|l|}{$edg(X_{1},X_{2})$} & \multicolumn{1}{|l|}{$O(\log(k))$%
}\\\hline
\multicolumn{1}{|l|}{$St$, $Link(X_{1},X_{2})$} & \multicolumn{1}{|l|}{$O(k)$%
}\\\hline
\multicolumn{1}{|l|}{$Path(X_{1},X_{2})$, $DirCycle$, $Clique$} &
\multicolumn{1}{|l|}{$O(k^{2})$}\\\hline
\multicolumn{1}{|l|}{$Conn,Cycle$} & \multicolumn{1}{|l|}{$O(\log
(k).\min\{n,k.2^{O(k)}\})$}\\\hline
\end{tabular}

\begin{center}
Table 2: Sizes of states for some automata running on terms in $T(F_{k})$.
\end{center}

All automata are P-FA because computing the transitions involves only
polynomial-time calculations. For the automata checking $Conn$\ and $Cycle$,
the upper-bound $O(\log(k).n)$ ($n$ is the number of vertices of the input
graph) shows that they are P-FA.

Properties $Sgl$, $St$, $Conn$, $DirCycle$, $Cycle$ and $Clique$ are relative
to the whole graph $G$.\ However, we need frequently their relativizations to
sets of vertices, for example $St[X]$ meaning that the induced subgraph $G[X]$
is stable (cf.\ the examples in the Introduction).\ However, from an FA over
$F_{\infty}$ that decides $St$, one gets by taking an appropriate inverse
image\footnote{We recall from Section 1.3 that it is based on the relabelling
$h$: $F_{\infty}^{(1)}\rightarrow F_{\infty}$ such that, for every
$\mathbf{a}\in\mathbf{C}$ we have $h((\mathbf{a},0)):=\boldsymbol{\varnothing
,}$ $h((\mathbf{a},1)):=\mathbf{a}$ and $h(f):=f$ for all other operations of
$F_{\infty}$.\ The same inverse image works for relativizing any property.} an
FA over $F_{\infty}^{(1)}$ that decides $St[X]$.

\bigskip

We defined in Section 1.2 the notions of good and irredundant
terms.\ Proposition 35\ in the appendix gives a polynomial-time algorithm that
transforms a term into an equivalent good and irredundant one.\ We can build a
P-FA $\mathcal{GI}$ that checks if the input term is good and irredundant.\ If
$\mathcal{A}$ is a deterministic FA, then an FA constructed with Proposition
15\ from the product of $\mathcal{A}$ and $\mathcal{GI}$ gives correct results
on good irredundant terms and rejects the others.\ It has the same type (P,
FPT or XP) as $\mathcal{A}$.

A last technical point concerns notation.\ When dealing with terms $t$ over
effectively given signatures, we denote by $\#\overline{X}.P(\overline{X})$
the mapping associating with a term $t$ the number of tuples $\overline{X}%
$\ of sets of positions that satisfy property $P$ in $t$.\ In the present
section, we will denote in the same way the mapping associating with a graph
$G$ the number of tuples of sets of vertices that satisfy $P$, and also the
mapping $t\longmapsto\#\overline{X}.P(\overline{X})(G(t))$ for $t\in
T(F_{\infty})$, that we wish to compute by FA. The same convention will apply
to $\mathrm{MSp}\overline{X}.P(\overline{X})$, $\mathrm{Sp}\overline
{X}.P(\overline{X})$ etc.

\subsection{Counting induced subgraphs}

\bigskip

Let $H$ be a connected undirected graph.\ An induced subgraph of an undirected
graph $G$ is $H$-\emph{induced} if it is isomorphic to $H$.\ We can use FA to
count and enumerate the $H$-induced subgraphs of a given graph. The property
of a set $X\subseteq V_{G}$\ that $G[X]\simeq H$ is MS\ expressible.\ Hence,
automata that compute the functions $\#X.G[X]\simeq H$ and $\mathrm{Sat}%
X.G[X]\simeq H$ will give us the desired algorithms. The property $G[X]\simeq
H$ implies that $X$ has fixed cardinality $|V_{H}|$.\ Hence, we can apply
Corollary 18 and the following remark. However, a direct construction yields
in general a smaller FA.\ 

For example let $H$ be the graph $\mathit{House}$, i.e., the graph $K_{5}%
$\ with vertex set $[5]$ minus the four edges $1-4,1-5,3-4$ and $2-5$. We let
\ $\overline{X}=(X_{1},X_{2},X_{3},X_{4},X_{5})$ \ and $P(\overline{X})$ stand for:

$edg(X_{1},X_{2})\wedge edg(X_{1},X_{3})\wedge edg(X_{2},X_{3})\wedge
edg(X_{2},X_{4})\wedge edg(X_{4},X_{5})\wedge$

$edg(X_{3},X_{5})\wedge\lnot edg(X_{1},X_{4})\wedge\lnot edg(X_{1}%
,X_{5})\wedge\lnot edg(X_{3},X_{4})\wedge\lnot edg(X_{2},X_{5}).$

A P-FA over $F_{k}^{\mathrm{u}}$ with $O(k^{2})$ states for $edg(X,Y)$ is
constructed in \cite{BCID}, Section 5.1.2 and \cite{CouEng}, Section 6.3. From
Propositions 15 and 16, we get for $P(\overline{X})$ a P-FA that uses
$O(k^{20})$ states on terms in $T(F_{k}^{\mathrm{u}(5)})$, but a specific
construction yields a P-FA using $O(k^{5})$ states on these terms. By
Corollary 18, we get FA that compute $\#\overline{X}.P(\overline{X})$ and
$\mathrm{Sat}\overline{X}.P(\overline{X}).$\ However, the number of
$\mathit{House}$-induced subgraphs of $G(t)$ is\ only half of $\#\overline
{X}.P(\overline{X})(t)$ because $\mathit{House}$ has one automorphism apart
from identity.\ Hence, the FA that computes $\#\overline{X}.P(\overline
{X})(t)$\ does some useless computations. We can avoid this drawback by
replacing $P(\overline{X})$ by $P(\overline{X})\wedge X_{2}<X_{3}$ where
$<$
is the lexicographic order on positions of the input term. An FA defining
$<$
is easy to build.\ The role of this condition is to select a single 5-tuple
for each $\mathit{House}$-induced graph. This linear order on $V_{G(t)}$
depends on the term $t$ and the definition by Dewey words of the
vertices.\ However, the value $\#\overline{X}.P(\overline{X})$\ is the same
for all terms: it is \emph{order-invariant}\ (cf. \cite{Cou96} on this notion
and \cite{EKS} for its applications to model-checking).

The same improvement applies to the enumeration problem in order to avoid
duplications in the enumeration of $House$-induced subgraphs. But even without
using any linear order, Theorem 17 and Corollary 18\ yield P-FA\ that compute
the functions $\#X.G[X]\simeq H$ and $\mathrm{Sat}X.G[X]\simeq H$ for each
fixed graph $H$.

\subsection{Edge counting and degree}

For a p-graph $G$ and $X\subseteq V_{G}$, we denote by $\beta_{X}$ the mapping
that gives, for each label $a$ the number of $a$-ports in $X$. If $X=V_{G}$,
we denote it by $\beta_{G}$. We denote by $\Lambda\lbrack k,n]$ the set of
mappings $\beta:[k]\rightarrow\lbrack0,n]$ such that $\Sigma_{i\in\lbrack
k]}\beta(i)\leq n.$ This set has cardinality\ $\binom{n+k}{k}$ (by an easy
bijective proof), hence $\Theta(n^{k})$ for fixed $k$. We will bound it by
$(n+1)^{k}.$

All automata in this section will be constructed so as to work correctly on
good irredundant terms\footnote{It is not hard to see that a term $t$ in
$T(F_{\infty}^{(s)})$ is good (resp.\ irredundant) if and only if $pr_{s}(t)$
is.}.\ Irredundancy is useful for counting edges and we recall that the size
of a good term $t\in T(F_{k}^{(s)})$ is $O(n.k^{2})$ where $n$ is the number
of vertices of $G(t)$. Hence, computation times can be bounded in function of
$n$.

\bigskip

(4.2.1) \textit{Counting the edges of induced subgraphs}

Given a directed graph $G$ and $X\subseteq V_{G}$, we let$\ e(X)$ be the
number of edges of $G[X]$.\ This value is not the cardinality of a set
$Y\subseteq V_{G}$ satisfying a property $P(X,Y)$ by an obvious cardinality
argument. However, we will compute it by an attributed FA\ $\mathcal{B}$\ over
$F_{\infty}^{(1)}$ (cf. Remark 22).\ 

We let\ $\mathcal{B}:=\mathcal{A}\ltimes\mathcal{D}$ where $\mathcal{A}$ has
set of states $[\mathbb{N}_{+}\rightarrow\mathbb{N}]_{f}$ ($\mathbb{N}_{+}$ is
the set of port labels), $q_{\mathcal{B}}(t\ast X)=(\beta_{X},e(X))$ for every
$t\ast X\in T(F_{\infty}^{(1)})$ where $\beta$ and $e$ are relative to $G(t)$.
The transitions of $\mathcal{A}$ are as follows:

\begin{quote}
$\oplus\lbrack\beta,\beta^{\prime}]\rightarrow\lambda x\in\mathbb{N}%
_{+}.(\beta(x)+\beta^{\prime}(x)),$

$\overrightarrow{add}_{a,b}[\beta]\rightarrow\beta$,

$relab_{a\rightarrow b}[\beta]\rightarrow\beta^{\prime}$ where $\beta^{\prime
}(a):=0,$ $\beta^{\prime}(b):=\beta(a)+\beta(b)$ and $\beta^{\prime}%
(x):=\beta(x)$ if $x\notin\{a,b\}$,

$(\mathbf{a},i)\rightarrow\lambda x\in\mathbb{N}_{+}.($\texttt{if} $x=a$
\texttt{then} $i$ \texttt{else} $0)$, where $i\in\{0,1\}$.
\end{quote}

We now define an ($F_{\infty}^{(1)}\times Q_{\mathcal{A}})$-algebra
$\mathcal{D}$ (cf. Remark 22). Its domain is $\mathbb{N}$ and its operations are:

\begin{quote}
$(\oplus,\beta,\beta^{\prime})_{\mathcal{D}}(m,m^{\prime}):=m+m^{\prime},$

$(\overrightarrow{add}_{a,b},\beta)_{\mathcal{D}}(m):=m+\beta(a).\beta(b)$,

$(relab_{a\rightarrow b},\beta)_{\mathcal{D}}(m):=m,$

$(\mathbf{a},i)_{\mathcal{D}}:=0$.
\end{quote}

The definition of $(\overrightarrow{add}_{a,b},\beta)_{\mathcal{D}}$ is
correct because we assume $t$ irredundant. The value $e(X)$ is the second
component of the state reached by\ $\mathcal{B}:=\mathcal{A}\ltimes
\mathcal{D}$ at the root of $t\ast X\in T(F_{\infty}^{(1)})$.\ Let $t\ast X\in
T(F_{k}^{(1)})$ denote a graph $G(t)$ with $n$ vertices (and $X\subseteq
V_{G}(t)$).\ Then $q_{\mathcal{B}}(t\ast X)=(\beta_{X},e(X))\in\Lambda\lbrack
k,x]\times\lbrack0,x(x-1)]\subseteq\Lambda\lbrack k,n]\times\lbrack0,n(n-1)]$
where $x:=\left\vert X\right\vert $. There are less than $(n+1)^{k+2}$ such
states and they have size $O(k.\log(n))$. Transitions and outputs can be
computed in time $O(k.\log^{2}(n))$ and so, $\mathcal{B}$ is a P-FA. (The
$\log^{2}(n)$ factor comes from the multiplication of two positive integers in
$[0,n]$).

An algorithm of \cite{BGP}\footnote{The algorithms of this article assume
implicitly that the input terms are irredundant.\ Since the preprocessing that
makes a term irredundant takes linear time, the given upper bounds to
computation times are correct. This article also gives tight lower bounds to
these computation times under the exponential time hypothesis.} computes the
function $\ \mathrm{Min\_e}X.(\left\vert X\right\vert =p)$, i.e., the minimum
number of edges of an induced subgraph having $p$ vertices.\ This is called
the \textsc{sparse} $p$-\textsc{subgraph} problem. This algorithm takes time
$n.p^{O(k)}$ on terms in $T(F_{k}^{\mathrm{u}})$.\ We can obtain it as an
instance of our constructions by applying Case (c) of Proposition 21 and
Definition 24.\ The construction we will describe works for directed graphs
and, by an easy adaptation, for undirected ones.

We let $\mathcal{A}_{Card=p}$ be the deterministic FA over $F_{\infty}^{(1)}$
that checks the equality $\left\vert X\right\vert =p.$ We let $\mathcal{B}%
_{p}:=(\mathcal{A}_{Card=p}\times\mathcal{A})\ltimes\mathcal{D}$ (we omit some
easy formal details) be the attributed FA\ that computes $e(X)$ for sets $X$
of cardinality at most $p$. Let $t\ast X\in T(F_{k}^{(1)})$.\ The state
$q_{\mathcal{B}_{p}}(t\ast X)$ is $(\left\vert X\right\vert ,\beta_{X},e(X))$
if $\left\vert X\right\vert \leq p$ and $(Error,\beta_{X},e(X))$
otherwise.\ Clearly, $(\left\vert X\right\vert ,\beta_{X},e(X))\in
\lbrack0,p]\times\Lambda\lbrack k,p]\times\lbrack0,p(p-1)].$ The states
$(Error,\beta_{X},e(X))$ can be merged into a unique $Error$ state. The
accepting states are those of the form $(p,\beta,m)$ and the computed value is
$m=e(X)$ if the given set $X$ has cardinality $p$. The number of states
$(\left\vert X\right\vert ,\beta_{X},e(X))$\ is less than $(p+1)^{k+3}$, these
states have size $O(k.\log(p))$, the computation time of a transition is
$O(k.\log^{2}(p))$ and $\mathcal{B}_{p}$ is a P-FA\footnote{Its parameter is
the bound $k$ on clique-width, but it is also a P-FA for $k+p$ as parameter.}.

For computing $\ \mathrm{Min\_e}X.(\left\vert X\right\vert =p),$ we make
$\mathcal{D}$ into a minimizing algebra (cf. Definition 24) by using the
natural order on $\mathbb{N}$.\ Then, $\boldsymbol{0}_{\mathcal{D}}=0$,
$m\squplus m^{\prime}:=\min\{m,m^{\prime}\}$. We take then $\mathcal{C}%
:=pr_{1}(\mathcal{B}_{p})$ whose nondeterminism degree is less than
$(p+1)^{k+3}$ on a term in $T(F_{k}).$\ The construction of Case (c) of
Proposition 21 gives a deterministic FPT-FA\ $\mathcal{C}^{\prime}$, whose
computation time is $O(\left\vert t\right\vert .k.\log^{2}(p).p^{2k+6}%
)=O(n.k^{3}.\log^{2}(p).p^{2k+6})$ on input $t\in T(F_{k})$ where $n$ is the
number of vertices of $G(t)$.

\bigskip

More generally, we define $e(\overline{X}):=e(X_{1})+...+e(X_{s})$ and we want
to compute the function $\mathrm{Min}\_e\overline{X}.P(\overline{X})$ where
$P(\overline{X})$ is defined by a deterministic FA $\mathcal{A}_{P}$\ over
$F_{\infty}^{(s)}$.\ We extend the construction given above for
$\mathrm{Min\_e}X.(\left\vert X\right\vert =p)$: for each $i=1,...,s$, we let
$\mathcal{A}_{i}$ compute $\beta_{X_{i}}$ (it is an inverse image of
$\mathcal{A}$) and we build an attributed FA\ $\mathcal{B}_{P}:=(\mathcal{A}%
_{1}\times...\times\mathcal{A}_{s}\times\mathcal{A}_{P})\ltimes\mathcal{D}$
such that $q_{\mathcal{B}_{P}}(t\ast\overline{X})=(\beta_{X_{1}}%
,...,\beta_{X_{s}},e(\overline{X})).$ Then, we make $\mathcal{D}$ into a
minimizing algebra as above and we obtain in the same way a deterministic
FA\ that computes $\mathrm{Min}\_e\overline{X}.P(\overline{X})$. Its type, FPT
or XP, depends on $\mathcal{A}_{P}$.\ 

\bigskip

(4.2.2) \textit{Counting the edges between disjoint sets of vertices}

We consider directed graphs.\ We generalize the notion of outdegree of a
vertex by defining $\ e(X_{1},X_{2})$ as the number of edges from $X_{1}$ to
$X_{2}$ if $X_{1}$ and $X_{2}$ are disjoint sets of vertices and as $\bot$
otherwise. Hence $e(\{x\},V_{G}-\{x\})$ is the outdegree of $x$ in $G$. To
compute this function similarly as in (4.2.1), we define an attributed
FA\ $\mathcal{B}:=\mathcal{A}\ltimes\mathcal{D}$ over $F_{\infty}^{(2)}$. Its
set of states is $\{(Error,0)\}\cup([\mathbb{N}_{+}\rightarrow\mathbb{N}%
]_{f}^{2}\times\mathbb{N})$ and we want that, for $t\ast(X_{1},X_{2})\in
T(F_{\infty}^{(2)})$:

\begin{quote}
$q_{\mathcal{B}}(t\ast(X_{1},X_{2}))=(Error,0)$ if $X_{1}\cap X_{2}%
\neq\emptyset$, and

$q_{\mathcal{B}}(t\ast(X_{1},X_{2}))=((\beta_{X_{1}},\beta_{X_{2}}%
),e(X_{1},X_{2}))$ otherwise.
\end{quote}

The transitions and the algebra $\mathcal{D}$ are easy to define. On a term in
$T(F_{k}^{(2)})$ that denotes a graph with $n$ vertices, each state belongs to
the set $\{(Error,0)\}\cup(\Lambda\lbrack(k,n]^{2}\times\lbrack0,(n-1)^{2}])$
of cardinality less than $(n+1)^{2k+2}$ hence, has size $O(k.\log(n))$.
Transitions and outputs can be computed in time $O(k.\log(n)^{2})$. Hence,
$\mathcal{B}$ is a\ P-FA.

\bigskip

(4.2.3) \textit{Maximum directed cut}

For a directed graph $G$, we want to compute the maximal number of edges from
a subset $X$\ of $V_{G}$\ to its complement, hence the maximal value of
$e(X,X^{c})$.\ This problem is considered in \cite{LKM,GHO}. The deterministic
FA of Section (4.2.2), adapted by Proposition 16 to check $e(X,X^{c})$ uses
less than $(n+1)^{2k+2}$ states on a term in $T(F_{k}^{(1)})$ denoting a graph
$G$ with $n$ vertices. By the method used in Section (4.2.1), we get an
algorithm that computes the maximal value of $e(X,X^{c}),$ for $X\subseteq
V_{G},$\ in time $O(n^{4k+a})$ for some constant $a$. The article \cite{GHO}
gives an algorithm taking time $O(n^{4.2^{r(G)}+b})$ where $r(G)$ is the
\emph{bi-rankwidth} of the considered graph $G$. We recall that $r(G)/2\leq
cwd(G)\leq2.2^{r(G)}$ \cite{KanRao}.\ Hence, our method gives an algorithm of
comparable time complexity.

\subsection{Regularity of a graph}

\bigskip

The regularity of an undirected graph is not MS expressible because the
complete bipartite graph $K_{n,m}$ is regular if and only if $n=m$ and we can
apply the arguments of Proposition 5.13 of \cite{CouEng} for proving this claim.

That a graph is not regular can be expressed by the formula $\exists
X,Y.(P(X,Y)\wedge Sgl(X)\wedge Sgl(Y))$ where $P(X,Y)\ $\ is the property
$\ e(X,X^{c})\neq e(Y,Y^{c}).$ By the construction of (4.2.2) and Propositions
15 and 16 it$\ $is P-FA decidable.\ We can apply Proposition 11(1) to get a
P-FA\ for checking that a graph is not regular, hence also a P-FA\ that checks
regularity. However, we can construct directly a simpler P-FA without using an
intermediate nondeterministic automaton.

Let $G$ be defined by an irredundant term $t$. The key fact is that if in
$G(t)/u,$ two $a$-ports $x$ and $y$ have degrees $d$ and $d^{\prime},$ then
they have in $G$ degrees $d+p$ and $d^{\prime}+p$ for some $p\geq0$.\ The
reason is that if an operation at position $w$ above $u$ adds an edge between
$x$ and some vertex $z$, then it also adds an edge between $y$ and $z$ because
the labels of $x$ and $y$ are the same in $G(t)/w$ and the irredundancy
condition implies that there is no edge between $y$ and $z$. Hence, the
degrees of $x$ and $y$ are increased by the same value above $u$. If the
degrees are different in $G(t)/u$, they are so in $G$. We recall that $\pi
(G)$\ is the set of port labels of the vertices of $G$ and that $\beta_{G}(a)$
is the number of its $a$-ports. The notation is as in Section 4.2.\ The set of
states of $\mathcal{A}_{Reg}$ is defined as $\{Error\}\cup([\mathbb{N}%
_{+}\rightarrow(\mathbb{N}\cup\{\bot\})]_{f}\times\lbrack\mathbb{N}%
_{+}\rightarrow\mathbb{N}]_{f})$ and we want that, for every term $t\in
T(F_{\infty})$:

\begin{quote}
$q_{\mathcal{A}_{Reg}}(t)=Error$ \ if two $a$-ports of $G(t)$ have different
degrees; 
\end{quote}

otherwise,

\begin{quote}
$q_{\mathcal{A}_{Reg}}(t)=(\partial_{G(t)},\beta_{G(t)})$ where, for every $a$
in $\pi(G(t)),\partial_{G(t)}(a)$ is the common degree of all $a$-ports of
$G(t)$ and is $\bot$\ if there is no $a$-port.
\end{quote}

In the run on a term $t\in T(F_{k})$ such that $G(t)$ has $n$ vertices, less
than $(n+1)^{2k}$ states occur and these states have size $O(k.\log(n))$. In
the transition table (Table 3), $(\partial,\beta)$\ denotes a state that is
not $Error$.\ Hence, if $(\partial,\beta)$\ is accessible, we have
$\partial(a)=\bot$ if and only if $\beta(a)=0$.\ We denote respectively by
$\boldsymbol{0}$ and $\underline{\boldsymbol{\bot}}$ the constant mappings
with values $0$ and $\bot$. We take $\max\{\bot,n\}=n$ (for the transitions on
$\oplus$). It is clear that the transitions can be computed in time
$O(k.\log(n))$. Hence, we have a P-FA $\mathcal{A}_{Reg}$.

\bigskip%
\begin{tabular}
[c]{|c|c|}\hline
Transitions & Conditions\\\hline\hline
\multicolumn{1}{|l|}{$\boldsymbol{\varnothing}\rightarrow(\underline
{\boldsymbol{\bot}},\boldsymbol{0})$} & \multicolumn{1}{|l|}{}\\\hline
\multicolumn{1}{|l|}{$\mathbf{a}\rightarrow(\partial,\beta)$} &
\multicolumn{1}{|l|}{$\partial(x):=$\ \texttt{if} $x=a$\texttt{ then}
$0$\texttt{ else} $\bot,$}\\
\multicolumn{1}{|l|}{} & \multicolumn{1}{|l|}{$\beta(x):=$\ \texttt{if}
$x=a$\texttt{ then} $1$\texttt{ else} $0.$}\\\hline
\multicolumn{1}{|l|}{$add_{a,b}[(\partial,\beta)]\rightarrow(\partial^{\prime
},\beta)$} & \multicolumn{1}{|l|}{If $\beta(a)=0$ or $\beta(b)=0$ then
$\partial^{\prime}:=\partial,$}\\
\multicolumn{1}{|l|}{} & \multicolumn{1}{|l|}{else $\partial^{\prime
}(a):=\partial(a)+\beta(b)$,}\\
\multicolumn{1}{|l|}{} & \multicolumn{1}{|l|}{$\partial^{\prime}%
(b):=\partial(b)+\beta(a)$\ and}\\
\multicolumn{1}{|l|}{} & \multicolumn{1}{|l|}{$\partial^{\prime}%
(x):=\partial(x)$ for $x\notin\{a,b\}$.}\\\hline
\multicolumn{1}{|l|}{$relab_{a\rightarrow b}[(\partial,\beta)]\rightarrow
(\partial,\beta)$} & \multicolumn{1}{|l|}{$\beta(a)=0.$}\\\hline
\multicolumn{1}{|l|}{$relab_{a\rightarrow b}[(\partial,\beta)]\rightarrow
Error$} & \multicolumn{1}{|l|}{$\beta(a)\neq0$, $\beta(b)\neq0\ $and
$\partial(a)\neq\partial(b).$}\\\hline
\multicolumn{1}{|l|}{$relab_{a\rightarrow b}[(\partial,\beta)]\rightarrow
(\partial^{\prime},\beta^{\prime})$} & \multicolumn{1}{|l|}{The previous cases
do not apply,}\\
\multicolumn{1}{|l|}{} & \multicolumn{1}{|l|}{$\partial^{\prime}%
(a):=\bot,\partial^{\prime}(b):=\partial(a),$}\\
& \multicolumn{1}{|l|}{$\beta^{\prime}(a):=0,\beta^{\prime}(b):=\beta
(b)+\beta(a),$}\\
\multicolumn{1}{|l|}{} & \multicolumn{1}{|l|}{$\beta^{\prime}(x):=\beta
(x),\partial^{\prime}(x):=\partial(x)$ for $x\notin\{a,b\}.$}\\\hline
\multicolumn{1}{|l|}{$\oplus\lbrack(\partial_{1},\beta_{1}),(\partial
_{2},\beta_{2})]\rightarrow Error$} & \multicolumn{1}{|l|}{$\partial
_{1}(a)\neq\partial_{2}(a)$ for some $a$ such that}\\
\multicolumn{1}{|l|}{} & \multicolumn{1}{|l|}{$\beta_{1}(a)\neq0$ and
$\beta_{2}(a)\neq0$.}\\\hline
\multicolumn{1}{|l|}{$\oplus\lbrack(\partial_{1},\beta_{1}),(\partial
_{2},\beta_{2})]\rightarrow(\partial,\beta)$} & \multicolumn{1}{|l|}{The
previous case does not apply,}\\
\multicolumn{1}{|l|}{} & \multicolumn{1}{|l|}{$\partial(x):=\max\{\partial
_{1}(x),\partial_{2}(x)\}$ and}\\
\multicolumn{1}{|l|}{} & \multicolumn{1}{|l|}{$\beta(x):=\beta_{1}%
(x)+\beta_{2}(x)$ for all $x$.}\\\hline
\end{tabular}

\begin{center}
Table 3:\ Transitions of $\mathcal{A}_{Reg}$
\end{center}

To take an example, we can get by Theorem 27 an XP-FA\ that computes
$\mathrm{MaxCard}X.Reg[X]$.\ It is of type XP because the nondeterminism
degree of $pr_{1}(\mathcal{A}_{Reg[X]})$ is XP-bounded by $(n+1)^{2k}$ (and
not FPT-bounded as one can check).

To specialize the problem, we let $Reg_{d}[X]$ mean that $G[X]$ is
$d$-\emph{regular}, i.e., has all vertices of degree $d$. The article
\cite{BGP} gives an algorithm for checking the existence of a $d$-regular
induced subgraph.\ We can replace $\mathcal{A}_{Reg}$\ by an FA\ $\mathcal{B}%
_{d}$ with set of states $\{Error\}\cup([\mathbb{N}_{+}\rightarrow
(\mathbb{[}0,d+1]\cup\{\bot\})]_{f}\times\lbrack\mathbb{N}_{+}\rightarrow
\mathbb{[}0,d+1]]_{f})$ and we get the algorithm of time complexity
$n.d^{O(k)}$ given in \cite{BGP} to check if the given graph with $n$ vertices
defined by a term in $T(F_{k})$ has a regular induced subgraph of degree
$d$.\ This article also gives algorithms for computing $\mathrm{MaxCard}%
X.Reg_{d}[X]$, $\mathrm{MinCard}X.(Reg_{d}[X]\wedge X\neq\emptyset)$
and\ $\#X.Reg_{d}[X]$, all of time complexity $n.d^{O(k)}.$ We can derive them
from Theorem 27, similarly as in Section (4.2.1).

The property $\exists X.(Card_{\leq p}(X)\wedge Reg[X^{c}])$ expresses that
the considered graph becomes regular if we remove at most $p$ vertices.\ It is
P-FA decidable by Corollary 18\ and the remark at the end of Section 3.1. The
property that the graph can be partioned into at most two regular subgraphs,
expressed by $\exists X.(Reg[X]\wedge Reg[X^{c}])$ is XP-FA\ computable, with
time complexity $O(n^{8k+a})$ for some constant $a$, similar to the case of
maximum directed cut.

\subsection{Partition problems}

Many partition problems consist in finding an $s$-tuple $\overline{X}%
=(X_{1},...,X_{s})$ satisfying $Partition(\overline{X})\wedge P_{1}%
(X_{1})\wedge...\wedge P_{s}(X_{s})$ where $P_{1},...,P_{s}$ are properties of
sets of vertices that can be MS expressible or, more generally, defined by FA.
We may also wish to count the number of such partitions, or to find one that
minimizes or maximizes the cardinality of $X_{1}$ or the number $e(\overline
{X}):=e(X_{1})+...+e(X_{s})$ (cf. Section (4.2)). We have discussed above the
partitioning of a graph into two regular induced subgraphs.\ Vertex coloring
problems are of this type with $P_{i}(X_{i})$ being $St[X_{i}]$ and a fixed
number $s$ of allowed colors (cf. the introduction).\ 

If the properties $P_{i}(X_{i})$ are MS expressible, then the partition
problem $\mathcal{P}$ expressed by the MS\ sentence $\exists\overline
{X}.Partition(\overline{X})\wedge P_{1}(X_{1})\wedge...\wedge P_{s}(X_{s})$ is
decided by\ an FPT-FA by Theorem 29.\ If the properties $P_{i}(X_{i})$ are
decided by FPT-FA or XP-FA, then $\mathcal{P}$ is decided by\ an FPT-FA or
XP-FA, provided the conditions of Theorem 27\ on the degree of nondeterminism
are satisfied.\ Counter-example 13\ shows that these conditions cannot be
avoided.\ We now examine some coloring problems.

\bigskip

(4.4.1) \textit{Coloring problems}

We let $Col(\overline{X})$ abbreviate the MS property $Partition(\overline
{X})\wedge St[X_{1}]\wedge...\wedge St[X_{s}]$. The function $\#\overline
{X}.Col(\overline{X})$ counts the number of $s$-colorings\footnote{This number
is $\chi_{G}(s)$ where $\chi_{G}$ is the chromatic polynomial of $G$. So, for
some graphs with known chromatic polynomial, we could check the correctness of
our computations.}.\ It is thus FPT-FA\ computable by Theorem 27. Another
number of possible interest is, if $G$ is $s$-colorable, $\mathrm{MinCard}%
X_{1}.\exists X_{2},...,X_{s}.Col(\overline{X})$ which is 0 if $G$ is
$(s-1)$-colorable; otherwise, it indicates how close $G$\ is to be
$(s-1)$-colorable. By Theorem 27, this number is computable by an FPT-FA.

\bigskip

There are other definitions of approximate $s$-colorings.\ \ One of them is
the notion of $(s,d)$-\emph{defective coloring}, expressed by the MS sentence:\ 

\begin{quote}
$\exists\overline{X}.(Partition(\overline{X})\wedge Deg_{\leq d}[X_{1}%
]\wedge\ ...\ \wedge Deg_{\leq d}[X_{s}]).$
\end{quote}

For fixed $s$, we can consider the problem of determining the smallest $d$ for
which this property holds. This number is at most $\lceil n/s\rceil$ \ for a
graph with $n$ vertices.

The property $Deg_{\leq d}[X]\ $meaning that each vertex of $X$ has degree at
most $d$ in $G[X]$\ is decided by an FPT-FA whose number of states on a term
in $T(F_{k}^{\mathrm{u}(1)})$ is $O(d^{2k})$. It follows that the existence of
an $(s,d)$-defective coloring can be checked, for a graph with $n$ vertices,
in time $O(n.d^{4s.k+a})$ for some constant $a$.\ \ By checking the existence
of an $(s,d)$-defective coloring for successive values of $d$ starting from 1,
one can find the minimal value of $d$ in time $O(n^{4s.k+a+1})$ hence
$O(n^{8s.2^{rwd(G)}+a+1})$ which is similar to the time bound
$O(n^{4s.2^{rwd(G)}+b})$ given in \cite{GHO} (because for every undirected
graph $G$, we have $cwd(G)\leq2^{rwd(G)+1}-1).$

Another possibility is to define $\mathit{MD}(\overline{X}):=(MaxDeg[X_{1}%
],...,MaxDeg[X_{s}])$\ \ and to compute the set: $\mathrm{SetVal}\overline
{X}.\mathit{MD}(\overline{X})\upharpoonright Partition(\overline{X}),$ \ from
which the existence of an $(s,d)$-defective coloring can easily be determined.
Since the automaton for $MaxDeg$ uses $O(n^{2k})$ states for a graph with $n$
vertices defined by a term in $T(F_{k})$, we get for $\mathit{MD}(\overline
{X})$ the bound $O(n^{2k.s})$ and, by Lemma 7 and Theorem 27, the bound
$O(n^{4s.k+c})$ for some constant $c$, which is of same order as by the first method.\ 

\bigskip

(4.4.2) \ \textit{Graph partition problems with numerical constraints}

Some partition problems consist in finding an $s$-tuple $\overline{X}$ satisfying:

\begin{quote}
$Partition(\overline{X})\wedge P_{1}(X_{1})\wedge\ ...\ \wedge P_{s}%
(X_{s})\wedge R(|X_{1}|,...,|X_{s}|)$,
\end{quote}

where $P_{1},...,P_{s}$ are properties of sets and $R$ is a \textbf{P}%
-computable arithmetic condition. An example is the notion of
\ \emph{equitable s-coloring}: $P_{i}(X_{i})$ is $St[X_{i}]$ for each $i$ and
$R(|X_{1}|,...,|X_{s}|)$ expresses that\ any two numbers $|X_{i}%
|$\ and\ $|X_{j}|$ differ by at most 1. The existence of an equitable
3-coloring is not trivial: it holds for the cycles but not for the graphs
$K_{n,n}$ for large $n$. The existence of an equitable $s$-coloring is
W[1]-hard for the parameter defined as $s$ plus the tree-width \cite{Fell2},
hence presumably not FPT for this parameter. Our constructions yield, for each
integer $s$, an FPT-FA for checking the existence of an equitable $s$-coloring
for clique-width as parameter.\ We obtain the answer from $\mathrm{Sp}%
\overline{X}.(Partition(\overline{X})\wedge St[X_{1}]\wedge\ ...\ \wedge
St[X_{s}])$ \ that is computable by an FPT-FA.

\subsection{Connected components}

\bigskip

The empty graph is defined as connected and a connected component as nonempty.
In \cite{BCID}, we have discussed in detail connectedness, denoted by $Conn$,
and we come back to this important graph property. We show that the general
constructions of\ Theorem 27 can be improved in some cases. We consider
undirected graphs.

\bigskip

(4.5.1) \textit{Number and sizes of connected components.}

We denote by $\kappa(G)$ the number of connected components of a graph $G$, by
$\kappa(G,p)$ the number of those with $p$ vertices, by $MinComp(G)$
(resp.\ $MaxComp(G)$) the minimum (resp.\ maximum) number of vertices of a
connected component of $G$.\ We will compute these values by FA.

\bigskip

The MS\ formula $CC(X)$ defined as $Conn[X]\wedge X\neq\emptyset\wedge\lnot
Link(X,X^{c})$ expresses that $X$\ is the vertex set of a connected
component.\ Hence, we have:

\begin{quote}
$\kappa(G)=\#X.CC(X)(G),$

$\kappa(G,p)=\mathrm{MSp}X.CC(X)(G)(p)$,

$MinComp(G)=\mathrm{MinCard}X.CC(X)(G)$ and $\ $

$MaxComp(G)=\mathrm{MaxCard}X.CC(X)$.
\end{quote}

These values can be computed by FPT-FA\ constructed by using Propositions 15,
16 and Theorem 27 in the following way: we build a deterministic FA
$\mathcal{A}$ to decide $Link(X,X^{c})$; it uses at most $2^{2k}$ on terms in
$T(F_{k}^{\mathrm{u}(1)}).$ The nondeterminism degree of $pr_{1}(\mathcal{A})$
on a term in $T(F_{k}^{\mathrm{u}})$ is bounded by $2^{2k}$. The corresponding
bound for the deterministic FA that decides $Conn[X]$ is $2^{2^{k}}$
(\cite{BCID}). Then, we can use the above mentioned results. However, we can
construct a smaller FA by modifying the FA $\mathcal{A}_{Conn}$ of \cite{BCID}.\ 

\bigskip

We can compute $\kappa(G)=\#X.CC(X)(G)$ for $G$ not empty as follows. The
formula $\lnot Link(X,X^{c})$ expresses that $X$\ is the vertex set of a
(possibly empty) union of connected components.\ Hence, $\#X.\lnot
Link(X,X^{c})(G)=2^{\kappa(G)}$.\ The construction of the FA\ computing
$\#X.\lnot Link(X,X^{c})(G)$ is clearly easier than that for $\#X.CC(X)(G)$.
This FA\ allows even to check if $G$ is connected (this property is equivalent
to $\#X.\lnot Link(X,X^{c})(G)=2)$. However, it is an FPT-FA, whereas we noted
above (cf.\ the comments about Table 2) that the automaton $\mathcal{A}%
_{Conn}$ that checks connectedness is a P-FA.

\bigskip

We can alternatively construct directly a \emph{deterministic}
FA\ $\mathcal{A}_{\kappa}$ to compute $\kappa(G).$ Its states are sets of
pairs $(L,m)$ such that $\emptyset\neq L\in\mathcal{P}_{f}(\mathbb{N}_{+})$
and $m$ is an integer.\ For every term $t$ in $T(F_{\infty}^{\mathrm{u}}),$ we
want that:

\begin{quote}
$q_{\mathcal{A}_{\kappa}}(t)=\{(L,m)\mid L$ and $m$ is the number of connected components

\qquad\qquad\qquad\qquad\qquad of $G(t)$ of type $L\}$.
\end{quote}

The transitions are easy to write; the output function is then defined by:

\begin{quote}
$Out_{\mathcal{A}_{\kappa}}(q):=\Sigma\llbracket m\mid(L,m)\in q\rrbracket .$
\end{quote}

If $t\in T(F_{k}^{\mathrm{u}})$ and $G(t)$ has $n$ vertices, the size of a
state on $t$ is $O(n.\log(k))$ and so, $\mathcal{A}_{\kappa}$ is a P-FA.

\bigskip

We now explain why this automaton is better than the one constructed by using
Theorem 27. We recall from \cite{BCID} that the states of $\mathcal{A}_{Conn}$
are such that:

\begin{quote}
$q_{\mathcal{A}_{Conn}}(t)=(L,L)$ with $L\in\mathcal{P}_{f}(\mathbb{N}_{+}%
)$\ if $G(t)$ is not connected and all its connected components have type $L$, otherwise,

$q_{\mathcal{A}_{Conn}}(t)$ is the set of types of the connected components of
$G(t)$.
\end{quote}

The graph $G(t)$ is connected if and only if the state at the root is $\{L\}$
or the empty set (because the empty graph is connected).\ (It is clear that
$\mathcal{A}_{Conn}$ is a homomorphic image of $\mathcal{A}_{\kappa}$.)\ Note
that $\mathcal{A}_{Conn}$ yields more information than just the connectedness
of $G(t)$: it computes also the set of types of the connected components. By
Propositions 15 and 16, we get for property $CC(X)$ an automaton
$\mathcal{A}_{CC(X)}$ such that, for every $t$ and $X$:

\begin{quote}
$q_{\mathcal{A}_{CC(X)}}(t\ast X)$ \ is $Error$ if there is an edge between
$X$ and its complement;

otherwise, $X$ is a union of connected components of $G(t)$, and

$q_{\mathcal{A}_{CC(X)}}(t\ast X)$\ records the set of types, let us denote it
by $\sigma(X)$, of these connected components.
\end{quote}

\bigskip

We simplify for clarity: the state $q_{\mathcal{A}_{CC(X)}}(t\ast X)$ contains
more than the set $\sigma(X)$.\ This is why we write that "it records ..." and
not "it is $\sigma(X)$".\ Then, let \ $\mathcal{A}_{\kappa}^{\prime}$ be
constructed from $\mathcal{A}_{CC(X)}$ by Theorem 27 so as to compute
$\kappa(G(t))$.\ For each term $t$, the state $q_{\mathcal{A}_{\kappa}%
^{\prime}}(t)$ records, for each set $\alpha$\ of sets of labels, the number
of sets $X$ such that $\sigma(X)=\alpha.$ This is more than needed: the state
$q_{\mathcal{A}_{\kappa}}(t)$ records only information about the connected
components of $G(t)$, not about all unions of connected components.\ If for
example $G(t)$ is the graph:

\begin{center}
$a-b\qquad a-b\qquad b-c\qquad c-d$
\end{center}

then $q_{\mathcal{A}_{\kappa}}(t)=\{(ab,2),(bc,1),(cd,1)\}$ whereas
$q_{\mathcal{A}_{\kappa}^{\prime}}(t)$ records $\{(ab,3),(bc,1),$
$(abc,3),(cd,1),(bcd,1),(abcd,6)\}$.

\bigskip

We have tested these automata on a connected graph $G=add_{a,b}(H)$ of
clique-width 3 with 17 vertices such that $H$ has 8 connected components, each
with 2 or 3 vertices.\ The quickest automaton\ on a term defining $G$
is\textbf{ }$\mathcal{A}_{\kappa}$ (taking 0.0012\ s), followed by
$\mathcal{A}_{Conn}$ (0.0014 s) and $\mathcal{A}_{\#X.\lnot Link(X,X^{c})}$
(0.33 s) whereas $\mathcal{A}_{\#X.CC(X)}$ takes 39 s. It is interesting to
note that using unbounded integers in $\mathcal{A}_{\kappa}$\ makes the
computation quicker than by using $\mathcal{A}_{Conn}$ although $\mathcal{A}%
_{Conn}$ is finite on terms in $T(F_{3})$.

\bigskip

(4.5.2) \ \emph{Counting components by their size.}

We now compute $\mathrm{MSp}X.CC(X)(G)$. First we observe that for each
integer $p$, $\mathrm{MSp}X.CC(X)(G)(p)$ is computable from the values
$\mathrm{MSp}X.\lnot Link(X,X^{c})(G)(p^{\prime})$ for all $p^{\prime}%
\in\lbrack0,p]$. (We made above a similar observation for the computation of
$\kappa(G)$ from $\#X.\lnot Link(X,X^{c})(G)).$ However, as in this previous
case, we can construct a P-FA $\mathcal{B}$\ derived from $\mathcal{A}_{Conn}$
(and generalizing the previous $\mathcal{A}_{\kappa}$) such that, for every
term $t\in T(F_{\infty}^{\mathrm{u}})$:

\begin{quote}
$q_{\mathcal{B}}(t)$ is the set of triples $(L,p,m)$ such that $L$ is a
nonempty set of\ port labels, $m,p\in\mathbb{N}_{+}$ and $m$ is the number of
connected components of $G(t)$ of type $L$ having $p$ vertices.
\end{quote}

\bigskip

If $t\in T(F_{k}^{\mathrm{u}})$ and $G(t)$ has $n$ vertices, then
$n=\Sigma_{(L,p,m)\in q_{_{\mathcal{B}}}(t)}m.p$.\ Hence, $q_{\mathcal{B}}(t)$
can be described by a word of length $O(n.\log(k))$ (even if numbers are
written in unary; the factor $\log(k)$ corresponds to the coding of labels).
\ Here are the transitions:

$\boldsymbol{\varnothing}\rightarrow\emptyset,$

$\mathbf{a}\rightarrow\{(\{a\},1,1)\},$

$\oplus\lbrack q,q^{\prime}]\rightarrow q^{\prime\prime}$ where $q^{\prime
\prime}$ is the set obtained by replacing iteratively in the multiset $q\sqcup
q^{\prime}$ any pair $\{(L,p,m),(L,p,m^{\prime})\}$ by the unique triple
$(L,p,m+m^{\prime})$,

$relab_{h}[q]\rightarrow q^{\prime}$: for each set $L$, we let $h(L)$ be the
set obtained from $L$ by replacing $a$ by $h(a)$; then $q^{\prime}$ is the set
of triples $(L^{\prime},p,m^{\prime})$ \ such that:

\begin{quote}
$L^{\prime}:=h(L)$ for some $(L,p,m)\in q$,

$m^{\prime}:=\Sigma\llbracket m\mid L^{\prime}=h(L)$ $\ $and \ $(L,p,m)\in
q\rrbracket .$
\end{quote}

Finally, we describe the transitions $add_{a,b}[q]\rightarrow q^{\prime}$.
There are two cases.

\emph{Case 1}: $a$ or $b$ is not present in $q$ or they are both present in
$q$ but in a unique triple of the form $(L,p,1)$ (with $a,b\in L$).\ Then
$q^{\prime}:=q$.

\emph{Case 2}: Case 1 does not apply. We let:

\begin{quote}
$q^{\prime\prime}$ be the set of triples in $q$ that contain neither $a$ nor
$b$,

$L^{\prime}:={%
{\displaystyle\bigcup}
\{}L\mid(L,p,m)\in q-q^{\prime\prime}\}$,$\ $

and $q^{\prime}:=q^{\prime\prime}\cup\{(L^{\prime},p^{\prime},1)\}$ where
$p^{\prime}:=\Sigma\llbracket p.m\mid(L,p,m)\in q-q^{\prime\prime}\rrbracket.$
\end{quote}

We illustrate this case with an example:

\begin{quote}
$q=\{(\{a\},2,1),(\{a\},1,4),(\{a,b,c\},4,1),(\{b,d\},3,2),(\{c,d\},3,4)\},$

$q^{\prime}=\{(\{a,b,c,d\},16,1),(\{c,d\},3,4)\},$
\end{quote}

where 16 is obtained as $2.1\ +1.4+4.1\ +3.2$ because the connected components
of types $\{a\}$, $\{a,b,c\}$ and $\{b,d\}$ get fused into a unique one (of
type $\{a,b,c,d\}$).

For computing $\mathrm{MSp}X.CC(X)$ we take the output function:

\begin{quote}
$Out_{\mathcal{B}}(q):=\mu$ \ such that $\mu(p):=\Sigma\llbracket m\mid
(L,p,m)\in q\rrbracket $ for $p\in\mathbb{N}_{+}.$
\end{quote}

It is clear that the transitions and the output function can be computed in
time poly$(\parallel t\parallel)$. Hence, $\mathcal{B}$ is a P-FA. From
$\mathrm{MSp}X.CC(X)(G)$ we get $\kappa(G,p)$ for each $p$.

\bigskip

(4.5.3) \textit{Tools for separation problems.}

For dealing with separation problems, it is useful to compare the cardinality
of a set of vertices $X$ to the number of connected components of $G[X^{c}%
]$\ and to the maximal cardinality of a connected component of $G[X^{c}]$ that
we denote by $MaxCardCC(G[X^{c}])$. For this purpose, we define for a graph
$G$:

\begin{quote}
$\alpha(G)=\{(|X|,\kappa(G[X^{c}]))\mid X\subseteq V_{G}\},$

$\beta(G)=\{(|X|,MaxCardCC(G[X^{c}]))\mid X\subseteq V_{G}\}.$
\end{quote}

From $\alpha(G),$ one can determine, for given integers $p$ and $q,$ if there
exists a set $X$ of cardinality at most $p$ whose deletion splits the graph in
at least $q$ connected components. Similarly, from $\beta(G)$ one can
determine if there is such a set $X$ whose deletion splits the graph in
connected components of size at most $q$.

Let $P(X,U)$ mean that $U$ has one and only one vertex in each connected
component of $G[X^{c}]$ and $Q(X,Y)$ mean that $Y$ is the vertex set of a
connected component of \ $G[X^{c}]$.\ These properties are
MS\ expressible.\ Then $\alpha(G)=\mathrm{Sp}(X,U).P(X,U)(G)$\ and $\beta(G)$
can be computed from $\mathrm{Sp}(X,Y).Q(X,Y)$. Hence,\ by Example 14(a),
Propositions 15, 16 and Theorem 27, these two values are computable by FPT-FA.\ 

\subsection{Undecidability and intractability facts.}

Let $R$ be an $s$-ary \textbf{P}-computable numerical predicate (integers
being given in binary notation).\ We denote by $\mathrm{MS}+R$ the extension
of monadic second-order logic with the atomic formulas $R(|X_{1}%
|,...,|X_{s}|)$. We have seen such formulas in Section \ (4.4.2). We wish to
examine when the model-checking problem\footnote{We are interested in the
\emph{data-complexity} of the model-checking problem for a language
$\mathcal{L}$.\ For each fixed sentence in $\mathcal{L}$ that describes some
property of interest, we consider an algorithm whose input is a word or a term
that may describe a graph.} for $\mathrm{MS}+R$ is FPT or XP.\ Actually we
will only consider the case of words over finite alphabets, so the question
reduces to whether it is \textbf{P}-decidable. We first discuss undecidability
results.\ There is no implication between (un)decidability results on the one
hand and complexity results on the other, but decidability and FPT results for
terms and for graphs of bounded clique-width are proved with the same
tools.\ Undecidability results are actually easier to prove and they help to
foresee the difficulties regarding complexity. We let $Eq(n,m)$ mean $n=m$;
this binary relation defines a semi-linear set of pairs of integers. A unary
predicate $R$ on $\mathbb{N}$\ is identified with the corresponding set.

\bigskip

\textbf{Proposition 31}: One cannot decide if a given sentence of
$\mathrm{MS}+Eq$ or $\mathrm{MS}+R$ where $R\subseteq\mathbb{N}$ is not
ultimately periodic is true in some word over a fixed finite alphabet.

\bigskip

\textbf{Proof}: The case of $\mathrm{MS}+Eq$ is proved in Proposition 7.60\ of
\cite{CouEng} and the other one in \cite{Bes}. $\square$

\bigskip

We now consider the model-checking problem.

\bigskip

\textbf{Definition 32}: \emph{Separating sets of integers.}

\ 

Let $R\subseteq\mathbb{N}$, $p,n\in\mathbb{N}$ such that $n>p.$ We say that
$R$\ \emph{separates on} $[0,n]$ \emph{the integers in} $[0,p]$ if, for every
$x,y\in\lbrack0,p]$:

\begin{quote}
$x\neq y$ if and only if there exists $z\in\mathbb{N}$ such that
$x+y+z\in\lbrack0,n]$ and,

either $\ x+z\in R$ and $y+z\notin R$ or $y+z\in R$ and $x+z\notin R.$
\end{quote}

We say that an infinite set $R\subseteq\mathbb{N}$\ is \emph{separating}\ if
there exists $n_{0}$\ such that, for every $n>n_{0}$, $R$\ separates on
$[0,n]$ the integers in $[0,\lfloor\log(n)\rfloor]$. The sets $\{n!\mid
n\in\mathbb{N}\}$, $\{2^{n}\mid n\in\mathbb{N}\}$ and that of prime numbers
are separating.\ An ultimately periodic set of integers is not
separating.\ The set $D:=\{a_{n}\mid n\in\mathbb{N}\}$ such that $a_{0}=1$,
$a_{n+1}=2^{a_{n}+3}$ is not ultimately periodic and not separating either.
(To see this, observe that $D$ does not separate $a_{n}+1$\ and $a_{n}+2$\ on
$[0,a_{n+1}-1].)$

\bigskip

If $R$\ separates on $[0,n]$ the integers in $[0,p]$, then, for any two
disjoint subsets $X$ and $Y$ of $[n]$ of cardinality at most $p$ we have:

\begin{quote}
$|X|=|Y|$ if and only if:

$[n]\models\forall Z.[Z\cap(X\cup Y)=\emptyset\Longrightarrow(R(|X\cup
Z|)\Longleftrightarrow R(|Y\cup Z|))].$
\end{quote}

This means that the equipotence of small sets can be expressed in
$\mathrm{MS}+R$.

\bigskip

\textbf{Proposition 33}: Let $R$ be a separating subset of $\mathbb{N}$.\ If
$\mathbf{P}\neq\mathbf{NP}$, the model-checking problems for $\mathrm{MS}+Eq$
and $\mathrm{MS}+R$ are not \textbf{P}-decidable.

\bigskip

\textbf{Proof}: We first consider $\mathrm{MS}+Eq$.\ We use a method similar
to that of Counter-example 13. Let $P$ be an instance of SAT\ in conjunctive
normal form whose variables are $x_{1},...,x_{n}$. Let $w(P)$ be the word
representing $P$ with $x_{i}$ written as $x$ followed by the binary writing of
$i$ (with no leading 0).\ For example, if $P$ is $(x_{1}\vee x_{2}\vee\lnot
x_{3})\wedge(x_{3}\vee\lnot x_{4}\vee\lnot x_{5})$ then $w(P)$ is the word
$(x1\vee x10\vee\lnot x11)\wedge(x11\vee\lnot x100\vee\lnot x101)$ over the
alphabet $A:=\{(,),\vee,\wedge,\lnot,x,0,1\}.$ The factors of this word that
belong to $\{0,1\}^{\ast}$ have length at most $1+\lfloor\log(n)\rfloor$. The
word $w(P)$ is represented by the logical structure $S(P):=\langle
\lbrack|w(P)|],\leq,(lab_{a})_{a\in A}\rangle$ such that $lab_{a}(i)$ holds if
and only if $w(P)[i]=a$.

We build a formula $\varphi(U)$ of $\mathrm{MS}+Eq$, written with $\leq$ \ and
the unary relations $lab_{a}$ for $a\in A$, such that $P$ as above has a
solution if and only if $S(P)\models\exists U.\varphi(U)$.\ The set $U$
defines a set of occurrences of $x$ in the word $w(P)$ whose corresponding
variable $x_{i}$ takes value $True$. We require that, either all occurrences
of a variable $x_{i}$ or none of them has value $True.$ We express this
condition by a formula $\varphi_{1}(U)$\ of $\mathrm{MS}+Eq$ where
$Eq(|X|,|Y|)$ is only used for sets $X$\ and $Y$ of consecutive occurrences of
0 and 1's. We sketch its construction.\ If $i<j$, we denote by $S(P)[i,j]$ the
factor of $S(P)$ from position $i$ to position $j$.\ We construct a formula
$\theta(i,j,i^{\prime},j^{\prime})$ expressing that $S(P)[i,j]$ is a prefix of
$S(P)[i^{\prime},j^{\prime}]$: it says that for each $u\in\lbrack i,j]$\ there
is $u^{\prime}\in\lbrack i^{\prime},j^{\prime}]$\ such that $u^{\prime
}-i^{\prime}=u-i$ and $S(P)[u^{\prime}]=S(P)[u]$.\ This formula uses $Eq$.
Then, by using $\theta$, we construct $\varphi_{1}(U)$ saying that $U$ is a
set of occurrences of $x$ and that, for each $u\in U$, if $j$ is maximal such
that $S(P)[u,j]\in x\{0,1\}^{\ast}$, if $u^{\prime}$ is another occurrence of
$x$ such that $S(P)[u,j]=S(P)[u^{\prime},j^{\prime}]$\ where $j^{\prime}$ is
maximal such that $S(P)[u^{\prime},j^{\prime}]\in x\{0,1\}^{\ast}$, then
$u^{\prime}\in U$.

The formula $\varphi(U)$\ is taken of the form $\varphi_{1}(U)\wedge
\varphi_{2}(U)$\ where $\varphi_{2}(U)$\ is a first-order formula expressing
that the truth values of $x_{1},...,x_{n}$ defined by $U$\ satisfying
$\varphi_{1}(U)$\ form a solution of $P$.\ 

If the sentence $\exists U.\varphi(U)$ could be checked in words $w\in
A^{\ast}$ in time $\mathrm{poly}(|w|)$, then each instance $P$\ of SAT could
be checked in time $\mathrm{poly}(|w(P)|)$ and we would have $\mathbf{P}%
=\mathbf{NP}$.

We now translate $\varphi_{1}(U)$ into a sentence $\varphi_{3}(U)$ of
$\mathrm{MS}+R$ such that $\exists U.(\varphi_{3}(U)$ \ $\wedge\varphi
_{2}(U))$ is equivalent to $\exists U.\varphi(U)$ in every structure
$S(P)$.\ It is clear that $2n<|w(P)|$ as all variables $x_{1},...,x_{n}$ occur
in $P$. Hence, any sequence of 0 and 1's in $w(P)$ has length bounded by
$1+\lfloor\log(n)\rfloor=\lfloor\log(2n)\rfloor\leq\lfloor\log(|w(P)|)\rfloor
$.\ The equality tests $Eq(|X|,|Y|)$ used in $\varphi_{1}(U)$ can be expressed
in terms of $R$ that we assume separating. Hence, the satisfiability of $P$ is
expressed in $S(P)$\ by a sentence of $\mathrm{MS}+R$, and so, the
model-checking problem for $\mathrm{MS}+R$ is not \textbf{P}-solvable in
polynomial time either. $\square$

\bigskip

\textbf{Questions 34}: (1) Can one replace in the previous proposition "$R$ is
separating" by "$R$ is not ultimately periodic"? It might happen that the
model-checking problem for $\mathrm{MS}+R$ where $R$ is very sparse (like the
above set $D$) is \textbf{P}-decidable on words.

(2) Is the model-checking problem for $\mathrm{MS}+Eq$ \textbf{NP}-decidable
on words? The same question can be raised for $\mathrm{MS}+R$ where $R$ is a
semi-linear subset of $\mathbb{N}^{k},k\geq2$.

\bigskip

\section{Implementation}

\bigskip

The system AUTOGRAPH\footnote{See
http://dept-info.labri.u-bordeaux.fr/\symbol{126}idurand/autograph}, written
in LISP (and presented in the conference paper \cite{BCID13a}) is intended for
verifications of graph properties and computations of functions on
graphs.\ Its main parts are as follows.

(1) A library of basic fly-automata over $F_{\infty}$ for the following
properties and functions:

\qquad(1.1) $X\subseteq Y$, $X=\emptyset$, $Sgl(X)$, $Card_{\leq p}(X)$,
$Card_{p,q}(X)$, $Partition(\overline{X})$ and the function $Card(X)$ (they
concern arbitrary sets),

\qquad(1.2) $edg(X,Y)$ and $lab_{a}(X)$, the atomic formulas of MS\ logic over p-graphs,

\qquad(1.3) some MS\ expressible graph properties: stability, being a clique,
$Link(X,Y)$, $Path(X,Y)$, connectedness, existence of directed or undirected
cycles, degree at most $d$, etc. cf.\ Table 2 and \cite{BCID} and finally,

\qquad(1.4) some graph properties and functions on graphs that are not
MS\ expressible: regularity, number of edges between two sets, maximum degree.

(2) A library of procedures that transform or compose fly-automata: these
functions implement the constructions of Propositions 10, 15, 16 and Theorems
17 and 27.

\bigskip

AUTOGRAPH includes no parser for the formulas $\varphi$\ expressing properties
and functions.\ The translation of these formulas into LISP programs that call
the basic FA and the composition procedures is easily done by hand because,
since we have FA for many basic graph properties, the formulas that specify
the problems are not too complicated. Some automata (in particular for cycles,
regularity and other degree computations) are defined so as to work correctly
on irredundant terms. A preprocessing can verify whether a term is
irredundant, and transform it into an equivalent irredundant one if it is
not\footnote{Another type of preprocessing defined in \cite{BCID} consists in
\emph{annotating} the given term.\ }. Whether input terms are good or not may
affect the computation time, but not the correctness of the outputs.

\bigskip

By using FA, we could find\footnote{AUTOGRAPH is written in Common Lisp and
run on a MacBook Pro laptop with processor 2.53\ GHz Intel Core Duo and a
4\ GB memory.} that Petersen's graph has 12960\ 4-colorings, and verify the
correctness of this result by using the chromatic polynomial.\ We found also
that McGee's graph has 57024\ \ acyclic 3-colorings in less than 30 minutes.

\bigskip

AUTOGRAPH\ has a method for enumerating (that is, for listing) the sets
$\mathrm{Sat}\overline{X}.P(\overline{X})$, by using an existing
FA\ $\mathcal{A}$ for $P(\overline{X})$. A specific enumeration program is
generated for each term (see \cite{Dur}).\ Running it is also interesting for
accelerating the verification that $\exists\overline{X}.P(\overline{X})$ is
true, because the computation can stop as soon as the existence of some
satisfying tuple $\overline{X}$ is confirmed.\ More precisely, the
nondeterministic automaton $pr(\mathcal{A})$ is not run deterministically
(cf.\ Definition 1(c)), but its potentially accepting runs are constructed
"one by one". In this way, we could check in 2 seconds that McGee's graph
is\ acyclically 3-colorable.\ This technique works for $\exists\overline
{X}.P(\overline{X})$ but not for $\forall\overline{X}.P(\overline{X})$,
$\#\overline{X}.P(\overline{X})$, $\mathrm{MSp}\overline{X}.P(\overline{X})$
etc... because these properties and functions are based on a complete
knowledge of $\mathrm{Sat}\overline{X}.P(\overline{X})$.

\bigskip

\emph{Using terms with shared subterms.}

Equal subterms of a "large" term $t$ can be fused and $t$ can be replaced by a
\emph{directed acyclic graph} (a \emph{dag}). The construction from $t$ of a
dag where any two equal subterms are shared can be done in linear time by
using the minimization algorithm of deterministic acyclic finite automata
presented in \cite{Rev}.\ Deterministic FA can run on such dags in a
straightforward manner. We have tested that on 4-colorable graphs defined
recursively by $G_{n+1}=t(G_{n},G_{n})$ where $t\in T(F_{7},\{x,y\})$ (a term
with two variables $x$ and $y$ denoting p-graphs; $G_{n}$ has $O(2^{n})$
vertices and edges). We checked that these graphs are 4-colorable by using the
term in $T(F_{7})$ \ and the dag resulting from the recursive definition.\ The
computation times are in Table 4.

\begin{center}
\bigskip%

\begin{tabular}
[c]{|l|l|l|}\hline
$n$ & term & dag\\\hline\hline
6 & 11 mn & 1 mn, 6 s\\\hline
9 & 88 mn & 1 mn, 32 s\\\hline
20 &  & 4 mn\\\hline
28 &  & 40 mn\\\hline
30 &  & 2 h, 26 mn\\\hline
\end{tabular}

Table 4: Computations using dags instead of terms.
\end{center}

\bigskip

This method raises a question: can one transform a term in $T(F_{k})$ into an
equivalent one in $T(F_{k^{\prime}})$ for some $k^{\prime}$ not much larger
than $k$, whose associated minimal dag (the one with a maximal sharing of
subterms) has as few nodes as possible?\ 

\section{Conclusion}

We have given logic based methods for constructing FPT and XP\ graph
algorithms based on automata.\ Our constructions allow several types of
optimizations: different logical expressions of a property can lead to
different automata having different observed computation times and direct
constructions of FA are sometimes better than the general ones resulting from
Theorem 27. We also have cases where FPT-FA\ are easier to implement and
practically more efficient\ than certain equivalent P-FA, and similarly for
XP-FA\ and FPT-FA. \emph{Can one identify general criteria for the possibility
of such optimizations and improvements?}

\bigskip

\emph{Do we need optimal terms? }Graphs are given by terms in $T(F_{\infty})$
and no \emph{a priori} bound on the clique-width must be given since all
FA\ are over $F_{\infty}$. As an input graph is given by a term $t$ over
$F_{k}$ with $k\geq cwd(G)$, one may ask how important it is that $k$ be close
to $cwd(G)$. Every graph with $n$ vertices is denoted by a term in $T(F_{n})$
where each vertex has a distinct label and no relabelling is made.\ Such a
term, if it is irredundant, has size $O(n^{2}.\log(n)).$ Hence, as input to a
P-FA, it yields a polynomial time computation. This is of course not the case
with an FPT- or XP-FA.

\bigskip

\emph{Edge quantifications}. The logical representation of graphs used in this
article does not allow edge set quantifications in MS\ formulas.\ MS formulas
written with edge set quantifications (MS$_{2}$ formulas in short) are more
expressive than MS\ formulas, and more functions based on them, such as
$\#\overline{X}.\varphi(\overline{X}),$ can be defined. An easy way to allow
edge set quantifications is to replace a graph $G$ by its \emph{incidence
graph} $Inc(G)$ where edges are made into new vertices and adjacency is
replaced by incidence. The clique-width of $Inc(G)$ is at most $2.twd(G)+4$
for $G$ directed and at most $twd(G)+3$ for $G$ undirected, where $twd(G)$ is
the tree-width of\ $G$ (\cite{Bou}).\ MS\ formulas over $Inc(G)$ allow
quantifications over sets of edges of $G$ and correspond to MS$_{2}$ formulas.
Hence, the constructions of FA\ presented in this article work for the
expression of properties and functions based on MS$_{2}$ formulas and
tree-width (but not clique-width) as parameter \cite{CouLAGOS}.\ Other
constructions based on a variant of tree-width are discussed in \cite{Cou12}.

\bigskip

\textbf{Acknowledgements}: We thank C.\ Paul and the referees for their many
useful comments.

\bigskip

\section*{Appendix}

\subsection*{Monadic second-order logic}

\bigskip

\emph{Representing graphs by logical structures.}

We define a simple graph $G$ as the relational structure $\langle
V_{G},edg_{G}\rangle$ with domain $V_{G}$ and a binary relation $edg_{G}$ such
that $(x,y)\in edg_{G}$ if and only if there is an edge from $x$ to $y$ (or
between $x$ and $y$ if $G$ is undirected). A p-graph $G$ whose type $\pi(G)$
is included in $\mathbb{N}_{+}$ is identified with the structure $\langle
V_{G},edg_{G},(lab_{a\,G})_{a\in\mathbb{N}_{+}}\rangle$ where $lab_{a\,G}$ is
the set of $a$-ports of $G$. Since only finitely many sets $lab_{a\,G}$\ are
not empty, this structure can be encoded by a finite word over a fixed finite
alphabet. We only consider properties of (and functions on) graphs rather than
of (and on) p-graphs, but the formal setting allows that.\ By considering a
graph as a relational structure, we have a well-defined notion of logically
expressible graph property.\ However, in the present article, we do not use
this relational structure: we handle graphs through terms in $T(F_{\infty})$
($F_{\infty}$\ is the signature of clique-width operations defined in Section
(1.2)) and we construct automata over $F_{\infty}$\ from MS\ formulas.

\bigskip

\emph{Monadic second-order formulas}

The basic syntax of monadic second-order formulas (MS formulas in short) uses
set variables $X_{1},...,X_{n},...$\ but no first-order variables. Formulas
are written without universal quantifications and they can use set terms
(cf.\ Section 1.3). These constraints yield no loss of generality (see, e.g.,
Chapter 5 of \cite{CouEng}).

To express properties of p-graphs we use the atomic formulas $X_{i}\subseteq
X_{j}$, $X_{i}=\emptyset$, $Sgl(X_{i})$ (meaning that $X_{i}$ denotes a
singleton set) and $Card_{p,q}(X_{i})$ (meaning that the cardinality of
$X_{i}$ is equal to $p$ modulo $q$, with $0\leq p<q$ and $q\geq2)$\footnote{We
do not distinguish monadic second-order formulas from \emph{counting } monadic
second-order formulas, defined as those using $Card_{p,q}(X_{i})$, because all
our results hold in the same way for both types. See Chapter 5 of
\cite{CouEng} for situations where the distinction matters.} where the
variables denote sets of vertices. We also use the atomic formulas
$edg(X_{i},X_{j})$ meaning that $X_{i}$ and $X_{j}$ denote respectively
$\{x\}$ and $\{y\}$ such that $x\rightarrow_{G}y$ and $lab_{a}(X_{i})$ meaning
that $X_{j}$\ denotes a singleton consisting of an $a$-port.

\begin{quote}

\end{quote}

It is convenient to require that the free variables of every formula and its
subformulas of the form $\exists X_{n}.\varphi$ are among $X_{1},...,X_{n-1}$.
This syntactic constraint yields no loss of generality (see Chapter~6 of
\cite{CouEng}) but it makes easier the construction of automata. In examples,
we use set variables $X,Y$, universal quantifications, and other obvious
notation to make formulas readable. A \emph{first-order existential
quantification} is a construction of the form $\exists X_{n}.(Sgl(X_{n}%
)\wedge\varphi(X_{1},...,X_{n}))$, also written $\exists x_{n}.\varphi
(X_{1},...,X_{n-1},\{x_{n}\})$ for readability.\ All quantifications of a
first-order formula have this form.\ First-order\ order formulas may have free
set variables and may be built with set terms. So, $\exists x_{2}%
.\varphi(X_{1},X_{1}^{c}-\{x_{2}\})$ is a first-order formula if $\varphi$
contains only first-order\ quantifications.\ 

A graph property $P(X_{1},...,X_{s})$ is $MS$\emph{ expressible} if there
exists an MS formula $\varphi(X_{1},...,X_{s})$ such that, for every p-graph
$G$ and for all sets of vertices $X_{1},...,X_{s}$ of this graph, we have
$\langle V_{G},edg_{G},(lab_{a\,G})_{a\in\mathbb{N}_{+}}\rangle\models
\varphi(X_{1},...,X_{s})$ if and only if $P(X_{1},...,X_{n})$ is true in $G$.

\subsection*{Good and irredundant terms}

We prove a technical result about terms over $F_{\infty}$, the signature of
clique-width operations. If $t$,$t^{\prime}\in T(F_{\infty})$, then $t\approx
t^{\prime}$ means that these terms define isomorphic p-graphs, $\pi(t)$ is the
set of port labels of $G(t)$, $\max\pi(t)$ is the maximal label in $\pi(t)$,
$\mu(t)$ is the set of port labels that occur in $t$ and $\max\mu(t)$ is the
maximal one in $\mu(t)$; we recall that port labels are positive integers.

\bigskip

\textbf{Proposition 35}: (1) The set of good and irredundant terms in
$T(F_{\infty})$ is P-FA recognizable.

(2) There exists a polynomial-time algorithm that transforms every term in
$T(F_{\infty})$ into an equivalent term that is good and irredundant.

\bigskip

\textbf{Proof}: (1) We have observed after Definition 7 that the set of good
terms is P-FA recognizable.\ By Proposition 8(2) of \cite{BCID} the set of
terms that are not irredundant is accepted by a nondeterministic
FA\footnote{This FA\ guesses a pair of occurrences of edge addition operations
showing that the considered term is not irredundant.} whose states on a term
$t$ are pairs of port labels in $\mu(t)$ and nondeterminism degree is at most
$\ |\mu(t)|^{2}$, hence $\mathrm{poly}(\Vert t\Vert)$.\ By determinizing it
and exchanging accepting states and nonaccepting ones, we get a P-FA
$\mathcal{A}$ that recognizes the set of irredundant terms.\ By taking the
product of $\mathcal{A}$ with the FA recognizing good terms, we get a P-FA (by
Proposition 15) that recognizes the good and irredundant terms.\ 

(2) Proposition 8 of \cite{BCID} gives, for each integer $k,$ a linear-time
algorithm that transforms a term $t$ in $T(F_{k})$ into an equivalent
irredundant one $t^{\prime}\in T(F_{k})$ such that $|t^{\prime}|=|t|$ and
$\Vert t^{\prime}\Vert\leq\Vert t\Vert$ by deleting occurrences of operations
that create redundancies. This algorithm attaches to each position of $t$ a
set of pairs of port labels from $\mu(t)$. These sets can be encoded in size
$|\mu(t)|^{2}.\log(k)\leq\mathrm{poly}(\Vert t\Vert)$. We obtain a
polynomial-time algorithm taking as input a term in $T(F_{\infty})$.

We now consider an input term $t$ that is irredundant and we transform it into
an equivalent one that is good and still irredundant. By induction on the
structure of $t\in T(F_{k}),$ we define:

\begin{quote}
a good term $\widehat{t}\in T(F_{k^{\prime}})$ such that $\pi(\widehat
{t}\,)=[\max\pi(\widehat{t}\,)]$ \ and $k^{\prime}\leq k$,

and a bijection $h_{t}:\pi(\widehat{t}\,)\rightarrow\pi(t)$ such that
$t\approx relab_{h_{t}}(\widehat{t}\,).$
\end{quote}

The inductive definition is shown in Table 5 where Condition (*) says the following:

\begin{quote}
$h_{t}$ is a bijection: $[|\pi(t)|]\rightarrow\pi(t)$ such that $h_{t}%
(i)=h_{t_{1}}(i)$ for $i\in\lbrack\max\pi(t_{1})]$; (clearly, $|\pi
(t)|\geq\max\pi(t_{1})).$
\end{quote}

\begin{tabular}
[c]{|c|c|c|c|}\hline
$t$ & $\widehat{t}$ & $h_{t}$ & Conditions\\\hline\hline
\multicolumn{1}{|l|}{$t$} & \multicolumn{1}{|l|}{$\boldsymbol{\varnothing}$} &
\multicolumn{1}{|l|}{$Id$} & \multicolumn{1}{|l|}{$\pi(t)=\emptyset$
(\emph{i.e.}, $G(t)=\varnothing$)}\\\hline
\multicolumn{1}{|l|}{$\mathbf{a}$} & \multicolumn{1}{|l|}{$\mathbf{1}$} &
\multicolumn{1}{|l|}{$1\rightarrow a$} & \multicolumn{1}{|l|}{}\\\hline
\multicolumn{1}{|l|}{$t_{1}\oplus t_{2}$} & \multicolumn{1}{|l|}{$\widehat
{t_{2}}$} & \multicolumn{1}{|l|}{$h_{t_{2}}$} & \multicolumn{1}{|l|}{$\pi
(t_{1})=\emptyset$}\\\hline
\multicolumn{1}{|l|}{$t_{1}\oplus t_{2}$} & \multicolumn{1}{|l|}{$\widehat
{t_{1}}$} & \multicolumn{1}{|l|}{$h_{t_{1}}$} & \multicolumn{1}{|l|}{$\pi
(t_{2})=\emptyset$}\\\hline
\multicolumn{1}{|l|}{$t_{1}\oplus t_{2}$} & \multicolumn{1}{|l|}{$\widehat
{t_{1}}\oplus relab_{\ell^{-1}\circ h_{t_{2}}}(\widehat{t_{2}}\,)$} &
\multicolumn{1}{|l|}{$h_{t}$} & \multicolumn{1}{|l|}{$\pi(t_{1})\neq
\emptyset,\pi(t_{2})\neq\emptyset$ and (*)}\\\hline
$\overrightarrow{add}_{a,b}(t_{1})$ & \multicolumn{1}{|l|}{$\widehat{t_{1}}$}
& \multicolumn{1}{|l|}{$h_{t_{1}}$} & \multicolumn{1}{|l|}{$\{a,b\}\nsubseteq
\pi(t_{1})$}\\\hline
$\overrightarrow{add}_{a,b}(t_{1})$ & $\overrightarrow{add}_{h_{t_{1}}%
^{-1}(a),h_{t_{1}}^{-1}(b)}(\widehat{t_{1}}\,)$ &
\multicolumn{1}{|l|}{$h_{t_{1}}$} & \multicolumn{1}{|l|}{$\{a,b\}\subseteq
\pi(t_{1})$}\\\hline
$relab_{h}(t_{1})$ & \multicolumn{1}{|l|}{$\widehat{t_{1}}$} &
\multicolumn{1}{|l|}{$h\circ h_{t_{1}}$} & \multicolumn{1}{|l|}{}\\\hline
\end{tabular}

\begin{center}
Table 5: Inductive construction of $\widehat{t}\in T(F_{\infty})$ and $h_{t}.$
\end{center}

\bigskip

It is clear that $\widehat{t}$\ and $h_{t}$ can be computed in polynomial time
from $t$.

\bigskip

\textit{Claim 1}: $\widehat{t}\in T(F_{k^{\prime}})$ for some $k^{\prime}\leq
k$, $\pi(\widehat{t}\,)=[\max\pi(\widehat{t}\,)]$, $\ h_{t}$ is a
bijection:$\pi(\widehat{t}\,)\rightarrow\pi(t)$ and $t\approx relab_{h_{t}%
}(\widehat{t}\,).$

\bigskip

\textit{Proof}: These facts are clear from the inductive construction and we
have $k^{\prime}=\max\mu(\widehat{t}\,)$. $\square$ \bigskip

\textit{Claim 2}: $\ \widehat{t}$ \ is irredundant.

\bigskip

\textit{Proof}: Because $t$ is assumed irredundant. $\square$

\bigskip

\textit{Claim 3}: $\ \widehat{t}$ \ is good. \bigskip

\textit{Proof}: Let $n$ be the number of vertices of $G(t)$, assumed to have
at least one edge. (The case of graphs without edges is easily treated
separately). The inductive construction shows that, for each subterm
$t^{\prime}$ of $\widehat{t}$, each label in $\pi(t^{\prime})$ labels some
vertex of $G(t^{\prime})$, hence $\max\mu(\widehat{t}\,)$ is at most the
number of vertices of $G(\widehat{t}\,)$, equal to $n$.

Again by induction, we can see that $\oplus$\ has $n-1$ occurrences
in$\ \widehat{t}$ (because$\ \widehat{t}$ has no occurrence of
$\boldsymbol{\varnothing}$ and $G(t)\simeq G(t^{\prime})$), and that the
symbols $relab_{h}$ have at most $2n-1$ occurrences (one can of course delete
those of the form $relab_{Id}$).

The number of operations $\overrightarrow{add}_{a,b}$ is at most
$(n-1).(k^{\prime2}-k^{\prime})$ because $\widehat{t}$ is irredundant by Claim
2.\ It follows that $|\widehat{t}\,|\leq n+n-1+2n-1+(n-1).(k^{\prime
2}-k^{\prime})\leq(k^{\prime}+1)^{2}.n+1$ as one checks by noting that
$k^{\prime}\geq2$ because $G(\widehat{t}\,)$ has edges.\ Hence $\widehat{t}$
is good. $\square$


\begin{thebibliography}{99}                                                                                               %


\bibitem {AHV}S.\ Abiteboul, R. Hull and V. Vianu, \emph{Foundations of
databases}. Addison-Wesley, 1995.

\bibitem {ALS}S. Arnborg, J. Lagergren and D.\ Seese, Easy problems for
tree-decomposable graphs, \emph{J. Algorithms} \textbf{12} (1991) 308-340.

\bibitem {AEIM}Y.\ Asahiro, H. Eto, T. Ito and E. Miyano, Complexity of
finding maximum regular induced subgraphs with prescribed degree,
\ \emph{Theoretical Computer Science}, \textbf{550} (2014) 21-35.

\bibitem {Bes}A.\ B\`{e}s, Expansions of MSO\ by cardinality relations,
\emph{Logical Methods in Computer Science} \textbf{9 }(4) (2013).

\bibitem {Bou}T. Bouvier, Graphes et d\'{e}compositions, Doctoral
dissertation, Bordeaux University, December 2014.

\bibitem {BGP}H.\ Broersma, P.\ Golovach and V.\ Patel, Tight complexity
bounds for FPT subgraph problems parameterized by the clique-width,
\emph{Theoretical Computer Science} \textbf{485} (2013) 69-84.

\bibitem {Cou86}B. Courcelle, Equivalences and transformations of regular
systems; applications to recursive program schemes and grammars,
\emph{Theoretical Computer Science} \textbf{42} (1986) 1-122.\ 

\bibitem {Cou96}B. Courcelle, The monadic second-order logic of graphs X:
Linear orders, \emph{Theoretical Computer Science} \textbf{160} (1996) 87-143.

\bibitem {Cou12}B. Courcelle, On the model-checking of monadic second-order
formulas with edge set quantifications, \emph{Discrete Applied Mathematics
}\textbf{160 }(2012) 866-887.

\bibitem {CouLAGOS}B. Courcelle, Fly-automata for checking monadic
second-order properties of graphs of bounded tree-width, \emph{Proceedings of
LAGOS\ 2015}, Beberibe, Brazil, to appear in \emph{Electronic Notes in
Discrete Mathematics}, 2015.

\bibitem {CouLAGOSa}B.\ Courcelle, Fly-automata for checking MSO$_{2}$ graphs
properties, full version of \cite{CouLAGOS}, 2015, http://arxiv.org/abs/1511.08605

\bibitem {BCID}B. Courcelle and I.\ Durand, \ Automata for the verification of
monadic second-order graph properties, \emph{J. Applied Logic} \textbf{10} (2012)\ 368-409.

\bibitem {BCID13}B. Courcelle and I.\ Durand, \ Model-checking by infinite
fly-automata, in \emph{Proceedings of 5-th Conference on Algebraic
Informatics} (CAI), \emph{Lec.\ Notes Comput.\ Sci.} \textbf{8080} (2013) \ 211-22.\ 

\bibitem {BCID13a}B. Courcelle and I.\ Durand, Infinite transducers on terms
denoting graphs, in \emph{Proceedings of the 6th European Lisp Symposium},
Madrid, June 2013.

\bibitem {FREC2014}B. Courcelle and I.\ Durand, Fly-automata, model-checking
and recognizability, Proceedings of the workshop Frontiers of Recognizability,
Marseille, 2014, http://arxiv.org/abs/1409.5368

\bibitem {CouEng}B. Courcelle and J. Engelfriet, \emph{Graph structure and
monadic second-order logic, a language theoretic approach}, Volume
\textbf{138} of \emph{Encyclopedia of mathematics and its application},
Cambridge University Press, 2012.

\bibitem {CMR}B. Courcelle, J. Makowsky and U. Rotics, Linear time solvable
optimization problems on graphs of bounded clique-width. \emph{Theory Comput.
Syst}. \textbf{33 }(2000) 125-150.

\bibitem {CouMos}B.\ Courcelle and M. Mosbah, Monadic second-order evaluations
on tree-decomposable graphs. \textit{Theor. Comput. Sci.} \textbf{109 }(1993) 49-82.

\bibitem {DF}\ R. Downey and M. Fellows, \textit{Parameterized complexity},
Springer, 1999.

\bibitem {DF2}R. Downey and M. Fellows, \textit{Fundamentals of parameterized
complexity}, Springer, 2013.

\bibitem {Dro}M. Droste, W.\ Kuich and H.\ Vogler (Eds.), \emph{Handbook of
weighted automata}, Springer, 2009.

\bibitem {Dur}I.\ Durand, Object enumeration, in \emph{Proc.\ of 5th Europeal
LISP Conference}, Zadar, Croatia, May 2012, pp.\ 43-57.

\bibitem {EKS}V.\ Engelmann, S. Kreutzer and S. Siebertz, First-order and
monadic second-order model-checking on ordered structures, in \textit{Proc. of
the 27th Symposium on Logic in Computer Science}, Dubrovnik, Croatia, 2012,
pp. 275-284.

\bibitem {Fell}M. Fellows, F. Rosamond, U. Rotics and S. Szeider, Clique-width
is NP-Complete. \emph{SIAM J. Discrete Math.} \textbf{23} (2009) 909-939.

\bibitem {Fell2}M. Fellows \emph{et al., }On the complexity of some colorful
problems parameterized by treewidth, \emph{Inf. Comput.} \textbf{209} (2011) 143-153.

\bibitem {FG}J. Flum and M. Grohe, \textit{Parameterized complexity theory},
Springer, 2006.

\bibitem {FK}E. Foustoucos and L. Kalantzi, The monadic second-order logic
evaluation problem on finite colored trees: a database-theoretic approach,
\emph{Fundam. Inform. }\textbf{92} (2009) 193-231.

\bibitem {FriGro}M.\ Frick and M.\ Grohe, The complexity of first-order and
monadic second-order logic revisited, \emph{Ann. Pure Appl. Logic}
\textbf{130} (2004) 3-31.

\bibitem {GHO}R. Ganian, P. Hlinen\'{y} and J. Obdrz\'{a}lek, A unified
approach to polynomial algorithms on graphs of bounded (bi-)rank-width,
\emph{Eur. J. Comb.} \textbf{34} (2013) 680-701.

\bibitem {GPW}G. Gottlob, R. Pichler and F. Wei, Monadic datalog over finite
structures of bounded treewidth. \emph{ACM Trans. Comput. Log.} \textbf{12 }(2010).

\bibitem {GroKre}M. Grohe and S. Kreutzer, Methods for algorithmic
meta-theorems, \ in \emph{Model theoretic methods in finite combinatorics},
\emph{Contemporary Mathematics,} \textbf{588}, American Mathematical Society, 2011.

\bibitem {KanRao}M. Kant\'{e} and M. Rao, The rank-width of edge-coloured
graphs, \emph{Theory of Computing Systems} \textbf{52} (2013) 599-644.

\bibitem {KLR}J. Kneis, A. Langer and P. Rossmanith, Courcelle's theorem - a
game-theoretic approach. \emph{Discrete Optimization} \textbf{8 }(2011) 568-594.

\bibitem {Kre}S. Kreutzer, Algorithmic meta-theorems, in \emph{Finite and
algorithmic model theory}, Cambridge University Press, 2011.

\bibitem {LKM}M. Lampis, G. Kaouri and V. Mitsou, On the algorithmic
effectiveness of digraph decompositions and complexity measures,
\emph{Discrete Optimization} \textbf{8} (2011) 129-138.

\bibitem {Lan}A.\ Langer, F. Reidl, P. Rossmanith and S. Sikdar, Practical
algorithms for MSO model-checking on tree-decomposable graphs, \emph{Computer
Science Review} \textbf{13-14} (2014) 39-74.

\bibitem {Rao}M.Rao, MSOL partitioning problems on graphs of bounded treewidth
and clique-width, \emph{Theor. Comput. Sci.} \textbf{377} (2007) 260-267.

\bibitem {Rei}K.\ Reinhardt, The complexity of translating logic to finite
automata, in \textit{Automata, logics, and infinite games: A guide to current
research}, E.\ Graedel \emph{et al. }eds., \textit{Lecture Notes in Computer
Science} \textbf{2500} (2002) 231-238.

\bibitem {Rev}D.\ Revuz, Minimisation of acyclic deterministic automata,
\textit{Theoret.\ Comput.\ Sci. }\textbf{92} (1992) 181-189.

\bibitem {Sei}H.\ Seidl, Finite tree automata with cost functions,
\textit{Theoret.\ Comput.\ Sci.} \textbf{126} (1994) 113-142.

\bibitem {VeaBjo}M. Veanes, N. Bjorner, Symbolic automata: the toolkit, in
\emph{Proceedings of TACAS 2012, Lec. Notes Comput. Sci.} \textbf{7214} (2012) 472-477.
\end{thebibliography}
\end{document}